\documentclass[a4paper,12pt]{article}

\usepackage[T1]{fontenc}

\usepackage[a4paper,top=2cm,bottom=2cm,left=2cm,right=2cm,marginparwidth=1cm]{geometry}

\usepackage{amsmath,amssymb,amscd}

\usepackage{graphicx}

\usepackage{color}

\usepackage{natbib}

\providecommand{\citep}[1]{\cite{#1}}
\providecommand{\citet}[1]{\cite{#1}}

\usepackage{hyperref}
\providecommand{\href}[2]{\texttt{#2}}
\providecommand{\url}[1]{\texttt{#1}}

\newcommand{\tref}[1]{(\ref{#1})}

\newcommand{\fref}[1]{Fig.~\ref{#1}}
\newcommand{\tableref}[1]{Table~\ref{#1}}
\newcommand{\sref}[1]{Section~\ref{#1}}

\newcommand{\pvalue}[1]{\textcolor{blue}{#1}}
\newcommand{\svalue}[1]{\textcolor{red}{\textit{#1}}}

\newcommand{\Amat}{\mathbf{\textsf{A}}}
\newcommand{\Gmat}{\mathbf{\textsf{G}}}
\newcommand{\Tmat}{\mathbf{\textsf{T}}}
\newcommand{\Ccal}{\mathcal{C}}

\newcommand{\tnote}[1]{}
\newcommand{\bnote}[1]{}
\newcommand{\tcomment}[1]{}

\newcommand{\blindreviewonly}[1]{#1}
\newcommand{\normalversiononly}[1]{}
\newcommand{\blindreview}[2]{#1}

\ifx\blindreviewon\undefined
    \renewcommand{\blindreviewonly}[1]{}
    \renewcommand{\normalversiononly}[1]{#1}
    \renewcommand{\blindreview}[2]{#2}
\fi

\begin{document}

%
%
%


\begin{center}
{\Large\textbf{Analysis of the Wikipedia Network of Mathematicians}}
\end{center}
\normalversiononly{
 \begin{center}
  \large {Bingsheng Chen},  {Zhengyu Lin},
  \href{http://www.imperial.ac.uk/people/t.evans}{Tim S.\ Evans}
 \\[0.5cm]
\href{http://complexity.org.uk/}{Centre for Complexity Science}, and \href{http://www3.imperial.ac.uk/theoreticalphysics}{Theoretical Physics Group},
\\
Imperial College London, SW7 2AZ, U.K.
\\
20th December 2018
\end{center}
}
\begin{abstract}
We look at the network of mathematicians defined by the hyperlinks between their biographies on Wikipedia. We show how to extract this information using three snapshots of the Wikipedia data, taken in 2013, 2017 and 2018.   We illustrate how such Wikipedia data can be used by performing a centrality analysis.  These measures show that Hilbert and Newton are the most important mathematicians.  We use our example to illustrate the strengths and weakness of centrality measures and to show how to provide estimates of the robustness of centrality measurements. In part, we do this by comparison to results from two other sources: an earlier study of biographies on the MacTutor website and a small informal survey of the opinion of mathematics and physics students at \blindreview{(name removed for double blind review) University.}{Imperial College London.}
\end{abstract}

\begin{center}
\end{center}

%

\normalversiononly{
\section*{Highlights}

\begin{itemize}
\item We identify the most important Mathematicians as Hilbert and Newton.
\item We show how to estimate uncertainty in social network measurements using different sources.
\item We use a simple model of noise to test the robustness of our network measurements.
\item We use a survey of students to compare against social network results.
\item Our results show how large scale crowdsourced information can provide useful insights into social science questions.
\end{itemize}


}

%
%
%

\section{Introduction}

The history of mathematics shows how mankind has passed ideas between eras and cultures \citep{BG04,S12e}. It illustrates how research in humanities topics is usually pieced together by experts using qualitative techniques.  However the arrival of the ability to record and analyse large data sets has opened up new approaches for research in humanities which can complement and support existing methods.  In this paper we look at one particular example, the way that Wikipedia can be used to leverage information about the relationships between mathematicians.

Wikipedia is a large set of web pages maintained by the crowd.  That is anyone may edit existing pages or add a new one.  There is some hierarchy, with some editors having more control over some protected pages, but largely quality assurance is intended to emerge through the consensus of the crowd. Not surprisingly, Wikipedia covers a vast range of topics.  As a result of this coverage and the fact that the data is open access, easily available and readily accessible, the information in Wikipedia has been mined in many projects. This includes several which look at biographies of individuals as we do, for example see  \citet{AKLV12,GMS15,EALKVS15,EL16,JKT16}.  Our work focusses on biographies of a specific profession, mathematics.

In this paper we ask a number of questions:-
\begin{itemize}
\item Can crowdsourcing produce a useful list of individuals in one field?
\item How can we use hyperlinks in Wikipedia biographies to produce a useful network?
\item Can links between biographies of individuals in one field be used to produce information on that field?
\item Specifically, can these links reveal something about the importance of individuals in mathematics?
\item How can we measure the uncertainties in network centrality measurements?
\end{itemize}

For the first question, we use Wikipedia's list of ``mathematicians'' to show how such crowd sourced lists can be effective.

From such a list, we will then show how the data on mathematicians can be extracted from Wikipedia and how a simple network is a useful representation of this data. To show such that the relationships between individuals encoded in this network is useful, we will measure the importance of mathematicians through network centrality measures. Centrality measures\footnote{David Schoch has produced a \href{http://schochastics.net/sna/periodic.html}{Periodic Table of Network Centrality} \citep{S16} which is a nice visualisation, classification and summary of the range of network centrality measures available. The small graphs in \citet{BH14} highlight properties of the more commonly used centrality measures.} are widely used to answer this type of question, for an overview see \citet{NMB05,N09b,BH14,S16,SVB17}.

We also place a great emphasis on the robustness of our results so another important element of our work is to show how we quantify the uncertainty in our measurements. 

So we will ultimately provide an answer to who is the most important person in mathematics.
Our positive results will serve to support our assumption that the hyperlinks in Wikipedia biographies contain information on the importance of the web sites they point to.

The Wikipedia data used in this paper, along with the code and our results of processing this data, is available online \blindreview{(citation removed for blind review).}{\citep{ECL17}.}

\section{Data and Methods}

The English language Wikipedia was the primary source for this project since it contains a large number of biographies of Mathematicians and a list of these biographies. It was also used because the data is open source, easily accessible and there is good supporting technical documentation.  Our work was primarily with Wikipedia data extracted in 2017 but we also have data from 2013 and 2018 available for comparison. Our data was extracted from the English Language Wikipedia, processed into a network, and finally analysed using various Python packages including NetworkX \citep{HSS08}.

We also used a results from an earlier project \blindreview{(citations to unpublished theses removed for double blind review)}{\citep{C11,H11b}} which analysed a second web site of biographies of mathematicians, the \href{http://www-history.mcs.st-and.ac.uk/}{MacTutor History of Mathematics archive} created by John J O'Connor
and Edmund F Robertson rather than being crowd sourced. We will use results from this earlier analysis of the MacTutor data to make comparisons with our Wikipedia results in \sref{sMacTutor}. The basic methods used \blindreview{(citations to unpublished theses removed for double blind review)}{by \citet{C11} and \citet{H11b}} to produce and analyse a network derived from the MacTutor biographies are very similar to those we used for our data on the Wikipedia biographies.

Finally, in \sref{sdatasurvey} we describe our informal survey of undergraduate students in the Mathematics and Physics departments at \blindreview{(name removed for double blind review) University}{Imperial College London}. This provided a fourth set of results and allowed further comparisons to be made.

\subsection{Extracting the Biographies of Mathematicians}

To start our analysis we must first define what we mean by a mathematician. We took the list of mathematicians on English Wikipedia as our fundamental definition of who is a ``mathematician''.
We started from twenty-six catalogue web pages on English Wikipedia, each of which lists all the biographies of mathematicians on the English language Wikipedia
whose family name starts with the same specified letter.  For instance the ``\href{https://en.wikipedia.org/wiki/List_of_mathematicians_(A)}{List of mathematicians (A)}'' page contains a link to \href{https://en.wikipedia.org/wiki/Odd_Aalen}{``Aalen, Odd''}. These twenty six index pages provided a list of \href{https://en.wikipedia.org/wiki/URL}{URLs} (Universal Resource Locator), here the addresses of English language Wikipedia biographies of individual mathematicians\footnote{Note in interpreting the 26 catalogue web pages, we used the position on the page to indicate which hyperlinks were to biographies of Mathematicians.  We did not use other links on these catalogue pages e.g.\ links to the Wikipedia home page.}. See \citet{EL16} for alternative approaches to finding sets of biographies based on the profession of individuals.

The vast majority of these Wikipedia `mathematician' pages are indeed biographies of individuals. Whether or not an expert would call them a  mathematician is, in some cases, debatable.  For instance many would classify \href{https://en.wikipedia.org/wiki/Kristen_Nygaard}{Kristen Nygaard} as a computer scientist or politician rather than a mathematician, yet he appears on Wikipedia's list of mathematicians.

There are also one or two pages in our list which are not dedicated to a single person.  For instance, the \href{https://en.wikipedia.org/wiki/Nicolas_Bourbaki}{Nicolas Bourbaki} page is for work produced by a variety of mainly French 20th-century mathematicians under a single pseudonym, while the individual contributions of the three \href{http://www-history.mcs.st-andrews.ac.uk/Biographies/Banu_Musa.html}{Ban\={u} M\={u}s\={a}} brothers is often difficult to distinguish so their work is often referred to as if it was authored by a single person.
Another example is the way that the \href{https://en.wikipedia.org/wiki/Noether_Lecture}{Noether Lecture} is listed as an individual mathematician under ``Lecture, Noether'' in the 2017 data but not in our 2013 data. However we did not find any other examples of problematic pages.

We choose to leave the data unchanged and we treat all pages in the crowdsourced list as if each was the biography of an individual mathematician. We aim to see if ``the wisdom of the crowd'' can, without further intervention, provide useful information on mathematicians whatever the strengths and weaknesses are of this crowdsourced list.

Given our list of the URLs of all the biographies of mathematicians on Wikipedia, each page was exported to an XML source file, which we provide elsewhere \blindreview{(citation removed for double blind review).}{\citep{ECL17}.}  From there, the hyperlinks between these biographical web pages were found. There is much more information in these biographies and in the XML source files we provide, for instance hyperlinks to relevant topics and in the text itself, but we did not use any additional information.

\subsection{Definition of the Network}

Each Wikipedia page, each mathematician, was represented by a unique vertex in our graph. We then add an unweighted, directed edge between a pair of vertices if there is at least one hyperlink in either direction between the two corresponding Wikipedia pages. Our hypothesis is that the hyperlinks between these biographies of mathematicians capture relationships between the academic work of these mathematicians and so these links reflect the way mathematics has developed. In particular an important assumption we make is that these links carry information about the importance of the mathematicians.

Our approach raises several important issues regarding the accuracy of, and interpretation of, our network. As our bibliographies are crowdsourced, we have little measure of the quality or accuracy of the links in the bibliography though there has been much discussion of this issue elsewhere; for examples see \citet{G05}, responses to that article, \citet{WCR13}, and references therein.    Our own impression is that the quality of our biographies is generally high \citep{C06,WH07}.
We are less sure about the range of mathematicians covered. It would be natural if some editors focus on a particular area, one university or one mathematical  field, which would produce an over representation or over emphasis of some lesser known mathematicians.  Perhaps the English language Wikipedia emphasises a Western viewpoint of the history of mathematics, and so we might have an under representation of mathematicians from certain backgrounds; for a discussion of gender issues and Wikipedia see \citet{WGJS15} while \citet{EALKVS15} is an example of a discussion of culture and Wikipedia.

Even if the biographies are accurate, our method may under represent links associated with a mathematician. For instance, rather than linking to another mathematician, a Wikipedia page may refer to a web page dedicated to a method or technique. For example, \href{https://en.wikipedia.org/wiki/Vladimir_Arnold}{Vladimir Arnold} solved Hilbert's thirteenth problem, but the Wikipedia biography of Arnold has no links to \href{https://en.wikipedia.org/wiki/David_Hilbert}{Hilbert}, it only links to a page dedicated to \href{https://en.wikipedia.org/wiki/Hilbert\%27s_thirteenth_problem}{Hilbert's thirteenth problem}. We could probe the whole of Wikipedia, say using the length of the shortest path between our mathematicians to measure strength of interaction, but as a first approximation we will assume this issue effects all mathematicians proportionally\footnote{That is we assume the more famous mathematicians are more likely to be effected but this effect produces a similar fractional decrease in the number of their edges.} and we will ignore it.

Another assumption we make is that our graph is unweighted so that all relationships between mathematicians are equally strong.  In reality, the influences between mathematicians are not equivalent and are hard to determine.  The biographies also provide personal rather than professional relationships between mathematicians. For example, \href{https://en.wikipedia.org/wiki/Isaac_Newton}{Issac Newton's} Wikipedia page mentions that \href{https://en.wikipedia.org/wiki/Charles_Hutton}{Charles Hutton} commented that his belief of Newton died as a virgin which does not of itself indicate a direct link between the work of the two mathematicians. It might be possible to perform an assessment of the nature of a hyperlink based on the text surrounding links, this this would either be too slow to do by hand, or would require sophisticated numerical tools beyond the scope of this paper. Instead, we wish to see how far we can go with simpler tools when the data is provided on a large scale.

One way to put some measure of the strength of a relationship could be to count the number of hyperlinks from one biography to another. Our feeling is that this may be a function of writing style and so may not be a useful measure. For instance  \href{https://en.wikipedia.org/wiki/Vladimir_Arnold}{Arnold's} Wikipedia page has five references to \href{https://en.wikipedia.org/wiki/Hilbert\%27s_thirteenth_problem}{Hilbert's thirteenth problem} and different writers could easily have given different number of links to \href{https://en.wikipedia.org/wiki/David_Hilbert}{Hilbert's} biography alongside the links to the Wikipedia page on the problem.

We have ignored the direction of the hyperlinks as it is not clear what meaning the direction might convey. Although the work of a later mathematician cannot influence the work of earlier researchers, the Wikipedia biographies can have hyperlinks in either direction with respect to time. For instance there exist hyperlinks between \href{https://en.wikipedia.org/wiki/Galileo_Galilei}{Galileo} (died 1642) and \href{https://en.wikipedia.org/wiki/Isaac_Newton}{Newton} (born 1642) on both pages. In addition, while dates of birth and death can indicate the direction of influence, most of the mathematicians in our data have overlapping lifetimes.

Finally pages may have internal references but these provided no useful meaning for us and were ignored.  This gave us no self-loops and so our network was a simple graph.

While there are many uncertainties surrounding the meaning of the relationships between mathematicians encoded in our network, the fundamental idea is that by using such a large number of mathematicians and links, the patterns we find on larger scales should capture genuine information about relationships between the academic work of these mathematicians.

One way to confirm that our network makes sense is to use our results to make comparisons with other studies. Within this paper we produce three networks, each based on a snapshot of Wikipedia taken in either 2013, 2017 or 2018.  These will are highly correlated but the variations will in part be to the uncertainties over various links which lead to changes by editors in the pages over four or five years. We will also refer to results from earlier unpublished studies \blindreview{(citation of unpublished theses removed for double blind review)}{\citep{C11,H11b}} based on another set of web based biographies, those derived from the \href{http://www-history.mcs.st-and.ac.uk/}{MacTutor} web site \citep{MacTutor}.  A final comparison will be made with an informal survey of students which we organised.

The other method we use to check the robustness of our results is to provide a simple model of noise in our data allowing us to see explicitly how sensitive our quantitative results are under this model. We will now move on to define our model of noise.

\subsection{Noise Model}\label{snoise}

As noted above, some edges in our network may be incorrect, perhaps because of a lack of expertise on behalf of some editors\footnote{Wikipedia is a website that allows users to edit its contents if the content is not protected. For all the mathematician pages used here, the highest level of protection is semi-protected, which still allow users to edit the page.}  or simply because of historical uncertainty. It is hard to determine the validity of over ten thousand edges but we expect that such a large number of relationships will ensure our results and conclusions are robust. To demonstrate this, we developed a simple model of the noise in the network in order to judge the uncertainty in our results.

We will simulate the process of editing a Wikipedia page as one of edge rewiring. We will remove a fraction $p$ of the edges, representing the decision by some editor that these were poor relationships. We will then assume that over the same period, editors will add roughly the same number of new hyperlinks to biographies.
Furthermore, we will also assume that the editors will be more likely to connect to biographies with many connections, so the two vertices connected by a new edge are chosen in proportion to the original degree, the degree of the vertex before any edges were changed $k_\mathrm{orig}$.

To a good approximation\footnote{Ignoring the effects such as the correlation between vertices connected by edges and the constraints of being a simple graph.}, the process of removing edges from a vertex starting with degree $k_\mathrm{orig}$ is a binomial with $k_\mathrm{orig}$ trials and a mean of $(1-p)k_\mathrm{orig}$.  Likewise and the process of adding edges back to this vertex is also roughly binomial with  $(2E)p$ trials and an expectation value of $(2E)p.k_\mathrm{orig}/(2E)$. For a vertex starting with degree $k_\mathrm{orig}$, the new degree, $k_\mathrm{new}$, is on average the same
\begin{equation}
 \langle k_\mathrm{new}  \rangle = (1-p) k_\mathrm{orig} +  p (2E)\cdot \frac{k_\mathrm{orig}}{2E}  =   k_\mathrm{orig}
 \, .
\end{equation}
The degree of a vertex will fluctuate in our model with a variance $\sigma(k_\mathrm{orig})$ given approximately by
\begin{equation}
 \sigma^{2}(k_\mathrm{orig})
  =  p (1-p) k_\mathrm{orig} + p 2E \frac{k_\mathrm{orig}}{2E} \left(1-\frac{k_\mathrm{orig}}{2E} \right)
  = p k_\mathrm{orig} \left( 2 - p - \frac{k_\mathrm{orig}}{2E} \right)
\end{equation}

Edges will be changed by editors for many reasons but without further information, we will use the variation in edges between given mathematicians between our 2013 and 2017 datasets to motivate our choice for $p$, the level of noise in our model.
We find that that for the same pair of vertices, $90.8\%$ of the edges in the 2013 data set were also found in the 2017 dataset.
Therefore, we will choose $p=0.1$ and use models where $10\%$ of edges have been rewired to estimate the level of uncertainty in our results. While the average degree of each vertex will be equal to the original degree, these parameter values give the standard deviation of the degree of a vertex to be  around $0.44 \sqrt{k_\mathrm{orig}}$ which is compatible with the numerical results shown in \fref{degstd} in the Appendix.


\subsection{General Ranking scheme for each individual centrality measure}

There is no perfect way to define a ranking scheme in any context, in part because there is no perfect way to combine several different ratings into one single score \citep{LM12}. In our case the different centrality measures are all of different numerical scales, some of which depend on normalisations in their definitions which are irrelevant constants in our context.  For this reason, and to simplify the presentation of our results, we chose to put all our measures on a scale from 0 to 100 using a simple linear rescaling, namely
\begin{align}
\mbox{Rescaled Centrality Measure}= \frac{\mbox{Original Centrality Measure}}{\mbox{Largest Original Centrality Measure Value}} \times 100
\end{align}
\begin{equation}
 C^\prime_v = \frac{C_v}{C_\mathrm{max}} \times 100 \, , \qquad
 C_\mathrm{max} = \max ( C_v | v \in \mathcal{V}) \, ,
 \label{rescale}
\end{equation}
where $C_i$ is the original centrality measure for mathematician $i$, and the $C_\mathrm{max}$ is the largest of those values.  We do this separately for each of the centrality measures we consider and all our results will be expressed in terms of these rescaled measures. For each measure, this linear rescaling preserves the order of mathematicians as defined by that measure, but it also preserves the relative differences in the centrality scores of mathematicians.


\subsection{Informal Survey}\label{sdatasurvey}

The final source of information on the importance of mathematicians comes from a very different source. We carried out an informal survey of undergraduate mathematics and physics students at \blindreview{(name removed for double blind review) University.}{Imperial College London.} The survey contained two compulsory questions: one was the current year of student, the second asked for their top three mathematicians.  Participants were given a list of the top twenty mathematicians obtained from our social network analysis. At the same time, participants could nominate different mathematicians if they were not on the list provided. We divided the sample by year to see if increasing mathematical knowledge at University had a noticeable effect on the outcome. The survey was sent via email and the information was gathered using an online form.

\section{Results \& Discussion}

\subsection{Basic Network Parameters}

In this project, we applied the method described above to two sets of Wikipedia data; first based on pages downloaded on 13th November 2013, the second set taken on 20th June 2017 and finally the last set taken on 22nd September 2018.

The resulting network for the 2013 Wikipedia data gave us $6050$ vertices/mathematicians. These biographies provided a total of $15120$ hyperlinks which led to $9701$ undirected unweighted edges in our graph. The largest connected component contained $4096$ ($67.7\%$) of the mathematicians with $9573$ undirected edges between the mathematicians in the largest component. Of the $1954$ ($32.3\%$) mathematicians outside the largest connected component, almost all isolated from each other though they typically have more links to other non-mathematician Wikipedia pages.  For instance the second largest connected component contains just five mathematicians in total\footnote{Five Norwegian statisticians make up the second largest connected component. They are \href{https://en.wikipedia.org/wiki/Erling_Sverdrup}{Erling Sverdrup} plus four recipients of a prize named after Sverdrup: Dag Tj{\o}stheim, Tore Schweder, Nils Lid Hjort, and Odd Aalen. Again this illustrates how the links on the Wikipedia page may not indicate any direct mathematical connection.  However if later mathematicians are inspired or enabled by such a prize, perhaps links such as these are just as a useful measure of esteem, an indication of the influence and legacy of one mathematician, as any other type of link.}

The data set downloaded in 2017 was around a third bigger in terms of the number of Wikipedia pages and links.  Likewise the largest component had about 30\% more edges and vertices.  However despite this large change in the scale, many other properties showed very small change between 2013 and 2017 as shown in \tableref{para}: the largest component still contained just over two thirds of the nodes and the average degree, both overall and in the largest component, grew by a few percent.
The network built from the 2018 Wikipedia data showed further rises in the number of nodes (mathematicians) and edges (hyperlinks) over previous years but the growth was broadly comparable.


The lack of change in the average degree of the largest component prompted us to use our simple model for noise as described in section \sref{snoise}. This keeps the number of nodes and edges the same\footnote{The small change in the number of edges is due to the creation of a few self-loops which were eliminated.  This effect was very small and the computational implementation was not corrected to eliminate this feature.} while the degree of each fluctuates by about 5\% for the nodes of largest degree. The mean degree of each node over the 1000 sample networks is roughly equal to the that in the 2017 data. Our noise model does not keep other features of the data and we see small differences between 2017 data and those produced by our noise model in some of the other measures such as average path length.

\begin{table}[h]
\centering
\begin{tabular}{r||c|cc|c}
Quantity                & 2013    & 2017    & \%       & 2017                  \\
                        &         &         & Increase & After Rewiring        \\ \hline \hline
Mathematicians/Vertices & $6050$  & $7677$  & +26.9\%  &    $7677$             \\ \hline
Hyperlinks              & $15120$ & $20247$ & +33.9\%  &   $20247$             \\ \hline
Undirected Edges        & $9701$  & $12796$ & +31.9\%  &   $12789.9 \pm 2.5$ \\ \hline
Average Degree          & $3.21$  & $3.33$  &  +3.7\%  &    $3.33$             \\ \hline
Vertices in LCC         & $4096$  & $5325$  & +30.0\%  &  $5212.2 \pm 17.3$    \\ \hline
Edges in LCC            & $9573$  & $12627$ & +31.9\%  & $12656.0 \pm 9.6$     \\ \hline
Average Degree in LCC   & $4.71$  & $4.74$  &  +0.6\%  &    $4.86 \pm 0.01$    \\ \hline
Network Diameter        & $13$    & $14$    &  +7.7\%  &    $13.9 \pm 0.9$     \\ \hline
Average Path Length     &  $5.07$ & $5.14$  &  +1.4\%  &    $4.89 \pm 0.01$    \\ \hline
Clustering Coefficient  &  $0.13$ & $0.12$  &  -7.7\%  &    $0.09 \pm 0.002$   \\
\end{tabular}
\caption{Network parameters for the 2013 and 2017 dataset, the percentage change between 2013 and 2017 data, and the mean values found for an ensemble of 1000 rewired 2017 data sets (with one standard deviation uncertainty quoted) as defined by our noise model of \sref{snoise} with $p=0.1$. The LCC is the largest connected component.}
\label{para}
\end{table}



\subsection{Degree}\label{ssdegree}

\begin{figure}[htb!]
\centering
\includegraphics[width=0.75\textwidth]{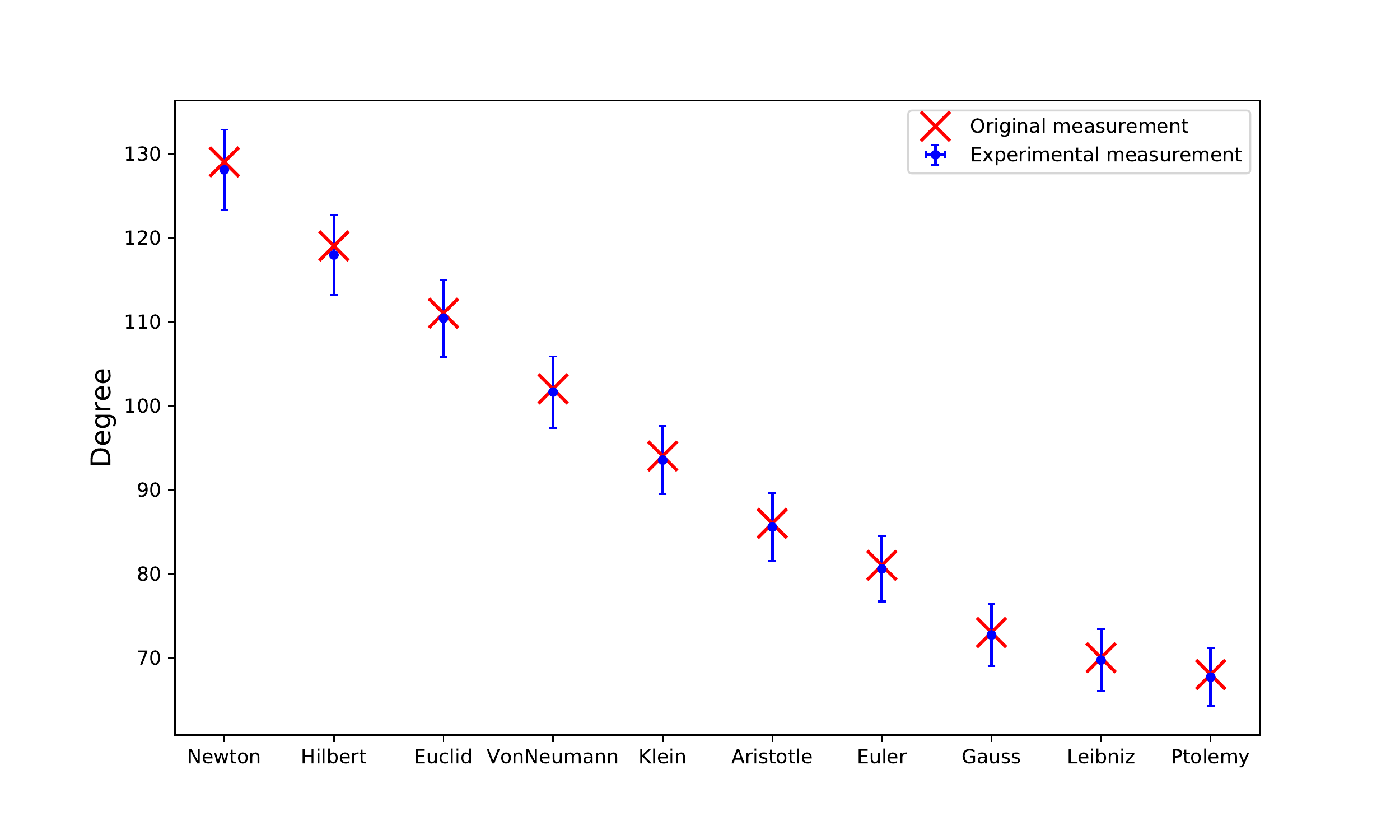}
\caption{The variation in degree under the noise model for the ten mathematicians whose Wikipedia biographies have the largest degree in the 2017 data (crosses).  The circles give the mean degree for the same mathematicians as measured over 1000 simulations using our noise model of \sref{snoise} where the error bars are specified by one standard deviation.}
\label{zipf}
\end{figure}

The degree of a node is the number of edges connected to that node.  As a crude measure of importance, the more biographies which are connected Newton's biography, the more likely it is that Newton's work played an important role in either developing existing ideas or in laying the foundations for later work. The results for our measurements of the degree for the ten mathematicians in 2017 with largest degree, showing the uncertainty estimates from our noise model, are shown in \fref{zipf} (see \fref{zipfv2} for 2018 results).

\begin{figure}[htb!]
\centering
\includegraphics[width=0.45\textwidth]{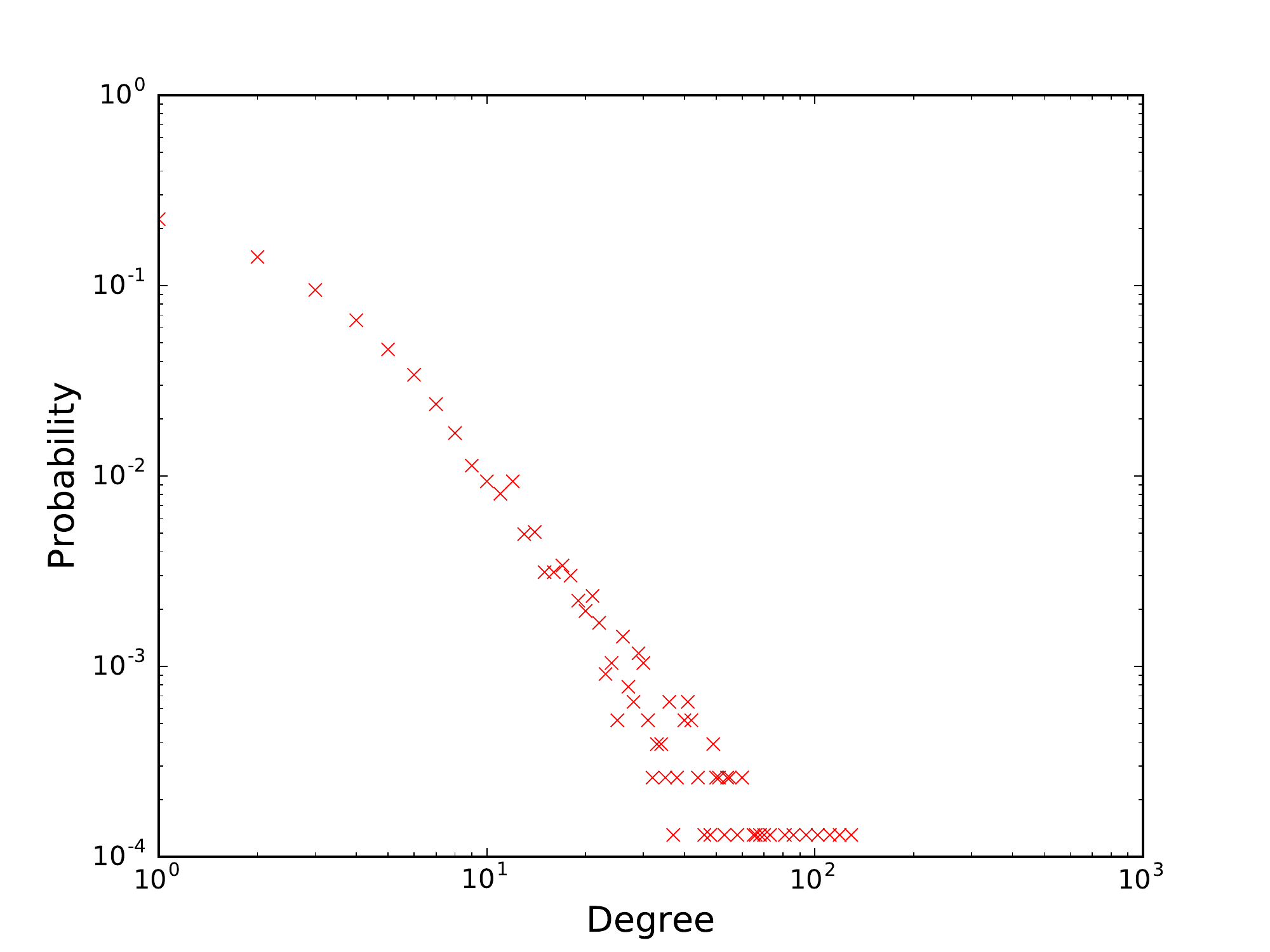}
\vspace*{0.05\textwidth}
\includegraphics[width=0.45\textwidth]{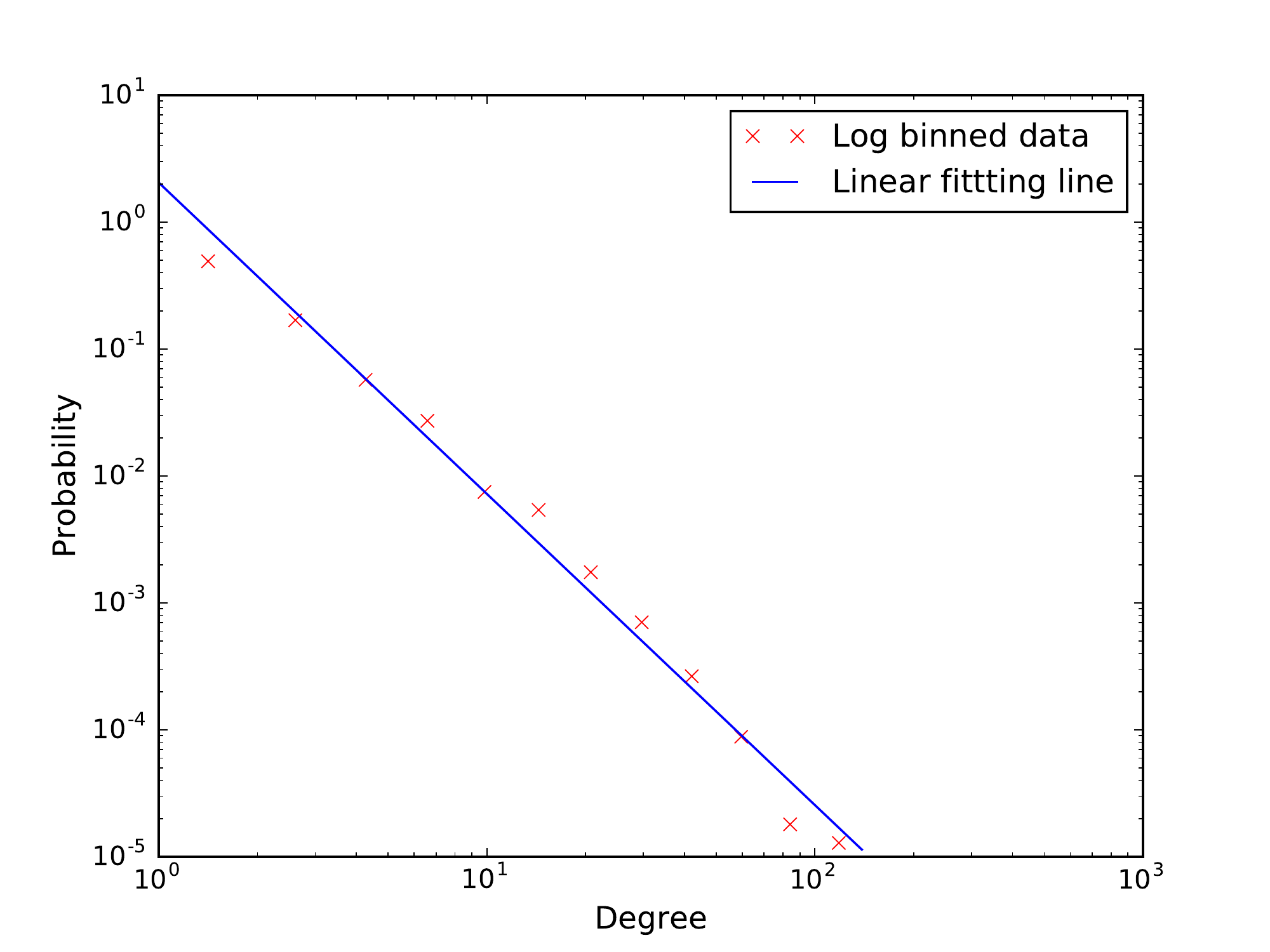}
\caption{On the left, the degree distribution for the 2017 network of mathematicians.  The same data binned with logarithmic bins (using log binning with the ratio of consecutive bin edges set to be 1.5) is shown on the right and a best fit straight line to this data has been added (slope is $-2.5 \pm 0.1$).}
\label{degreedis}
\end{figure}

It is worth noting that, as expected from analysis of many websites, the degree distribution of our network of mathematicians has a fat tail as shown for 2017 in \fref{degreedis} (for 2018 see \fref{degreedisv2}).


\begin{figure}[htb!]
\centering
\includegraphics[width=0.6\textwidth]{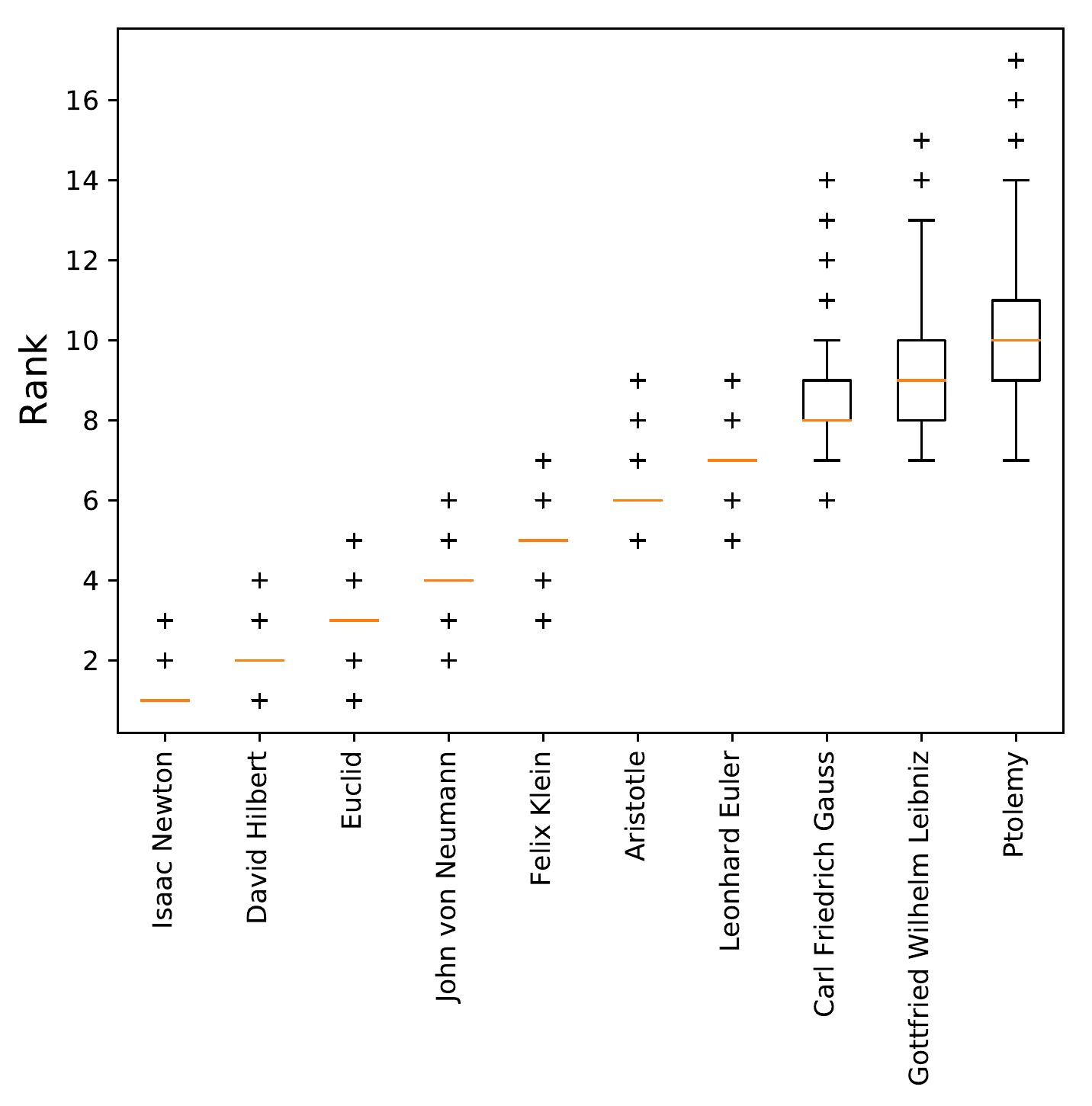}
\caption{Whisper box plot for degree rank of mathematicians from 1000 simulations of our noise model from \sref{snoise} applied to the 2017 data.
The lower and upper edges of blue box show the $25$ percentile ($Q_1$) and the $75$ percentile ($Q_3$) of the rank of each mathematician, the red line in the middle of the box is the median. Given the small variation here, these lines often coincide. The black lines, at the end of the whiskers connected to the box, are defined to be at $Q_1 - 1.5 (Q_3-Q_1)$ and  $Q_3 + 1.5 (Q_3-Q_1)$. The remaining black crosses beyond the whiskers indicate outliers beyond the whiskers.}
\label{degreebox}
\end{figure}

The robustness of the ranking of mathematicians by their degree in the 2017 network is essential if we are to judge how important the differences in their degree ratings. To do this we compare our 2017 data against the results of 1000 simulations using our noise model of \sref{snoise}. The fluctuations, spread, skewness and outliers of ranks in simulations can be visualized in a box-and-whisker plot\footnote{The integer nature of degree and the small variation in rank values for the degree of the top ten mathematicians means that in most cases features of the Whisker and Box plot of degree coincide.  However we will use the same definition for the whiskers and box in later plots where this type of visualisation is more useful.} in \fref{degreebox} (for equivalent results for 2018 data see \fref{degreeboxv2}).

\clearpage



\subsection{Closeness}\label{sscloseness}

Closeness of a node is the inverse of the average shortest path length from that node to all other nodes \citep{B50,HSS08,N09b} (see equation \tref{close} in Appendix for the formal definition used here). We will only use the largest component in our work with closeness.
Unlike the degree, this centrality measure probes the whole structure of the network, though it does so assuming that the only important routes are the shortest paths. The idea is that that mathematician with the largest centrality has the smallest average path length and so will, on average, be the closest to any other mathematician.  If two mathematicians are close then the likelihood is that the work of the two mathematicians is strongly interrelated or interdependent.

\begin{figure}[htb!]
\centering
\includegraphics[width=0.6\textwidth]{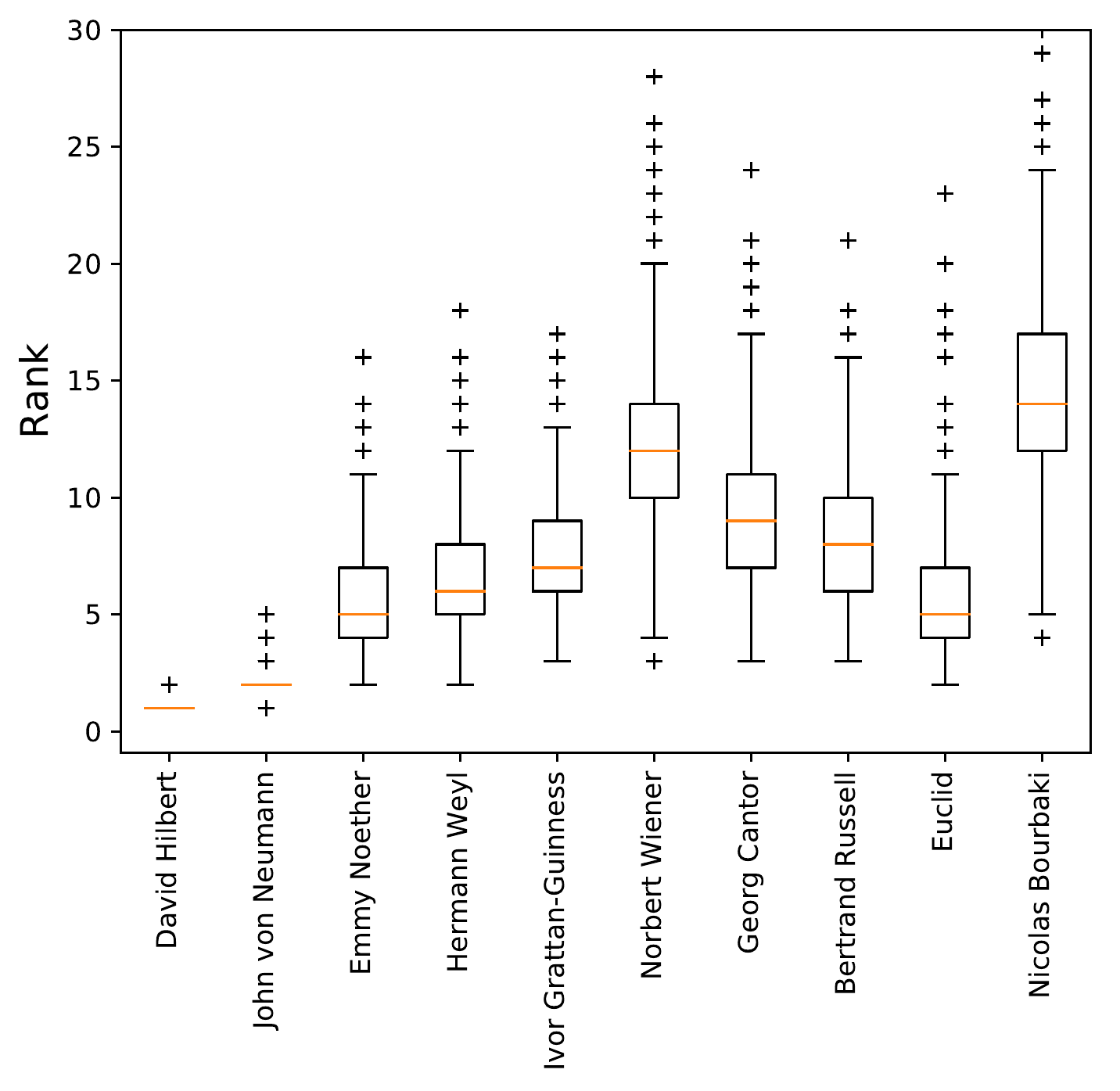}
\caption{Whisper box plot for the rank of mathematicians by closeness, for the ten mathematicians with largest closeness.  The closeness centrality is calculated for the largest component of the 2017 data and the uncertainties are estimated using 1000 simulations using the noise model of \sref{snoise} with $p=0.1$. The criteria used to place the boxes and other features of the plot are as in \fref{degreebox}.}
\label{clobox}
\end{figure}

The robustness of the closeness values was again estimated using our noise model and results for the top ten mathematicians in 2017 are shown in \fref{clobox} (see \fref{cloboxv2} for 2018 results).

\subsection{Betweenness}\label{ssbetweenness}

Betweenness centrality, like closeness, uses the length of the shortest path between nodes to try to measure importance. Betweenness of a node $v$ is the number of shortest paths which pass through that node, summing over the shortest paths between all possible pairs of distinct nodes $s$ and $t$ \citep{F77,B08d,HSS08,N09b}. See equation \eqref{bet} in the Appendix for a formal definition.

\begin{figure}[htb!]
\centering
\includegraphics[width=0.6\textwidth]{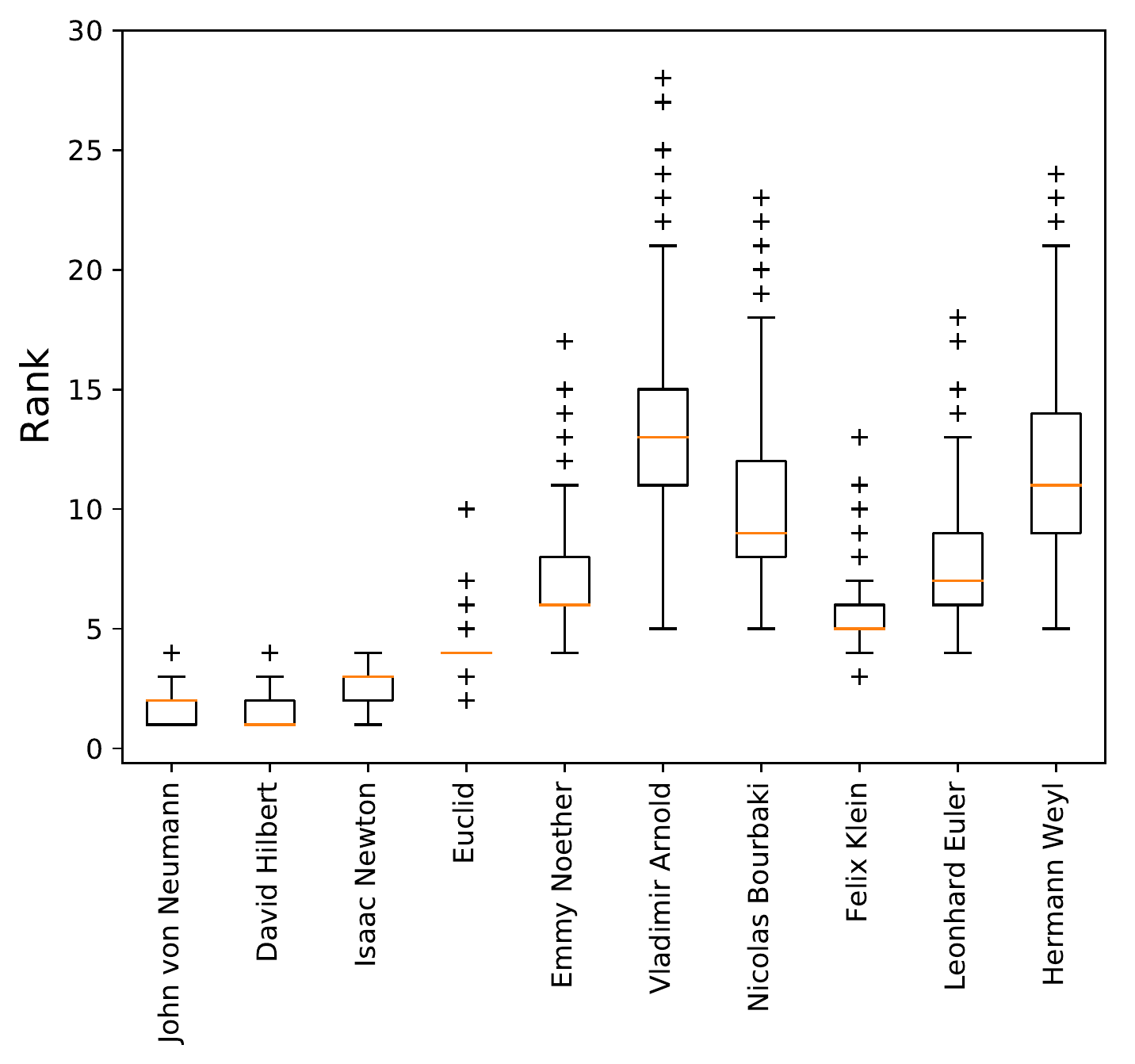}
\caption{Whisper box plot for rank by betweenness of the ten mathematicians with highest betweenness.  This is for the largest component of the 2017 data based on 1000 simulations using the noise model of \sref{snoise}.
The criteria used to place the boxes and other features of the plot are as in \fref{degreebox}.}
\label{betbox}
\end{figure}

The noise model of \sref{snoise} was again used to study the uncertainty in the ranking of mathematicians based on their betweenness ratings and the results are shown in \fref{betbox} (see \fref{betboxv2} for 2018 results).

There are several different fields within mathematics such as algebra, geometry, and analysis. If a mathematician works in many different areas, individual pieces of their work may reveal connections between different areas of maths. Such a mathematician is likely to have a high betweenness reflecting the important contribution of such work. For instance, \href{https://en.wikipedia.org/wiki/John_von_Neumann}{von Neumann} has the highest betweenness in our 2017 data with his Wikipedia biography suggesting he made significant contributions to many different areas of mathematics; eight Wikipedia pages on different fields of mathematics are listed on his Wikipedia biography along with further pages in other disciplines.

However, a biography can be connected to many other mathematicians in many different fields for other reasons. Some historians of mathematics have very high betweenness too.  For instance, this explains why \href{https://en.wikipedia.org/wiki/Ivor_Grattan-Guinness}{Ivor Grattan-Guinness} \citep{R15b} has the $16$-th highest betweenness in our 2017 data. The similar phenomenon was also observed in our other centrality measure based on the shortest path, closeness, where \href{https://en.wikipedia.org/wiki/Ivor_Grattan-Guinness}{Ivor Grattan-Guinness} was ranked fifth by closeness in our 2017 data.

\subsection{Eigenvector Centrality}\label{sseigenvector}

If a mathematician is connected to a minor mathematician, then one may think that this relationship is of lower value than that between two famous mathematicians, say that between Newton and Leibniz. If all your connections are to many unimportant mathematicians we might imagine that this of less value than having your work being valued and used by a few important mathematicians.  Eigenvector centrality \citep{HSS08,N09b} attempts to take the quality of your neighbours into account when assessing the importance of a node by being defined in terms of a process with feedback; the larger your eigenvector centrality measure of your neighbours, the larger your eigenvector centrality will be.  If a mathematician publishes a new theory, the spread of this work may be likened to a broadcasting process in that this may be reused many times by many people.  If that theory draws on results from many different mathematicians, this may indicate that the new work is of broad relevance and so of high impact.   Eigenvector centrality tries to represent this process as the long time limit of a simple broadcast process so the importance of a vertex emerges through the continual feedback provided by loops in the network. We perform our analysis on the largest component which then guarantees a unique value for each node in the largest component. Our formal definition is given in \sref{sformalcent} of the Appendix.


Unlike degree but like betweenness and closeness, eigenvector centrality probes the whole structure of the network.  However unlike betweenness and closeness, eigenvector centrality is not based on shortest paths in the network. It turns out that the eigenvector value for each node can be seen as the number of very long (technically infinite) walks of any type which pass through that vertex.

\begin{figure}[htb!]
\centering
\includegraphics[width=0.6\textwidth]{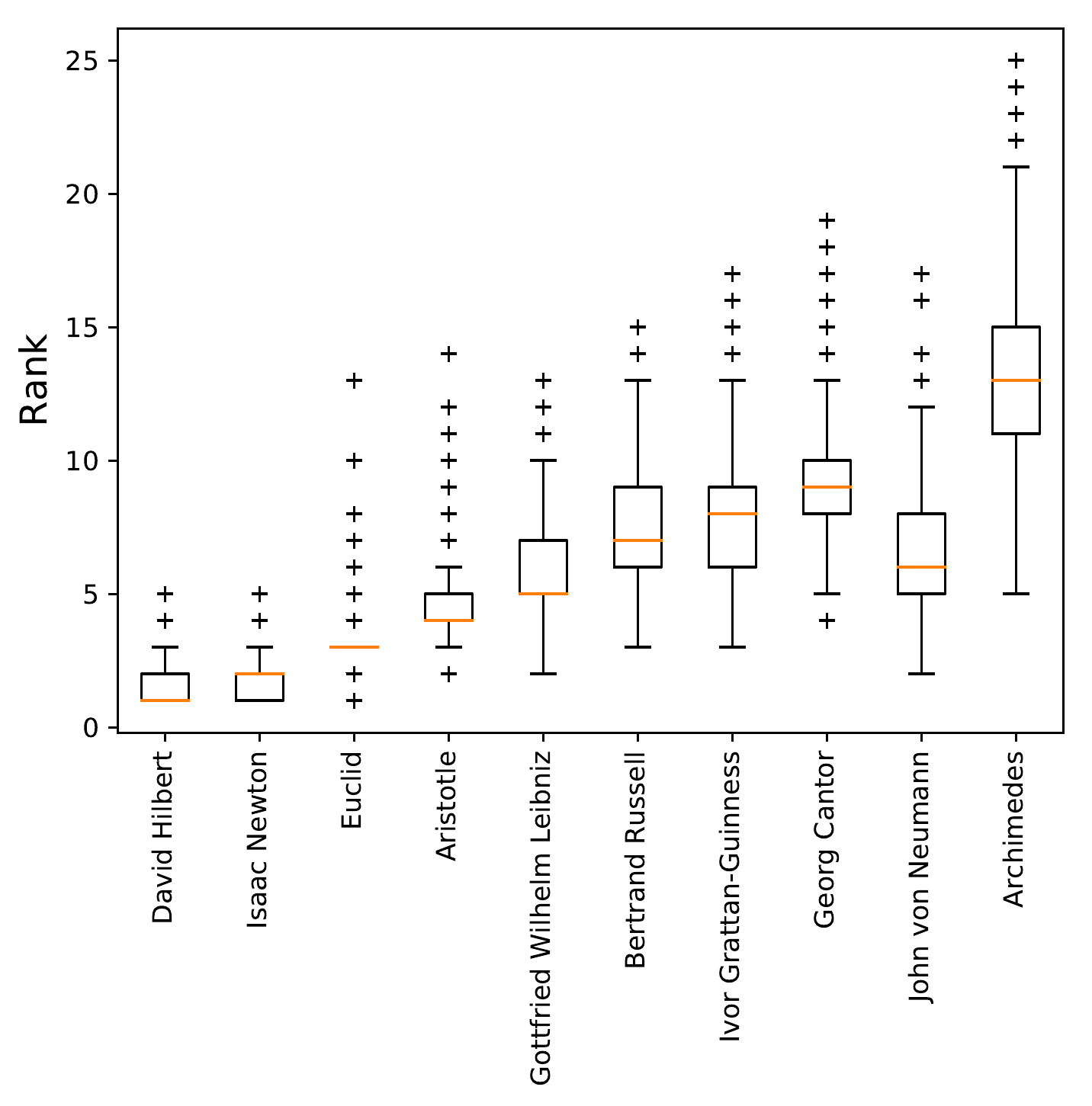}
\caption{Whisper box plot for rank of mathematicians derived from their Eigenvalue centrality.
This is for the largest component of the 2017 data based on 1000 simulations using the noise model of \sref{snoise}.
The criteria used to place the boxes and other features of the plot are as in \fref{degreebox}.}
\label{eigbox}
\end{figure}

The robustness of the ranking of mathematicians by their Eigenvector centrality value we estimated using our noise model and the results for 2017 are shown in \fref{eigbox} (see \fref{eigboxv2} for 2018 data).

\clearpage

\subsection{PageRank}\label{sspagerank}

PageRank is a centrality measure originally used to rank websites based on the network of hyperlinks linking websites \citep{BP98b,B08d,HSS08,N09b}.  The PageRank measure is derived from a simple process on a network. In the context of our web page biographies, the model pictures people surfing a website, and then either choosing a random link on each page visited and then following that link to the next page, or sometimes just jumping to a page chosen at random from all possible pages. While real individual users do not behave randomly, the success of search engines based on this method suggest that PageRank can, in some situations, capture the statistical behaviour of large numbers of users using web sites. As the Mathematician Wikipedia biographies are web pages, it is not unreasonable to assume that PageRank will be equally successful on our data. A more detailed definition of PageRank is given in \sref{sformalcent}.

\begin{figure}[htb!]
\centering
\includegraphics[width=0.6\textwidth]{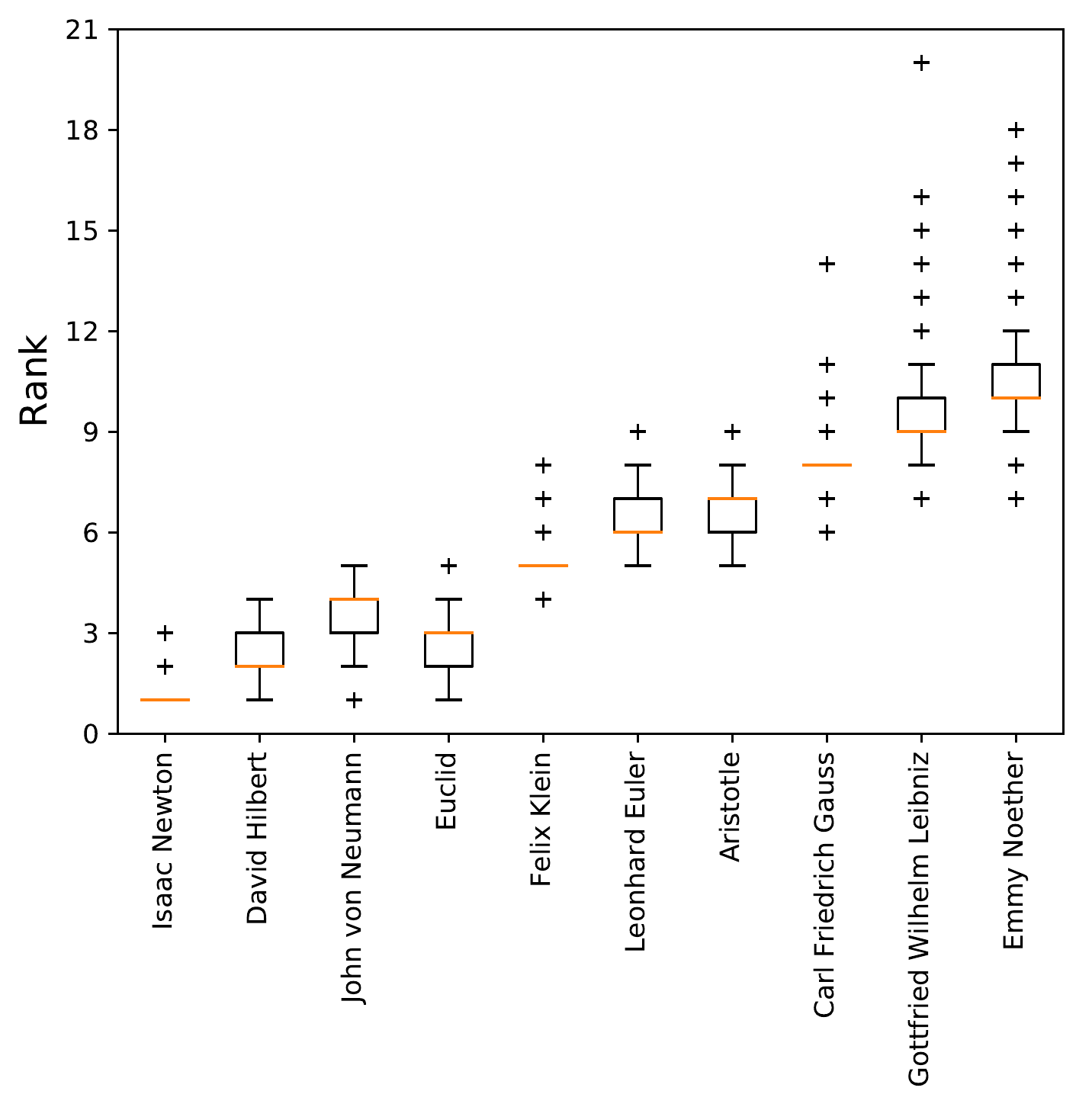}
\caption{Whisper box plot for rank of mathematicians derived from their PageRank ratings.
This is for the largest component of the 2017 data based on 1000 simulations using the noise model of \sref{snoise}.
The criteria used to place the boxes and other features of the plot are as in \fref{degreebox}.}
\label{pagebox}
\end{figure}

The robustness of the ranking of mathematicians by PageRank is shown in the box plot of \fref{pagebox} for the 2017 data (results for 2018 shown in \fref{pageboxv2}).
\clearpage

\subsection{Comparison of Different Centrality Measures}

Each different centrality measure defines `important' in a different way. While there are many aspects to importance, there are a very large number of different centrality measures, see \citet{S16} for a nice visualisation of this. So we should not be surprised if some of the many definitions of centrality measure pick up on similar aspects of centrality and so give similar results.  This we can see by looking at the correlations between centrality measures, a subject with a long history, for example see \citet{VCLC08}, \citet{SVB17} and references therein.

Since the centrality scores are not generally normally distributed, we will not rely on the Pearson Correlation coefficient to assess these correlations but we will also use the alternative Spearman's Rank Correlation Coefficient (the Pearson correlation applied to the ranked values of the centrality measures).

\begin{table}[htb!]
\centering
\begin{tabular}{l||c|c|c|c|c|c}
largest component                  & Degree          & PageRank          & Eigenvector          & Betweenness          & Closeness          & Average \\ \hline \hline
Degree               &          1.00   & \pvalue{0.98}     & \pvalue{0.82}        & \pvalue{0.86}        & \pvalue{0.57}      & \pvalue{0.95}    \\ \hline
PageRank             & \svalue{0.95}   & 1.00              & \pvalue{0.74}        & \pvalue{0.87}        & \pvalue{0.52}      & \pvalue{0.91}    \\ \hline
Eigenvector          & \svalue{0.63}   & \svalue{0.42}     & 1.00                 & \pvalue{0.70}        & \pvalue{0.56}      & \pvalue{0.87}    \\ \hline
Betweenness          & \svalue{0.88}   & \svalue{0.92}     & \svalue{0.49}        & 1.00                 & \pvalue{0.40}      & \pvalue{0.82}    \\ \hline
Closeness            & \svalue{0.70}   & \svalue{0.51}     & \svalue{0.93}        & \svalue{0.59}        & 1.00               & \pvalue{0.78}    \\ \hline
Average              & \svalue{0.85}   & \svalue{0.69}     & \svalue{0.90}        & \svalue{0.73}        & \svalue{0.96}      & 1.00
\end{tabular}
\caption{The correlation values for the mathematicians in the largest component. The upper right triangle contains the Pearson correlation values (in blue) while the lower left triangle contains the Spearman correlation values (in red italics). Note that for both cases the degree and PageRank are particularly well correlated as are Betweenness and Closeness.}
\label{tPearsonSpearmanlargest component}
\end{table}

Looking at the correlation values for the largest component in \tableref{tPearsonSpearmanlargest component}, we see that degree has a high correlation with many measures. On the other hand, closeness has a mediocre correlation with other measures almost all the time, typically around $0.5$, though that still represents a fair correlation.  In general we expect considerable correlation if all the centrality measures are influenced the same aspects of importance.

\begin{figure}[htb!]
\centering
\includegraphics[width=0.75\textwidth]{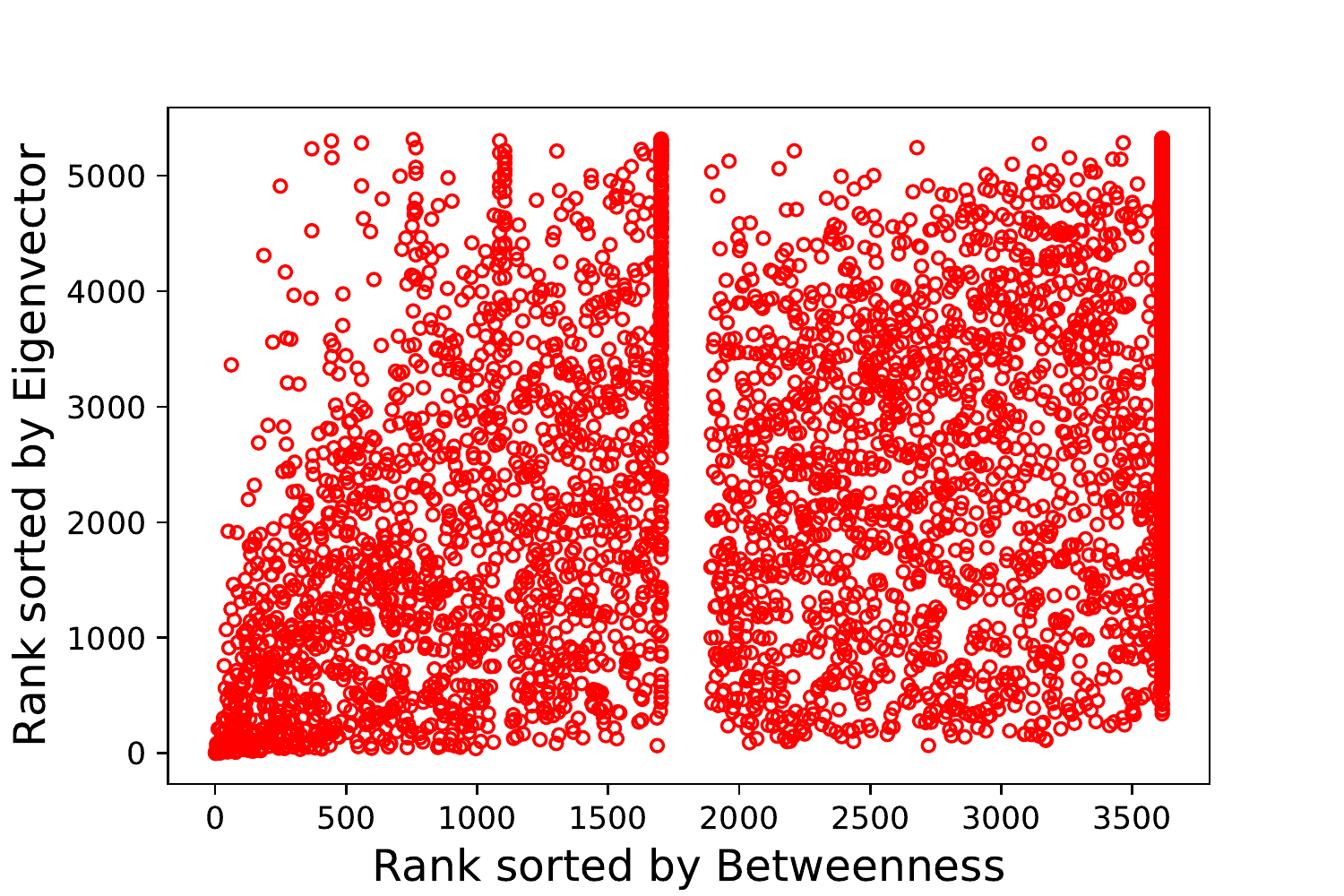}
\caption{Scatter plot of ranks in Betweenness and Eigenvector for the largest component from the 2017 data. The points appear quite widely distributed with little correlation yet the Spearman correlation is $0.49$. The issue is that around 40\% of the mathematicians have zero betweenness producing the vertical line of points around rank 3600.  Those mathematicians tend to have low eigenvector values (high rank by eigenvector) but this cannot be seen in this scatter plot.
}
\label{scatterbe}
\end{figure}

However interpreting such summary statistics is difficult here because of the correlation measures for the largest component are also strongly effected by the fat tail, the large numbers of mathematicians with low centrality values.  For instance, the betweenness value as a function of rank by betweenness is roughly a power law  distribution for the two thousand mathematicians but then the distribution shows a sharp cutoff. In particular, over two thousand mathematicians in the largest component have exactly zero betweenness.
The discrete values of betweenness and degree leading to many common values are an additional factor.  There are over a thousand mathematicians in the largest component with degree $1$ and betweenness $0$. The vertical gaps in \fref{scatterbe} illustrate the discreteness problem for betweenness. So some scatter plots appear to show a lack of correlation, as  \fref{scatterbe} suggests at first glance, but the correlation measures are high, pulled up by similar values for many low-valued mathematicians. Overall, we have to be very careful in interpreting these correlation measures and scatter plots for the largest component.

\begin{figure}[htb!]
\centering
\includegraphics[width=0.75\textwidth]{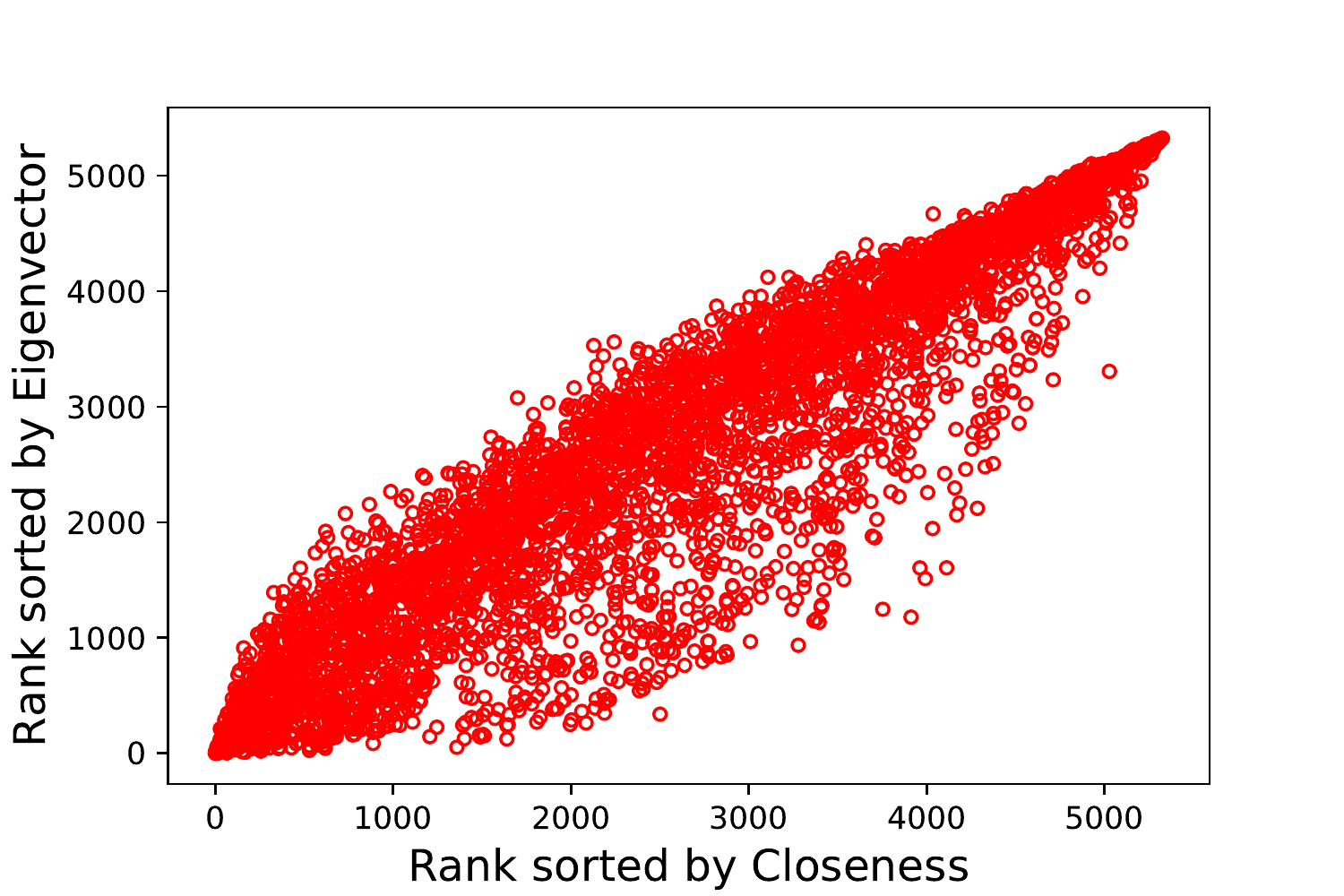}
\caption{Scatter plot of ranks in Eigenvector and Closeness for the largest component from the 2017 data. A strong positive correlation can be seen matching the Spearman correlation value found of $0.93$.}
\label{scatterdp}
\end{figure}

However, even for low ranked mathematicians, interesting results can be found by looking for outliers.  It is clear from \fref{scatterdp} and \fref{scatterbe} that while there is sometimes a general correlation or trend, there are many individual exceptions. This is where more sophisticated measures tailored to a particular context and question are needed, or perhaps simply where an expert opinion is required.
For example, \href{https://en.wikipedia.org/wiki/Solomon_Kullback}{Solomon Kullback} has a degree and a PageRank which on our scale are 2.3 and 10.2 respectively (out of 100) ranking him 2088 and 471 on each measure respectively. His position in government agencies probably limits his known links to mathematicians, hence the low degree, yet his PageRank suggests that his work links him to important developments in mathematics.  \href{https://en.wikipedia.org/wiki/Ivan_Rival}{Ivan Rival} has the same values with few links to mathematicians yet his role as editor of a key journal of discrete mathematics, ``Order'', may link him to particularly important mathematicians.  Perhaps an editor of a leading journal can have a major influence on mathematics.as indicated by the higher than expected PageRank in this case.

\begin{figure}[htb!]
\centering
\includegraphics[width=0.75\textwidth]{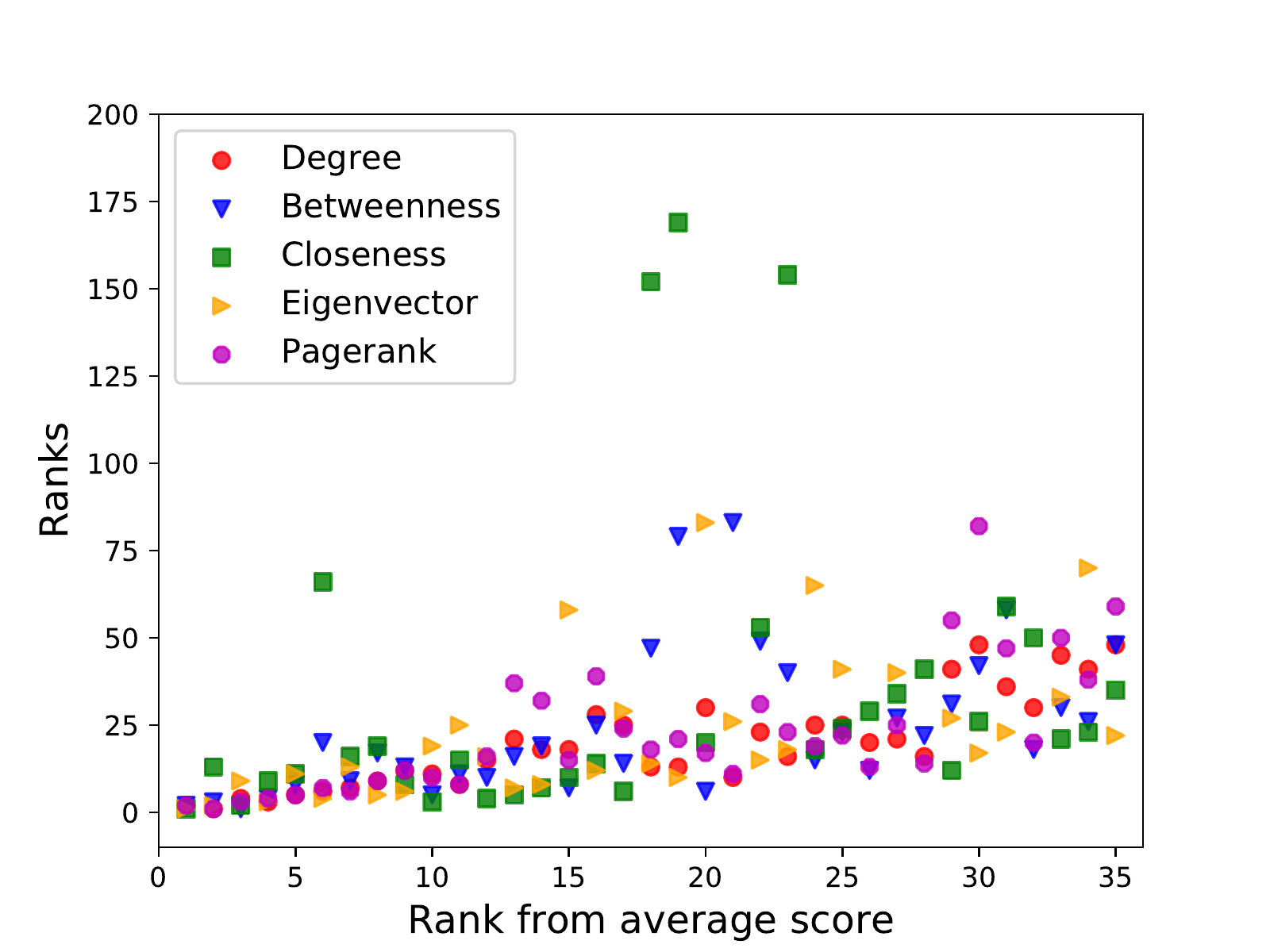}
\caption{A comparison of the rank of mathematicians under different centrality measures.  The horizontal axis is the rank of each mathematician by their average score; the top 35 are shown. Note that as the rank gets higher, there is a small but increasing variation in the ranks by different centrality measures for each mathematician.}
\label{overall35}
\end{figure}

Our data sets have a very large number of low rated mathematicians and we expect their properties to be particularly noisy. For instance, their fat-tailed degree distributions seen in \fref{degreedis} (and in \fref{degreedisv2})  show that changing even one hyperlink in their biographies is a large proportionate change in this measure.   So  when discussing correlations, it makes much more sense to look at a smaller group of highly rated mathematicians.  For instance if we restrict ourselves to the top thirty five mathematicians, we find ties in value of a centrality measure are rare even for integer valued degree. The correlation measures in this case are shown in \tableref{tPearsonSpearman35} and in \fref{overall35}. From the correlation matrix for the top thirty five mathematicians, we found that the degree measure is correlated very strongly with the PageRank centrality measure, a feature often seen with these two measures\footnote{For an undirected and connected graph as we have here for the largest component, if we set $\alpha=1$ in \tref{PageRank} we can show that PageRank is proportional to degree.}.  Closeness is still poorly correlated except with with the other measure based on shortest path measures, betweenness.

\begin{table}[htb!]
\centering
\begin{tabular}{l||c|c|c|c|c|c}
Top 35          & Degree          & PageRank          & Eigenvector          & Betweenness          & Closeness          & Average          \\ \hline \hline
Degree          & 1.00            & \pvalue{0.98}     & \pvalue{0.74}        & \pvalue{0.78}        & \pvalue{0.36}      & \pvalue{0.96}    \\ \hline
PageRank        & \svalue{0.92}   & 1.00              & \pvalue{0.61}        & \pvalue{0.84}        & \pvalue{0.40}      & \pvalue{0.94}    \\ \hline
Eigenvector     & \svalue{0.63}   & \svalue{0.39}     & 1.00                 & \pvalue{0.46}        & \pvalue{0.34}      & \pvalue{0.80}    \\ \hline
Betweenness     & \svalue{0.55}   & \svalue{0.71}     & \svalue{0.23}        & 1.00                 & \pvalue{0.74}      & \pvalue{0.87}    \\ \hline
Closeness       & \svalue{0.28}   & \svalue{0.30}     & \svalue{0.33}        & \svalue{0.77}        & 1.00               & \pvalue{0.57}    \\ \hline
Average         & \svalue{0.88}   & \svalue{0.80}     & \svalue{0.76}        & \svalue{0.71}        & \svalue{0.58}      & 1.00
\end{tabular}
\caption{The correlation values for the 35 top mathematicians as defined by the average of centrality scores in the 2017 data. The upper right triangle contains the Pearson correlation values (in blue) while the lower left triangle contains the Spearman correlation values (in red italics). Note that for both cases the degree and PageRank are particularly well correlated as are Betweenness and Closeness.}
\label{tPearsonSpearman35}
\end{table}


Since we have our noise model, we can also compare the robustness of different centrality measures.  The size of the fluctuations in different measures is shown for the rank of the top ten mathematicians in Figures \ref{degreebox}, \ref{clobox},
\ref{betbox}, \ref{eigbox}, and \ref{pagebox}, while the standard deviation in the actual centrality values is quoted for the top 35 in \tableref{tcentralitynoise}. Both show that the robustness of different centrality measures is very different. Looking at the top 35 mathematicians in \tableref{tcentralitynoise} we find that the average of the standard deviation divided by the mean is $0.17$ (betweenness), $0.14$ (Eigenvector), $0.069$ (PageRank), $0.066$ (degree), $0.016$ (Closeness), and $0.074$ (average score).

Thus closeness appears to be noticeably more robust than the other centrality measures. While this does not answer the question if it is a good measure of importance in all contexts, the reliability of closeness does make it a more useful measure.  On the other hand, betweenness is noticeably less stable than other measures, suggesting we should not rely on it as an indicator of importance. It is interesting that closeness and betweenness were well correlated for highly ranked mathematicians and that both rely draw the same set of shortest paths. However, closeness is an average over all shortest paths from one vertex, while betweenness counts just a few passing through a given vertex. So again, the instability of betweenness seems to come from its reliance on a few measurements. That aspect of betweenness is also why betweenness values are taken from a relatively small pool of likely rational (often integer) values leading to many ties, an issue we highlighted when discussing correlations above.


\subsection{Overall Ranking from Wikipedia Data}

The results of our measurements of five centrality measures on the network derived from the 2017 Wikipedia biographies of mathematicians are shown in \tableref{score2017}. The equivalent results for the 2013 and 2018 data sets may be found in \tableref{score2013} and \tableref{score2018} respectively.

\begin{table}[htb!]
\centering
\resizebox{\textwidth}{!}{\begin{tabular}{cccccccc}
\hline \hline
Name                      & Degree mark & Betweenness mark & Closeness mark & Eigenvector mark & PageRank mark & Average mark & Rank \\ \hline
David Hilbert             & $92.25$ & $94.58$     & $100$     & $100$       & $88.86$  & $95.14$      & $1$  \\
Isaac Newton              & $100$   & $70.66$     & $90.84$   & $98.02$     & $100$    & $91.9$       & $2$  \\
John von Neumann          & $79.07$ & $100$       & $97.34$   & $66.01$     & $85.62$  & $85.61$      & $3$  \\
Euclid                    & $86.05$ & $66.78$     & $92.4$    & $88.71$     & $84.19$  & $83.62$      & $4$  \\
Felix Klein               & $72.87$ & $43.71$     & $91.63$   & $58.53$     & $70.53$  & $67.45$      & $5$  \\  \hline
Aristotle                 & $66.67$ & $26.59$     & $84.39$   & $81.67$     & $63.01$  & $64.47$      & $6$  \\
Leonhard Euler            & $62.79$ & $40.25$     & $89.8$    & $57.36$     & $65.4$   & $63.12$      & $7$  \\
Gottfried Wilhelm Leibniz & $54.26$ & $29.01$     & $89.59$   & $76.44$     & $52.99$  & $60.46$      & $8$  \\
Bertrand Russell          & $50.39$ & $31.97$     & $92.5$    & $70.74$     & $48.3$   & $58.78$      & $9$  \\
Emmy Noether              & $51.16$ & $46.98$     & $93.96$   & $45.08$     & $52.28$  & $57.89$      & $10$ \\ \hline
Carl Friedrich Gauss      & $56.59$ & $38.69$     & $89.97$   & $40.51$     & $59.68$  & $57.09$      & $11$ \\
Hermann Weyl              & $44.96$ & $39.09$     & $93.88$   & $50.33$     & $45.33$  & $54.72$      & $12$ \\
Ivor Grattan-Guinness     & $39.53$ & $29.05$     & $92.8$    & $70.67$     & $35.56$  & $53.52$      & $13$ \\
Georg Cantor              & $41.86$ & $26.8$      & $92.64$   & $66.39$     & $37.58$  & $53.06$      & $14$ \\
Nicolas Bourbaki          & $41.86$ & $45.2$      & $91.73$   & $24.53$     & $45.66$  & $49.8$       & $15$ \\ \hline
Charles Sanders Peirce    & $37.21$ & $23.76$     & $90.47$   & $57.58$     & $35.28$  & $48.86$      & $16$ \\
Norbert Wiener            & $37.98$ & $31.64$     & $92.65$   & $35.59$     & $41.21$  & $47.82$      & $17$ \\
Galileo Galilei           & $46.51$ & $13.79$     & $80.66$   & $52.93$     & $43.61$  & $47.5$       & $18$ \\
Archimedes                & $46.51$ & $9.8$       & $80.03$   & $58.62$     & $41.9$   & $47.37$      & $19$ \\
Vladimir Arnold           & $34.11$ & $46.4$      & $89.47$   & $19.97$     & $44.51$  & $46.89$      & $20$ \\ \hline
Ptolemy                   & $52.71$ & $9.45$      & $75.02$   & $40.39$     & $49.44$  & $45.4$       & $21$ \\
Christiaan Huygens        & $38.76$ & $13.5$      & $85.02$   & $50.83$     & $37.83$  & $45.19$      & $22$ \\
Johannes Kepler           & $42.64$ & $15.23$     & $80.6$    & $45.99$     & $41.22$  & $45.14$      & $23$ \\
G. H. Hardy               & $37.98$ & $31.58$     & $89.74$   & $23.12$     & $42.86$  & $45.06$      & $24$ \\
Alan Turing               & $37.98$ & $24.83$     & $88.18$   & $29.74$     & $41.47$  & $44.44$      & $25$ \\ \hline
Michael Atiyah            & $41.09$ & $34.02$     & $87.25$   & $11.88$     & $47.78$  & $44.4$       & $26$ \\
Alfred Tarski             & $39.53$ & $23$        & $86.23$   & $31.27$     & $41.11$  & $44.23$      & $27$ \\
Alexander Grothendieck    & $42.64$ & $25.83$     & $85.76$   & $11.83$     & $45.71$  & $42.35$      & $28$ \\
Bernhard Riemann          & $31.01$ & $19.04$     & $90.88$   & $39.16$     & $28.72$  & $41.76$      & $29$ \\
George Boole              & $31.78$ & $12.24$     & $84.67$   & $42.09$     & $30.74$  & $40.3$       & $30$ \\  \hline
Andrey Kolmogorov        & $34.11$ & $27.43$ & $85.3$  & $11.77$ & $42.47$ & $40.21$ & $31$ \\
William Rowan Hamilton   & $29.46$ & $19.72$ & $88.66$ & $32.93$ & $29.63$ & $40.08$ & $32$ \\
Emil Artin               & $31.01$ & $23.6$  & $88.34$ & $21.27$ & $35.42$ & $39.93$ & $33$ \\
Alfred North Whitehead   & $27.91$ & $13.6$  & $86.22$ & $42.91$ & $27.5$  & $39.63$ & $34$ \\
Martin Gardner           & $31.78$ & $25.93$ & $84.56$ & $16.82$ & $38.81$ & $39.58$ & $35$\\
\hline
\end{tabular}}
\caption{Centrality scores for top 35 mathematicians from the 2017 data (without noise), each measure rescaled so the largest value is 100. Mathematicians are then ordered by the average mark.}
\label{score2017}
\end{table}

The simplest way to combine these different centrality ratings is to take the average of our centrality measures, remembering that each is rescaled according to \tref{rescale}.  We have not, however, then rescaled our ``average'' score and for that reason the highest average rating is less than 100.  We have indicated this in our tables of results, \tableref{score2017}, \tableref{score2013} and \tableref{score2018}. This simple average puts Hilbert as the most important mathematician with Newton only a short way behind.  The third most important mathematician according to this average rating is von Neuman who is some way behind in most ratings.

However this is where it becomes important to estimate the uncertainty in these results.  One way is to look at how different ways to combine ratings or rankings produce different results. There is no perfect way to do this and so there are many options \citep{LM12}.  We use the simplest approach; we will simply count who achieves the most number one rankings when considering each centrality measure individually. Doing that we see that with one exception (betweenness in 2017), either Hilbert or Newton always has the highest centrality measure in either the 2013 or 2017 data. By this way of looking for the best mathematician, there is little to choose between Newton and Hilbert as both are the highest in two of our five centrality measures in 2017.  In fact Newton has is top in three centrality measures in 2013 so by this scheme and data he could be deemed better than Hilbert.

\begin{figure}[htb!]
\centering
\includegraphics[width=0.75\textwidth]{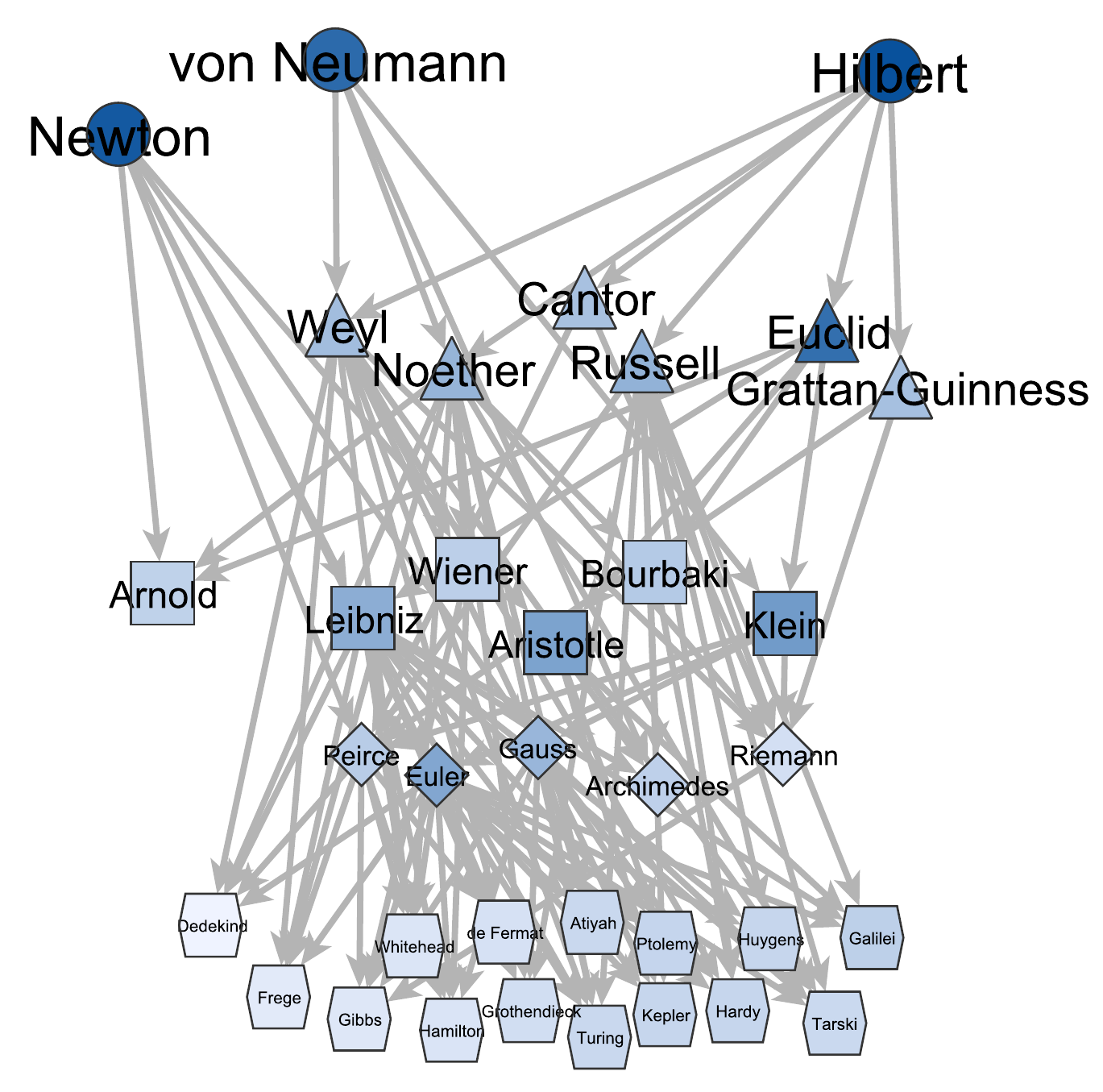}
\caption{The top 35 mathematicians (2017 data) based on five ratings: Degree,  Betweenness, Closeness, Eigenvector centrality and PageRank. Mathematician `A' is placed higher than mathematician `B' if each of their five ratings of `A' is higher than the corresponding rating for `B'.  The arrow of the line points from the higher to the lower ranked mathematician but only those essential for the logical relationships are shown. This is a Hasse diagram of the corresponding poset, equivalently it is the transitively reduced form of the corresponding directed acyclic graph. The shape of a node, size of label and the vertical location reflects the `height' of each node in the corresponding poset (see text for definition).  The colour reflects the average scaled rating of each mathematician.}
\label{dag17}
\end{figure}

We will use a variation of this approach and consider a second way to combine scores, one which produces a nice visualisation.  Formally we construct a partially ordered set, a poset, from the set of mathematicians and the relationship between their rankings between their rankings (see \citet{BMH94} \normalversiononly{and \citet{LE15b} } for examples from different contexts and further references).  In this case, for our set of mathematicians we say $A \succ B$ if each of the ratings for mathematician `A' is better than the corresponding rating for `B'. As Newton has a higher closeness than Newton but Newton has the higher degree of the two, we cannot assign any relationship between these two in this poset.  However Hilbert and Newton both have a higher rating than Euclid for all five centrality measures, so we can write this fact as $\mathrm{Newton} \succ \mathrm{Euclid}$ and $\mathrm{Hilbert} \succ \mathrm{Euclid}$. We can then identify the `top' nodes as those nodes $T$ for which there is no mathematician $A$ such that $A \succ T$.   In our case we find that for almost any set of ratings, we have that for the 2013 data we have just two top nodes: \href{https://en.wikipedia.org/wiki/Isaac_Newton}{Newton} and \href{https://en.wikipedia.org/wiki/David_Hilbert}{Hilbert} (see \fref{dag13}). For the 2017 case, as shown in \fref{dag17}, we find that in addition to these two we have a third mathematician at the top of our poset, \href{https://en.wikipedia.org/wiki/John_von_Neumann}{von Neumann}.  This is because his betweenness is the highest for the 2017 Wikipedia data as \fref{dag17} shows.

This poset structure also allows us to split our mathematicians into subgroups, each of which has similar ratings but where each group is lower rated than the previous group.  This is done by measuring the `height' of each mathematician within the poset.  To find the height of mathematician `A' you have to find the sequence mathematicians, a `chain', from one of the source tops nodes to mathematician `A', that is $\mathcal{M} = \{M_i\}$ where $M_i \succ M_{i+1}$, the first node is a top node $M_0=T$ and the last node is the mathematician of interest $M_\ell=A$. The height is the number of nodes in the longest chain minus one. Note the top nodes have height zero.  The result is shown in\footnote{For instance the height of Leibniz is 2 because of the chain $\{Hilbert, Euclid, Leibniz\}$. The longest chain from the other top node is just $\{Newton,Leibniz\}$.}  \fref{dag17}  for the 2017 Wikipedia data (see \fref{dag13} for the 2013 data).


On closer inspection, our noise model suggests that betweenness is one measure which is particularly sensitive to noise with mathematicians in the top 10 typically having a variance of 10\% in their betweenness scores. The visualisation in \fref{dag17} and indeed our simple averaging of results, takes no account of our estimation of uncertainty in the individual ratings.  Nevertheless, this way of displaying data provides a useful organisation of the data, with the height organising mathematicians into different tiers of importance.  As with all our measures, this is not the only answer but such an organisation can provide a good starting point for discussion and further investigation of the data.

Another way to look at the uncertainty in our rankings is to use the estimates provided by our noise model of section \sref{snoise} and the results are shown in \tableref{tcentralitynoise}. Now we see that average scores for Hilbert and Newton are within a standard deviation of each other, suggesting that this difference is not particularly significant and the Wikipedia data from 2017 cannot be used to place one above the other.  On the other hand, its does suggest, at least in terms of the average of the centralities and the uncertainty in those results, that the gap between von Neumann and the pair of Newton and Hilbert is significant.

The differences in the top 35 mathematicians between 2013 and 2017 is also intriguing.  In terms of who is in the list, most of the turnover
is in the last seven places. The last seven in 2013 have all dropped out, replaced by six newcomers (plus Boole) in this bottom part of the 2017 data,  see \tableref{score2017}.  This variation is a good measure of the uncertainty in the average ranking measure, that is when ranked around 30 you could easily move 7 places either way over the four years\footnote{The turnover in such ranked lists has been studied in other contexts but those techniques and suggested power-laws would require data over a longer period to be useful here \blindreview{\citep{BLHH07}.}{\citep{BLHH07,EG11}.} }. Equally, while there is less change at the top of our lists from 2013 to 2017, we still see small changes as high as the  fifth and sixth place.
Again the results from our noise model in \tableref{tcentralitynoise} confirms this behaviour, in this case showing our fifth place Klein and sixth place Aristotle are too close to be sure of their relative position. Similar variations can be seen when comparing to the 2018 data shown in the Appendix, e.g.\ \tableref{score2018}.



\begin{table}[htb!]
\centering
\resizebox{\textwidth}{!}{\begin{tabular}{cccccccc}
\hline \hline
Name                      & Degree      & Betweenness & Closeness   & Eigenvector  & PageRank     & Average & Rank \\ \hline
David Hilbert             & $92.03 \pm 4.7$  & $95.79 \pm 6.38$ & $99.99 \pm 0.05$ & $97.2 \pm 5.35$  & $88.21 \pm 5.03$ & $94.64 \pm 4.83$ & $1$  \\
Isaac Newton              & $99.83 \pm 0.8$  & $82.76 \pm 10$   & $94.76 \pm 1.38$ & $92.25 \pm 8.55$ & $99.96 \pm 0.37$ & $93.91 \pm 5.93$ & $2$  \\
John von Neumann          & $79.32 \pm 4.43$ & $93.87 \pm 7.64$ & $97.39 \pm 1.14$ & $65.78 \pm 6.62$ & $82.71 \pm 4.87$ & $83.81 \pm 5.42$ & $3$  \\
Euclid                    & $86.18 \pm 4.71$ & $65.38 \pm 8.54$ & $94.34 \pm 1.33$ & $82.05 \pm 8.47$ & $85.8 \pm 5.05$  & $82.75 \pm 6.23$ & $4$  \\
Felix Klein               & $73 \pm 4.1$     & $52.11 \pm 7.27$ & $93.23 \pm 1.21$ & $57.93 \pm 6.04$ & $70.95 \pm 4.28$ & $69.45 \pm 5.02$ & $5$  \\  \hline
Aristotle                 & $66.79 \pm 4$    & $37.39 \pm 5.88$ & $90.04 \pm 1.46$ & $72.57 \pm 8.32$ & $64.61 \pm 4.15$ & $66.28 \pm 5.28$ & $6$  \\
Leonhard Euler            & $62.91 \pm 3.8$  & $44.4 \pm 6.63$  & $91.71 \pm 1.41$ & $54.16 \pm 5.99$ & $64.89 \pm 4.19$ & $63.61 \pm 4.77$ & $7$  \\
Gottfried Wilhelm Leibniz & $54.42 \pm 3.45$ & $34.22 \pm 5.53$ & $92.04 \pm 1.31$ & $68.3 \pm 6.89$  & $53.4 \pm 3.66$  & $60.48 \pm 4.58$ & $8$  \\
Bertrand Russell          & $50.61 \pm 3.27$ & $34.9 \pm 5.8$   & $93.31 \pm 1.33$ & $63.91 \pm 6.36$ & $48.6 \pm 3.32$  & $58.27 \pm 4.42$ & $9$  \\
Emmy Noether              & $51.4 \pm 3.28$  & $46.53 \pm 6.96$ & $94.3 \pm 1.22$  & $45.68 \pm 5.08$ & $51.38 \pm 3.42$ & $57.86 \pm 4.43$ & $10$ \\ \hline
Carl Friedrich Gauss      & $56.75 \pm 3.5$  & $39.69 \pm 5.67$ & $90.83 \pm 1.31$ & $40.02 \pm 4.97$ & $58.98 \pm 3.82$ & $57.26 \pm 4.13$ & $11$ \\
Hermann Weyl              & $44.95 \pm 2.98$ & $37.34 \pm 5.72$ & $93.88 \pm 1.22$ & $48.28 \pm 5.07$ & $44.54 \pm 3.07$ & $53.8 \pm 3.96$  & $12$ \\
Ivor Grattan-Guinness     & $39.61 \pm 2.81$ & $27.84 \pm 4.89$ & $93.55 \pm 1.14$ & $62.98 \pm 5.75$ & $36.07 \pm 2.77$ & $52.01 \pm 3.84$ & $13$ \\
Georg Cantor              & $41.93 \pm 2.87$ & $26.18 \pm 4.72$ & $93.02 \pm 1.19$ & $59.98 \pm 5.78$ & $38.32 \pm 2.82$ & $51.89 \pm 3.83$ & $14$ \\
Galileo Galilei           & $46.7 \pm 3.12$  & $20.88 \pm 4.48$ & $86.37 \pm 1.67$ & $47.27 \pm 6.22$ & $45.03 \pm 3.32$ & $49.25 \pm 4.05$ & $15$ \\ \hline
Archimedes                & $46.69 \pm 3.12$ & $17.1 \pm 3.91$  & $85.68 \pm 1.64$ & $51.71 \pm 6.45$ & $43.53 \pm 3.08$ & $48.94 \pm 3.97$ & $16$ \\
Nicolas Bourbaki          & $42.05 \pm 2.95$ & $39.61 \pm 5.68$ & $91.49 \pm 1.33$ & $26.76 \pm 3.91$ & $44.04 \pm 3.18$ & $48.79 \pm 3.69$ & $17$ \\
Ptolemy                   & $52.84 \pm 3.28$ & $18.4 \pm 4.12$  & $83.18 \pm 2$    & $36.86 \pm 5.33$ & $51.08 \pm 3.4$  & $48.47 \pm 3.79$ & $18$ \\
Charles Sanders Peirce    & $37.12 \pm 2.53$ & $23.98 \pm 4.41$ & $91.66 \pm 1.29$ & $51.49 \pm 5.39$ & $35.43 \pm 2.59$ & $47.94 \pm 3.56$ & $19$ \\
Norbert Wiener            & $38.15 \pm 2.75$ & $30.18 \pm 4.65$ & $92.17 \pm 1.29$ & $35.1 \pm 4.25$  & $40.03 \pm 2.97$ & $47.13 \pm 3.4$  & $20$ \\ \hline
Johannes Kepler           & $42.77 \pm 2.84$ & $20.25 \pm 4.15$ & $86.02 \pm 1.61$ & $41.38 \pm 5.47$ & $41.96 \pm 2.99$ & $46.48 \pm 3.65$ & $21$ \\
Christiaan Huygens        & $38.71 \pm 2.75$ & $16.51 \pm 3.46$ & $87.38 \pm 1.46$ & $44.93 \pm 5.3$  & $38.21 \pm 2.88$ & $45.15 \pm 3.41$ & $22$ \\
Michael Atiyah            & $41.31 \pm 2.82$ & $33.39 \pm 5.07$ & $88.09 \pm 1.53$ & $15.39 \pm 3.14$ & $45.56 \pm 3.19$ & $44.75 \pm 3.35$ & $23$ \\
G. H. Hardy               & $38.17 \pm 2.69$ & $30.28 \pm 4.83$ & $89.77 \pm 1.37$ & $24.12 \pm 3.44$ & $41.14 \pm 2.96$ & $44.69 \pm 3.26$ & $24$ \\
Alexander Grothendieck    & $42.84 \pm 2.99$ & $30.55 \pm 5.48$ & $87.87 \pm 1.61$ & $16.32 \pm 3.23$ & $44.29 \pm 3.16$ & $44.37 \pm 3.52$ & $25$ \\ \hline
Alfred Tarski             & $39.63 \pm 2.85$ & $24.24 \pm 4.2$  & $87.32 \pm 1.42$ & $29.89 \pm 4.08$ & $40.5 \pm 3.01$  & $44.32 \pm 3.27$ & $26$ \\
Vladimir Arnold           & $34.23 \pm 2.5$  & $35.06 \pm 5.26$ & $89.51 \pm 1.54$ & $20.83 \pm 3.5$  & $41.48 \pm 3.07$ & $44.22 \pm 3.41$ & $27$ \\
Alan Turing               & $38.05 \pm 2.73$ & $24.04 \pm 4.27$ & $88.43 \pm 1.4$  & $29.02 \pm 4.08$ & $40.15 \pm 2.94$ & $43.94 \pm 3.25$ & $28$ \\
Bernhard Riemann          & $31.11 \pm 2.35$ & $18.27 \pm 3.48$ & $90.98 \pm 1.17$ & $37.02 \pm 4.26$ & $29.08 \pm 2.33$ & $41.29 \pm 2.92$ & $29$ \\
Nicolaus Copernicus       & $35.87 \pm 2.7$  & $11.64 \pm 3.13$ & $84.4 \pm 1.78$  & $40.02 \pm 5.68$ & $33.5 \pm 2.67$  & $41.09 \pm 3.45$ & $30$ \\ \hline
George Boole                                                    & $31.84 \pm 2.43$ & $15.05 \pm 3.06$ & $87.26 \pm 1.38$ & $37.45 \pm 4.73$ & $30.88 \pm 2.47$ & $40.49 \pm 3.02$ &$31$                                            \\
Pierre de Fermat                                                & $27.89 \pm 2.27$ & $14.34 \pm 3.14$ & $89.25 \pm 1.33$ & $44.16 \pm 4.8$  & $25.5 \pm 2.22$  & $40.23 \pm 2.99$  &$32$                                           \\
Andrey Kolmogorov                                               & $34.33 \pm 2.58$ & $26.15 \pm 4.15$ & $85.87 \pm 1.57$ & $13.78 \pm 2.76$ & $39.97 \pm 3$    & $40.02 \pm 2.93$  &$33$                                           \\
Emil Artin                                                      & $31.19 \pm 2.35$ & $23.16 \pm 3.98$ & $88.71 \pm 1.36$ & $22.33 \pm 3.24$ & $33.92 \pm 2.6$  & $39.86 \pm 2.84$   &$34$                                          \\
Gaetano Fichera                                                 & $38.92 \pm 2.59$ & $21.37 \pm 3.96$ & $85.01 \pm 1.6$  & $12.58 \pm 2.72$ & $39.49 \pm 2.72$ & $39.47 \pm 2.82$ &$35$\\
\hline
\end{tabular}}
\caption{Centrality scores for top 35 mathematicians derived from the the noise model described of \sref{snoise} applied to the 2017 data with $p=0.1$ for 1000 simulations. The mean value and one standard deviation is quoted for each centrality measure for each mathematician. As the scores for each run are always rescaled so that the largest value is 100, explaining why the value quoted for any one centrality measure is always less than 100. The column marked average gives the average over the five named centrality measures with associated standard deviation. Mathematicians are ordered in terms of this average and the ranks given  are in terms of this average over centrality values.}
\label{tcentralitynoise}
\end{table}

As always, it is the outliers that attract attention.  Having set the scale of the expected changes, the mathematician in the top 35 list from 2017 who shows the biggest change is \href{https://en.wikipedia.org/wiki/Nicolas_Bourbaki}{Nicolas Bourbaki} which is actually a pseudonym for a group of mainly French 20th-century mathematicians. He moved from 42nd in 2013 to 15th in 2017.  This suggests this English Wikipedia article has undergone unusual and substantial expansion e.g.\ the number of hyperlinks to other identified mathematicians has gone from 34 to 54 in these four years.

Finally we note the presence of a few names who perhaps did not contribute directly to specific developments in mathematics.  The historian of mathematics \href{https://en.wikipedia.org/wiki/Ivor_Grattan-Guinness}{Grattan-Guinness} \cite{R15b} has a high rank because he is linked to so many mathematicians.  However we also note that \href{https://en.wikipedia.org/wiki/Martin_Gardner}{Martin Gardner} is 35th in the 2013 list (and 36th in 2017 so just off our table).  His role in mathematics is as one of the best known popularisers of mathematics working in the English language in the second half of the 20th century, illustrating that you can make important contributions to mathematics in many different ways. How many mathematicians today were inspired by Gardener's work?

\subsection{Comparing Wikipedia and MacTutor Results}\label{sMacTutor}

The results we have obtained can be compared with those of a different data base, the \href{http://www-history.mcs.st-and.ac.uk/}{MacTutor History of Mathematics archive} created by John J O'Connor
and Edmund F Robertson. This is a web site of biographies of famous mathematicians, with hyperlinks between these biographies.  A network was constructed by again setting each biographical web page to be a vertex.  As for the Wikipedia biographies, a vertex almost always represented a single mathematician\footnote{The only two exceptions known were for the pages dedicated to the work of collectives of \href{http://www-history.mcs.st-and.ac.uk/Biographies/Bourbaki.html}{Nicolas Bourbaki} and the \href{http://www-history.mcs.st-and.ac.uk/Biographies/Banu_Musa.html}{Ban\={u} M\={u}s\={a} brothers} which were also represented by a single vertex.}.  A directed edge was assigned from one mathematician to another if there was at least one hyperlink between the two biographies.  The major difference between MacTutor and Wikipedia is that the MacTutor pages are not open to the public but are curated and written by O'Connor and Robertson. One result is that our MacTutor network has only  2249 vertices/mathematicians and 16980 directed edges between them, roughly a third of the size of our Wikipedia networks. This has been analysed by one of us \blindreview{(initials removed for double blind review)}{(TSE)} working with several other researchers but the results we quote here are based on the analysis of the data from late 2010 described in \blindreview{(citation of unpublished thesis removed for double blind review).}{\citet{C11}.} In particular, the centrality measures (for the directed network) for the top fifteen mathematicians are reproduced from \blindreview{(citation of unpublished thesis removed for double blind review)}{\citet{C11}} in \tableref{MacTutorTop15}.

The results for the top mathematicians are very similar to those we obtained from the Wikipedia data.  This is a further check of the robustness of our results.  It also suggests that for centrality measures there is not much difference in results between the use of directed and undirected networks based on hyperlinks between biographies. The similarity between results based on Wikipedia and MacTutor data is not so surprising. Both sets of biographies are written in English which might suggest common biasses.  Both web sites are free to view and so it is very likely that a writer for one website consciously or unconsciously drew on material from the other web site.

To get a rough idea, we if we average MacTutor the ranks across the four classic centrality measures we used for the Wikipedia data we find the following: Newton 1.5, Euclid 2.5, Hilbert 3.75, Riemann 4.2, and Euler 6.0. Two differences stand out when compared to the Wikipedia data. First Riemann is fourth by the MacTutor data but is 29th in the 2017 Wikipedia data.  Secondly, Wikipedia rates von Neumann as the third most important mathematician while the analysis of the Mactutor biography does not see him in the top ten. Again, given the similarity of other results this seems to highlight differences in the interests or expertise of the editors of these web sites, or perhaps in the procedures which lead to the public versions of the biographies.

\subsection{Comparison with Informal Survey}\label{ssurvey}

The final comparison we make is with our informal survey of undergraduate students studying mathematics or physics at \blindreview{(name removed for double blind review) University.}{Imperial College London.} The results of this survey are shown in \tableref{tableinformal}.
\begin{table}[h!]
 \begin{center}
 \begin{tabular}{ lccccc|c  }
Mathematician &  Year 1 & Year 2 & Year 3 & Year 4 \& Other & Total votes & 2017 Rank\\
 \hline
    Leonhard Euler  & 80 & 78 & 54 & 34 & 246 &  7 \\
    Friedrich Gauss & 57 & 52 & 46 & 26 & 181 & 11 \\
    Issac Newton    & 55 & 50 & 38 & 23 & 166 &  2 \\
    Euclid          & 48 & 43 & 29 & 16 & 136 &  4 \\
    Leibniz         & 25 & 19 & 12 &  4 &  60 &  8 \\
    David Hilbert   & 13 &  9 & 13 &  7 &  42 &  1 \\
    Aristotle       & 15 &  9 &  7 &  2 &  33 &  6 \\
    \blindreview{HoD}{Alesso Corti}    &  4 & 13 &  9 &  5 &  31 &  1019 \\
    Emmy Noether    &  6 &  0 & 10 & 10 &  26 & 10 \\
    von Neumann     &  2 &  3 &  8 &  7 &  20 &  3 \\
\end{tabular}
\end{center}
\caption{The top 10 mathematicians as derived from the results of our questionnaire sent in November 2016 to undergraduates in the Physics and Mathematics departments of \blindreview{(name removed for double blind review) University.}{Imperial College London.} The 2017 Rank is from the Wikipedia data.}
\label{tableinformal}
\end{table}

Nine of mathematicians mentioned come from the top eleven of the 2017 Wikipedia (see \tableref{score2017}) which at first sight appears to show a good general consistency between undergraduates and the web site data discussed above. After all, all but one mathematician nominated by participants is in our top twenty. Either our social network analysis is a good reflection of informed student opinion or it is merely a reflection of the way we constructed the survey since participants were given the list of top twenty names from our social network analysis to choose from (though they could add other names as one entry shows). Since eight of the nine mathematicians chosen by participants came from the top eleven of the list provided, we feel that this shows that participants were not unduly biased by the list provided otherwise we would have seen more names from those ranked below eleventh.

However it is also interesting to see that there is a very different order here as compared to that found with network analysis of the web sites. In particular, Hilbert and von Neuman are considerably underrated by undergraduates as compared to the Wikipedia and MacTutor rankings. This may reflect the direct impact or simply a lack of visibility in the undergraduate syllabus followed by these students.

It is also interesting that there is also good consistency between students in different years with the ranking of the top four (taking around 80\% of the votes) being identical. This suggests that the amount of training appears to have relatively little effect on the choice of best mathematician by these undergraduates.

Finally, the `joie de vivre' of these undergraduates is clearly evident as the head of the mathematics department \blindreviewonly{(indicated by `HoD', name removed for double blind review)} at the time of the survey appears as the most recent mathematician on this list.

\section{Discussion}

The simplest conclusion is that on the basis of our Wikipedia data we would suggest that the two most important mathematicians are Hilbert and Newton.  We have shown that we can also put a estimate of the uncertainty around such ratings by using simple models of the noise in the system.  Since we have Wikipedia data separated by four or five years, we have also been able to use the changes in the rankings of mathematicians over four or five years to get an estimate of the uncertainties in our ranking.  That means we are fairly confident in our results for the top four while for those ranked around thirty, we already suggest that an uncertainty of seven or so places is consistent with our analysis.

We've also shown how different sources can be used to provide further checks on the robustness of our conclusions. An independent web site created by John J O'Connor and Edmund F Robertson, MacTutor (\href{http://www-history.mcs.st-and.ac.uk/}{MacTutor History of Mathematics archive}, \citet{MacTutor}), gives broadly similar results \blindreview{(citations of unpublished theses removed for double blind review).}{\citep{C11,H11b}.}

Even our informal survey of undergraduates is fairly consistent with the results from the larger Wikipedia and MacTutor studies.  Where the comparisons are most interesting is in the differences.  In particular von Neumann is the third most important mathematician on the Wikipedia data but is far lower on the results from the survey and MacTutor.  Is this due to an over representation of Wikipedia editors who have an interest in computer science who particularly admire von Neumann's contributions in that field?  Conversely perhaps the \normalversiononly{British} higher educational system in mathematics, be it the teachers who are the editors of MacTutor or the students answering the survey, fail to give due weight to von Neumann's work because of its importance to a separate field, Computer Science. This shows that simple quantitative measures provide useful information but expert opinion is still required to understand many details.

It would be interesting to see how other lists of great mathematicians compare against the ones produced using our methods.  One approach would be to use well established measures of esteem to either rank mathematicians, for instance using bibliometric methods such as citation count or h-index. However the traditional bibliometric measures are flawed when making comparisons across large time scales and across many different topics. The list of mathematicians awarded prizes, such as \href{https://en.wikipedia.org/wiki/Fields_Medal}{the Fields medal}, could produce sets of great mathematicians, if no precise ranking. Such a list highlights some of the advantages of our approach.  Each prize has constraints in terms of subject about which one might argue.  Should \href{https://en.wikipedia.org/wiki/Edward_Witten}{Ed Witten}, who is typically described as a `physicist', have been awarded a Fields medal, the Nobel Prize of mathematics? Witten is, in fact, in our data but does not make our list of top 35 mathematicians. Other constraints apply to prizes too. The Fields medal is awarded only to those under the age of 40. Prizes are often only awarded only to living people and so prizes do not have the historical reach of our Wikipedia approach.  In fact, only two Field medal winners are in our top 35 mathematicians based on the 2017 data in \tableref{tcentralitynoise}, Atiyah and Grothendieck who were awarded the medal in 1966.  Of the others in our top 35 mathematicians, only Turing was eligible for the Fields medal illustrating a drawback of lists of winners of aged-limited prizes.

The selection process for most prizes is secret. \href{https://en.wikipedia.org/wiki/Alan_Turing}{Alan Turing}, who is typically ranked around 20th to 30th in our ratings, was young enough and recent enough to have been awarded a Fields medal but that did not happen. Yet Turing has a whole prize named after him, surely an even bigger measure of esteem. Prizes, like the many ad-hoc lists of great mathematicians produced by expert opinion, are created by hidden processes with unknown biases.  One might disagree with choices of those listed as being a mathematician on Wikipedia or with the links made between them, or indeed with the measure we have used to arrive at our conclusions. However at least our approach is completely open unlike most alternative approaches.

The comparison of our lists with the list of Fields medalists highlights that time has an important effect.  Most of our top 35 mathematicians did their work over a hundred years ago.  It seems likely that it was easier to have a larger impact on mathematics when the subject is young and our list reflects that. It would, though be interesting to compare like with like, perhaps comparing modern mathematicians using our methods and modern bibliometric methods.  Does our Wikipedia based rating for a mathematician lag behind the citation count of their work?
For modern mathematicians, it would be interesting to compare our rankings with those available from other sources: prizes, bibliometric measurements etc. However, that is a different project.  Name disambiguation is a serious problem in matching lists, the effect of time and field makes comparison of even modern mathematicians  difficult.  Our approach, perhaps even our data, could provide a starting point for such a project.

An important assumption in our work is that our definition of who is a ``mathematician'' is a good one.  For our main data set, we used the list of mathematicians provided on English Wikipedia.
Interestingly, our crowdsourced definition agrees with the the Fields medal committee in that it includes a `physicist' (Witten) in the list of mathematicians.
Our crowdsourced categorisation brings with it the strengths, and weaknesses, of that approach.
We can contrast this with the approach used to compile the list of mathematicians in the MacTutor database, which is based on the expert opinion of a pair of curators. It is another important result of our work that these two different approaches to define a collection of top mathematicians have provided comparable results. Our results provide further evidence that a crowdsourced approach to difficult questions can be an effective and reliable method.

It is worth challenging this assumption further.  Suppose we pick a mathematician at random from our data. Given our data is fat-tailed e.g.\ in terms of degree, we are very likely to pick a lowly ranked mathematician. The example of \href{https://en.wikipedia.org/wiki/Kristen_Nygaard}{Kristen Nygaard} is instructive as many of us would probably classify him as a computer scientist. Nygaard developed the core concepts of object-oriented programming for which he was awarded the Turing award, the Nobel prize or Fields medal equivalent for computer science. However, computer science often overlaps with mathematics, as Turing himself demonstrates. In addition, Nygaard had a masters in mathematics and worked for a time in operational research. The authors and referees might well use their expert opinions to exclude Nygaard from their optimal list of mathematicians in which case we might think our Wikipedia crowdsourced list of mathematicians contains mistakes.

This is not a good viewpoint in our opinion.  Different experts will always have different opinions.  It is easy to say two lists are different, it is difficult, if not impossible, to say if one list was better than another.  Most people would agree that Nygaard is at best a marginal case, mostly a computer scientist and only peripheral to the development of mathematics.  Many other people in a similar position to Nygaard could be in our list of mathematicians.  Equally there could be many others who are not in our list of mathematicians but who have a good case to be included. Essentially, any definition of a mathematician is uncertain, and that is a source of noise in any list.

However the example of \href{https://en.wikipedia.org/wiki/Kristen_Nygaard}{Kristen Nygaard} also shows why our method is so powerful. If someone like Nygaard is included, the number of links to other mathematicians in their Wikipedia biography gives a good indication of how central they are to mathematics. Nygaard's page has many links to people we could classify as computer scientists, many to those involved in Norwegian politics but only one to another mathematician in our database. If a mathematician is not particularly important to mathematics then our network representation will place that person placed in a peripheral position in the network.  Including, or indeed excluding, that person will have very little effect on our results. Our network method gives us some protection against uncertainties in the definition of a mathematician. People who are listed as mathematicians yet most of us would regard as marginal to the development of mathematics, are likely to have biographies with few if any connections to the largest connected component in our network, so such marginal cases will be lowly ranked and will have little effect on the rating of others.

Many of the problems in our data are most severe for lower ranked mathematicians.  Their biographies are likely to have been read and checked less often, they have fewer links so each link becomes relatively more more important for that mathematician. Again \href{https://en.wikipedia.org/wiki/Nygaard}{Kristen Nygaard} provides a nice example. His web page is extensive with many links but very few are to pages of mathematicians and it is likely that most readers are not interested in any of his links to mathematics. So those links are liable to be noisier and less reliable, just as his inclusion in the list of mathematicians at all is debatable.
The fat-tailed distribution of our measures shows that most mathematicians have low ratings.  Hence a small change in rating can produce a large change in rank.
This just emphasises that robustness checks are a vital part of any analysis, yet so often missing from discussions based on expert opinion alone. Our approach protects us against such uncertainty.

Another important conclusion is that our results support a key assumption we make, namely that the hyperlinks in Wikipedia biographies do contain useful information about the importance of individual mathematicians. Since our lists of top mathematicians look sensible, our own expert judgement, since they are relatively consistent over the three years of Wikipedia data we use, since they match well with a similar analysis of the MacTutor data, and since an informal survey is in rough agreement, it does suggest our method is reliable. This then supports our assumption that the hyperlinks reflect importance. In particular, we do not need to look at the context of each hyperlink to extract this information on importance. Of course, had we a reliable way to look at the context of each link, to reject those which were not useful (e.g.\ a link from mathematician A to mathematician B in text reading ``mathematician A never knew of the work by mathematician B''), then we might make our analysis more accurate and so reduce the uncertainties we place on our results.
The success of our assumption is not too surprising expected as search engines use the hyperlink structure to successfully  rank web pages as used in their recommendations to people searching the web. However it is interesting and non-trivial to see that in our specific context, network analysis of a large number of web sites can produce useful information about the most important mathematicians. Of course, there is much more information in these biographies than we use here and exploiting additional information is likely to improve the accuracy of the analysis, especially for less important mathematicians.

Our focus on data derived from crowdsourced biographies on Wikipedia means we should consider wider issues often raised in such a context. Other studies have looked at the accuracy of the information in Wikipedia, such as \citet{G05,WCR13,C06,WH07}, and the results seem generally positive.  Our use of a simple noise model makes some allowance for this issue. Issues over gender bias (for example see \citet{WGJS15}) or cultural bias (e.g.\ see \citet{EALKVS15}) may well be relevant here but we do not study them in any detail.  Since many of our most important mathematicians are historical figures\footnote{This is one reason why modern citation analysis cannot be used for our study.}, there is a further complication in that it may be hard to untangle inherent bias in the editors of Wikipedia biographies of historical mathematicians, from the bias present in the intermediary sources, such as those cited in the Wikipedia pages, or indeed biases inherent in societies in which these mathematicians lived.

For instance it is very obvious when looking at our data that there are very few Asian mathematicians in our results.  Is this a reflection of their true lack of influence on modern mathematics? One might hope that practical necessity or a greedy enthusiasm for greater knowledge, power, wealth can sometimes overcome cultural tensions.  Certainly examples such as the influence of Arab mathematics and astronomy on Western science contains many positive examples of this.

One aspect of our work makes us particularly vulnerable to cultural differences is our focus on individuals. If the original source of mathematical innovation was lost, deliberately or otherwise, we would fail to track this relationship through our biographies. This focus on the individual is an explicit bias on our data in terms of the view it provides on the history of mathematics.

In terms of cultural biases and naming of individuals, the example of the \href{http://www-history.mcs.st-andrews.ac.uk/Biographies/Banu_Musa.html}{Ban\={u} M\={u}s\={a}} brothers may be instructive where it is hard from the historical record to assign credit for a piece of work to one particular brother of the three. Another example of how we may lose track of individual historical mathematicians comes from the well known early text on Chinese mathematics, the Ji\v{u}zh\={a}ng Su\`{a}nsh\`{u}  (\href{https://en.wikipedia.org/wiki/The_Nine_Chapters_on_the_Mathematical_Art}{Nine Chapters on the Mathematical Art}).  This appears to be a compilation of knowledge built up over ten or more centuries and includes a Chinese version of Pythagoras.  Whether this was derived independently from the West or even if this Chinese work provided the basis for developments in the West is not known.  For our purposes, the lack of named individuals means this Chinese text is excluded from our data.

Judging the strength of this problem is difficult.  On the positive side we do see that later Chinese mathematicians are known and do appear in our data.  Both \href{https://en.wikipedia.org/wiki/Zhang_Heng}{Zhang Heng} (1st -2nd c.\ CE) and \href{https://en.wikipedia.org/wiki/Liu_Hui}{Liu Hui} (3rd c.\ BCE) are present and they have a relatively high rank (100 in our 2017 data) suggesting that the crowdsourcing of English Wikipedia may well be able to compensate for possible cultural bias from whatever various sources, just as has been shown for gender \citep{WGJS15}. It would be interesting  in terms of cultural biasses to perform similar analysis on the mathematicians listed in Wikipedia pages of other languages.

Another issue with our historical and often important mathematicians is that the artefacts recording mathematical results will not usually survive. For instance the Su\`{a}n sh\`{u} sh\={u} (\href{https://en.wikipedia.org/wiki/Book_on_Numbers_and_Computation}{Writings on Reckoning})  is a mathematical text found on bamboo strips in China (dated around 200 BCE).  Several of the strips have decayed and this reminds us that many such texts would not have survived (see chapter 3 of \citet{S12e} for further examples).  This text has a couple of names assumed to be the authors but, unlike the later Chinese mathematicians, we have no other knowledge of these people to allow us to connect their work to other developments. Thus these individuals, and others like them, play no role in our analysis.

Overall, what our work shows that there is, of course, no single answer to the simple question --- who is the most important mathematician. Social science is much harder than mathematics precisely because  questions have no single answer.  Our results from different years and different sources do not give the same answer reminding us that no researcher should ever take a single set of centrality measures at face value.  However we also show that it is possible to estimate uncertainties in these measures, as we have done with our noise model and by using different data sets. Armed with a sense of the uncertainty in such results, one can then look for patterns and genuine outliers.  Our work illustrates how to estimate uncertainty in social network measurements to gain further insights to add to existing debates.  It also emphasises that large-scale crowdsourced work can provide genuinely insights useful contributions.  A digital humanities approach such as ours does not replace high quality analysis of social science, it enhances that research.

Finally all the code and data used in this paper has been made available online \blindreview{(citation removed for double blind review).}{\citep{ECL17}.}

\normalversiononly{
\section*{Acknowledgement}

TSE would like to thank the many colleagues with whom he has worked on earlier studies of the \href{http://www-history.mcs.st-and.ac.uk/}{MacTutor History of Mathematics archive} \citep{MacTutor} data. In particular TSE thanks C.Clarke \citeyear{C11} and N.Hopkins \citeyear{H11b} whose work and reports provided the information on centrality measurements of MacTutor referred to in this paper.
}

\newpage

\newpage
\appendix
\setcounter{section}{0}
\renewcommand{\thesection}{\Alph{section}}
\setcounter{table}{0}
\renewcommand{\thetable}{\thesection\arabic{table}}
\setcounter{figure}{0}
\renewcommand{\thefigure}{\thesection\arabic{figure}}
\setcounter{equation}{0}
\renewcommand{\theequation}{\thesection\arabic{equation}}

\section{Appendix}

Additional information is provided in this appendix.

\subsection{Variance in Degree in Noise Model}

\begin{figure}[htb!]
\centering
\includegraphics[width=0.75\textwidth]{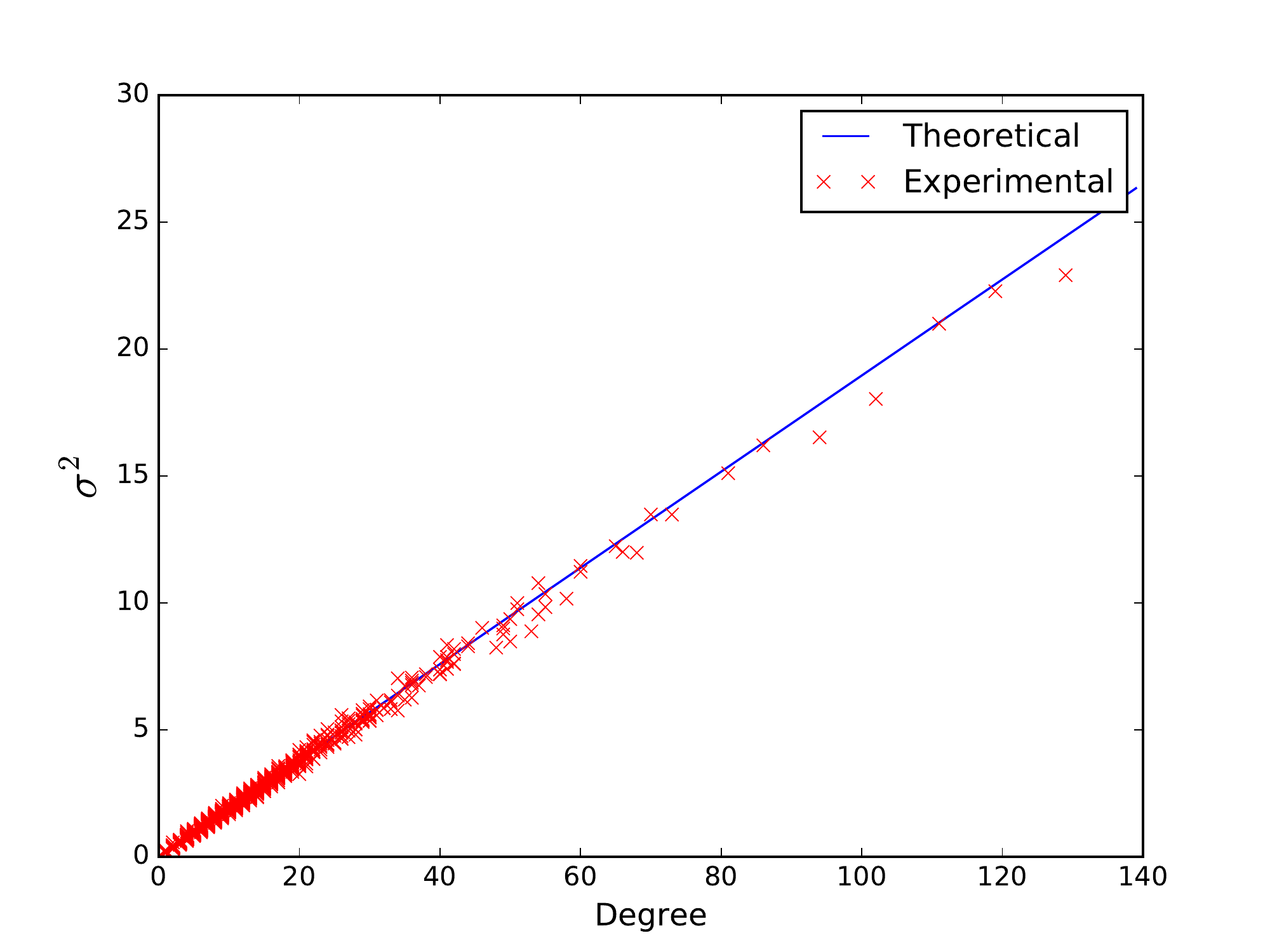}
\caption{Each cross indicates the standard deviation in degree of one node after $1000$ simulations. The theoretical result that $\sigma \approx 0.44 \sqrt{k_\mathrm{orig}}$ is compatible with this numerical result as the linear fit between variance and degree shows (an adjusted-r square value of $0.997$).}
\label{degstd}
\end{figure}

\subsection{Formal Definitions of Centrality Measures}\label{sformalcent}

The closeness $c_v$ for a vertex $v$ is defined to be \citep{B50,HSS08,N09b}
\begin{equation}
 c_v= \frac{(|\Ccal_v|-1)} {\sum_{u \in \Ccal_v \setminus v} d(u,v) } \, ,
 \label{close}
\end{equation}
where $d(u,v)$ is the length of the shortest path between vertex $u$ and some distinct vertex $v$ which is in the same component, $\Ccal_u$, as $u$. Note that we use a standard normalisation using $N$, the number of vertices, but this is irrelevant after our rescaling \tref{rescale}.

Our formal definition of betweenness $b_v$ of a vertex $v$ is\citep{F77,B08d,HSS08,N09b}
\begin{equation}
 b_v= \sum_{s,t\  \in \Ccal_v} \frac{\sigma (s,t|v)}{\sigma (s,t)} \, .
 \label{bet}
\end{equation}
Here $\Ccal_v$ is the set of vertices of the component containing vertex v, $\sigma(s,t)$ is the number of shortest paths available from vertex $s$ to $t$, and $\sigma(s,t|v)$ is the number of shortest paths from $s$ to $t$ which pass through vertex $v$. This takes account of cases where there are two or more shortest paths between a pair of nodes  $s$ and $t$.

Eigenvalue centrality derived from the  the adjacency matrix $\Amat$, which we define such that $A_{ij}$ is one (zero) if there is a link (no link) from vertex $i$ to vertex $j$. The Eigenvector centrality for a vertex $i$ is simply the $i$-th entry of the eigenvector of $\Amat$ associated with the largest eigenvalue \citep{N09b,HSS08}. We perform our analysis on the largest component which then guarantees a unique value for each node.

PageRank is defined in terms of a transfer matrix, $\Tmat$ where each entry, $T_{ji}$  represents the probability of a random walker at Vertex $i$ moving to vertex $j$ at the next time step. So we have that
\begin{equation}
 T_{ji}= \frac{1}{s_{i}^{(\mathrm{out})}} A_{ji} \, ,
 \quad \mathrm{where}
 \quad s^{\mathrm{(out)}}_{i}= \sum_j A_{ji} \, .
\end{equation}
An additional stochastic process also occurs. At each step, with probability $\alpha$, the random walker follows a link chosen at random as given by the transfer matrix $\Tmat$  but with probability $(1-\alpha)$ the current walk is deemed to end, or equivalently, we follow a new user or a new walk by starting at a randomly chosen vertex. The Markovian matrix $\Gmat$ which describes this process is given by
\begin{equation}
  G_{ij} = \alpha T_{ij}+ (1-\alpha) \frac{1}{N}\,
  \label{PageRank}
\end{equation}
where $N$ corresponds to total number of vertices and $\alpha$ is the damping factor, chosen to be $\alpha=0.85$ in this work. The probability that a random walker is at vertex $i$ in the long-time limit is proportional to the PageRank for that vertex and this is given by the $i$-th entry of the eigenvector associated with the largest Eigenvalue of the $\Gmat$.
This makes PageRank similar to Eigenvector but different to the other centrality measures considered in that PageRank probes the whole network structure using walks of all types.

\subsection{Additional Results}\label{appssaddres}

\subsubsection{MacTutor Results}

\begin{table}[htb!]
\centering
\small\begin{tabular}{c c c c c c c}
\hline\hline
Rank & Degree & Closeness & Betweenness & PageRank & \parbox[t][1ex][c]{10ex}{O(2nd)\\
 Clustering} & Word Count\\[0.5ex]
\\
\hline
1 & Newton & Newton & Euclid & Euclid & Hilbert & Euler\\[1ex]
2 & Hilbert & Hilbert & Newton & Newton & Newton & Galileo\\[1ex]
3 & Euclid & Riemann & Euler & Laplace & Euclid & Leibniz\\[1ex]
4 & Riemann & Euler & Riemann & Hilbert & Riemann & Newton\\[1ex]
5 & Euler & Euclid & Van der Waerden & Lagrange & Klein & Laplace\\[1ex]
6 & Klein & Cauchy & Weierstrass & Euler & Euler & Nash\\[1ex]
7 & Weierstrass & Gauss & Hilbert & Riemann & Weierstrass & Ptolemy\\[1ex]
8 & Poincare & Klein & Dieudonne & Gauss & Descartes & Tait\\[1ex]
9 & Gauss & Dirichlet & Cartan Henri & Klein & Leibniz & Kepler\\[1ex]
10 & Einstein & Laplace & Cauchy & Aristotle & Gauss & Aristotle\\[1ex]
11 & Cauchy & Lagrange & Hardy & Cauchy & Einstein & Lax Anneli\\[1ex]
12 & Lagrange & Poincare & Leibniz & Leibniz & Huygens & Copernicus\\[1ex]
13 & Laplace & Fourier & Dirichlet & Einstein & Lagrange & Euclid\\[1ex]
14 & Leibniz & Weierstrass & Weil & Jacobi & Aristotle & Polya\\[1ex]
15 & Hardy & Legendre & Fermat & Weierstrass & Poincare & Escher\\[1ex]
\hline
\end{tabular}
\caption{Centrality results for the top fifteen mathematicians in the directed  network based on the hyperlinks between biographies on the \href{http://www-history.mcs.st-and.ac.uk/}{MacTutor} \citep{MacTutor} database, data from 2011.  Copy of Table 4 from the appendix of \blindreview{(citation of unpublished thesis removed for double blind review).}{\citet{C11}.}}
\label{MacTutorTop15}
\end{table}

\clearpage
\subsubsection{Wikipedia 2013 Results}

\begin{figure}[htb!]
\centering
\includegraphics[width=0.75\textwidth]{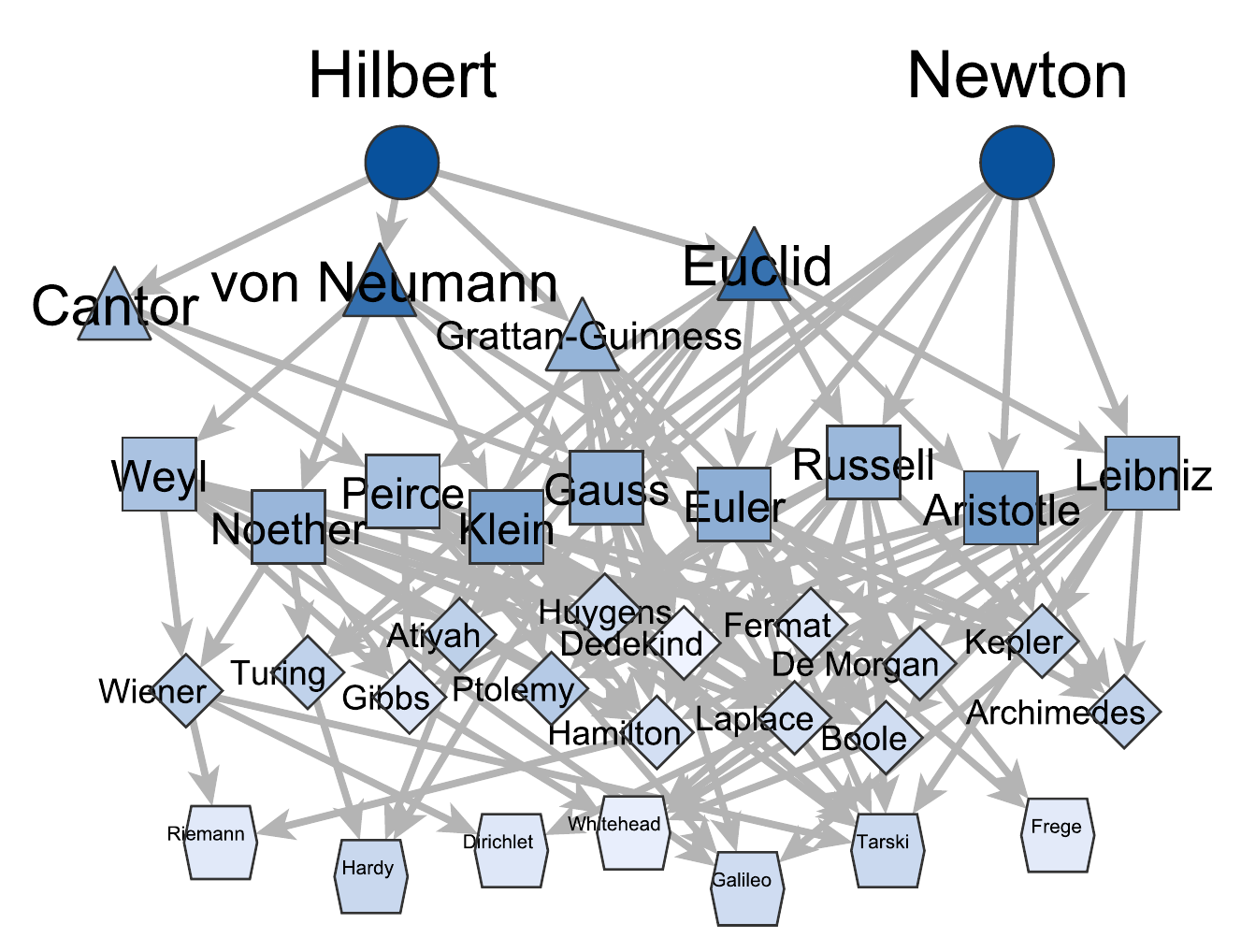}
\caption{The top 36 mathematicians (2013 data) by a scaled average of the five ratings: Degree,  Betweenness, Closeness, Eigenvector centrality and PageRank. Mathematician `A' is connected placed higher than mathematician `B' if each of their five ratings of 'A' is higher than the corresponding rating for `B'.  The arrow of the line points from the higher to the lower ranked mathematician but only those essential for the logical relationships are shown (a Hasse diagram of the corresponding poset, equivalently the transitively reduced form of the corresponding directed acyclic graph). The shape of a node, size of label and the vertical location reflects the `height' of each node in the corresponding poset (see text for definition).  The colour reflects the average scaled rating of each mathematician.}
\label{dag13}
\end{figure}

\begin{table}[htb!]
\centering
\resizebox{\textwidth}{!}{\begin{tabular}{cccccccc}
\hline \hline
Name                      & Degree & Betweenness & Closeness & Eigenvector & PageRank & Average mark & Rank \\ \hline
David Hilbert             & $87.39$     & $100$            & $100$          & $88.27$          & $85.64$       & $92.26$      & $1$  \\
Isaac Newton              & $100$       & $69.88$          & $90.59$        & $100$            & $100$         & $92.09$      & $2$  \\
John von Neumann          & $74.79$     & $92.67$          & $97.07$        & $58.73$          & $81.55$       & $80.96$      & $3$  \\
Euclid                    & $82.35$     & $60.63$          & $91.19$        & $85.62$          & $80.55$       & $80.07$      & $4$  \\
Aristotle                 & $66.39$     & $25.17$          & $83.43$        & $84.76$          & $62.28$       & $64.41$      & $5$  \\ \hline
Felix Klein               & $67.23$     & $33.82$          & $89.5$         & $51.9$           & $65.78$       & $61.65$      & $6$  \\
Leonhard Euler            & $56.3$      & $31.36$          & $88.08$        & $55.6$           & $58.45$       & $57.96$      & $7$  \\
Gottfried Wilhelm Leibniz & $51.26$     & $24.55$          & $88.27$        & $70.74$          & $50.29$       & $57.02$      & $8$  \\
Carl Friedrich Gauss      & $57.14$     & $35.62$          & $88.86$        & $39.64$          & $60.16$       & $56.28$      & $9$  \\
Ivor Grattan-Guinness     & $41.18$     & $38.02$          & $92.68$        & $70.88$          & $37.24$       & $56$         & $10$ \\ \hline
Emmy Noether              & $50.42$     & $38.29$          & $92.73$        & $40.92$          & $52.82$       & $55.04$      & $11$ \\
Bertrand Russell          & $46.22$     & $25.77$          & $90.22$        & $66.09$          & $44.69$       & $54.6$       & $12$ \\
Georg Cantor              & $42.86$     & $30.49$          & $92.51$        & $65.06$          & $39.24$       & $54.03$      & $13$ \\
Charles Sanders Peirce    & $40.34$     & $26.26$          & $90.63$        & $61.68$          & $39.08$       & $51.6$       & $14$ \\
Hermann Weyl              & $40.34$     & $34.8$           & $92.85$        & $44.82$          & $40.6$        & $50.68$      & $15$ \\ \hline
Ptolemy                   & $53.78$     & $9.82$           & $75.1$         & $50.47$          & $49.56$       & $47.75$      & $16$ \\
Norbert Wiener            & $36.97$     & $29$             & $92.12$        & $32.47$          & $40.58$       & $46.23$      & $17$ \\
Michael Atiyah            & $43.7$      & $36.68$          & $85.45$        & $10.21$          & $52.71$       & $45.75$      & $18$ \\
Johannes Kepler           & $41.18$     & $15.39$          & $81.1$         & $51.71$          & $39.15$       & $45.71$      & $19$ \\
Alan Turing               & $36.13$     & $29.35$          & $89.64$        & $27.83$          & $41.93$       & $44.98$      & $20$ \\  \hline
Archimedes                & $44.54$     & $9.22$           & $77.76$        & $53.15$          & $39.76$       & $44.88$      & $21$ \\
G. H. Hardy               & $35.29$     & $28.45$          & $88.73$        & $20.51$          & $40.96$       & $42.79$      & $22$ \\
Alfred Tarski             & $36.13$     & $21.1$           & $84.42$        & $32.26$          & $38.13$       & $42.41$      & $23$ \\
Augustus De Morgan        & $31.09$     & $12.96$          & $85.58$        & $45.63$          & $32.1$        & $41.47$      & $24$ \\
Christiaan Huygens        & $35.29$     & $10.63$          & $82.11$        & $43.32$          & $35.23$       & $41.32$      & $25$ \\  \hline
Galileo Galilei           & $36.97$     & $10.73$          & $79.68$        & $43.75$          & $35$          & $41.23$      & $26$ \\
George Boole              & $31.09$     & $11.72$          & $84.77$        & $47.2$           & $29.34$       & $40.82$      & $27$ \\
William Rowan Hamilton    & $28.57$     & $21.95$          & $87.96$        & $33.27$          & $29.17$       & $40.18$      & $28$ \\
Pierre-Simon Laplace      & $33.61$     & $13.39$          & $84.3$         & $34.12$          & $34.67$       & $40.02$      & $29$ \\
Srinivasa Ramanujan       & $32.77$     & $24.26$          & $86.36$        & $15.43$          & $40.06$       & $39.78$      & $30$ \\  \hline
Nicolaus Copernicus            & $35.29$     & $6.27$           & $77.25$        & $43.37$          & $32.53$       & $38.94$      & $31$  \\
Pierre de Fermat               & $26.05$     & $12.17$          & $86.16$        & $42.04$          & $23.67$       & $38.02$      & $32$  \\
Josiah Willard Gibbs           & $26.05$     & $13.41$          & $89.22$        & $33.77$          & $25.85$       & $37.66$      & $33$  \\
Lejeune Dirichlet & $31.93$     & $8.37$           & $83.54$        & $32.29$          & $30.11$       & $37.25$      & $34$  \\
Apollonius of Perga            & $31.93$     & $6.81$           & $78.37$        & $40.02$          & $28.78$       & $37.18$      & $35$ \\
\hline
\end{tabular}}
\caption{Centrality scores for top 35 mathematicians from 2013 data (without noise) given on a common scale with 100 for the largest value according to \tref{rescale}. Ordered in terms of their average score rating.}
\label{score2013}
\end{table}

\clearpage
\subsubsection{Wikipedia 2017 Results}

\begin{figure}[htb!]
\centering
\includegraphics[width=0.75\textwidth]{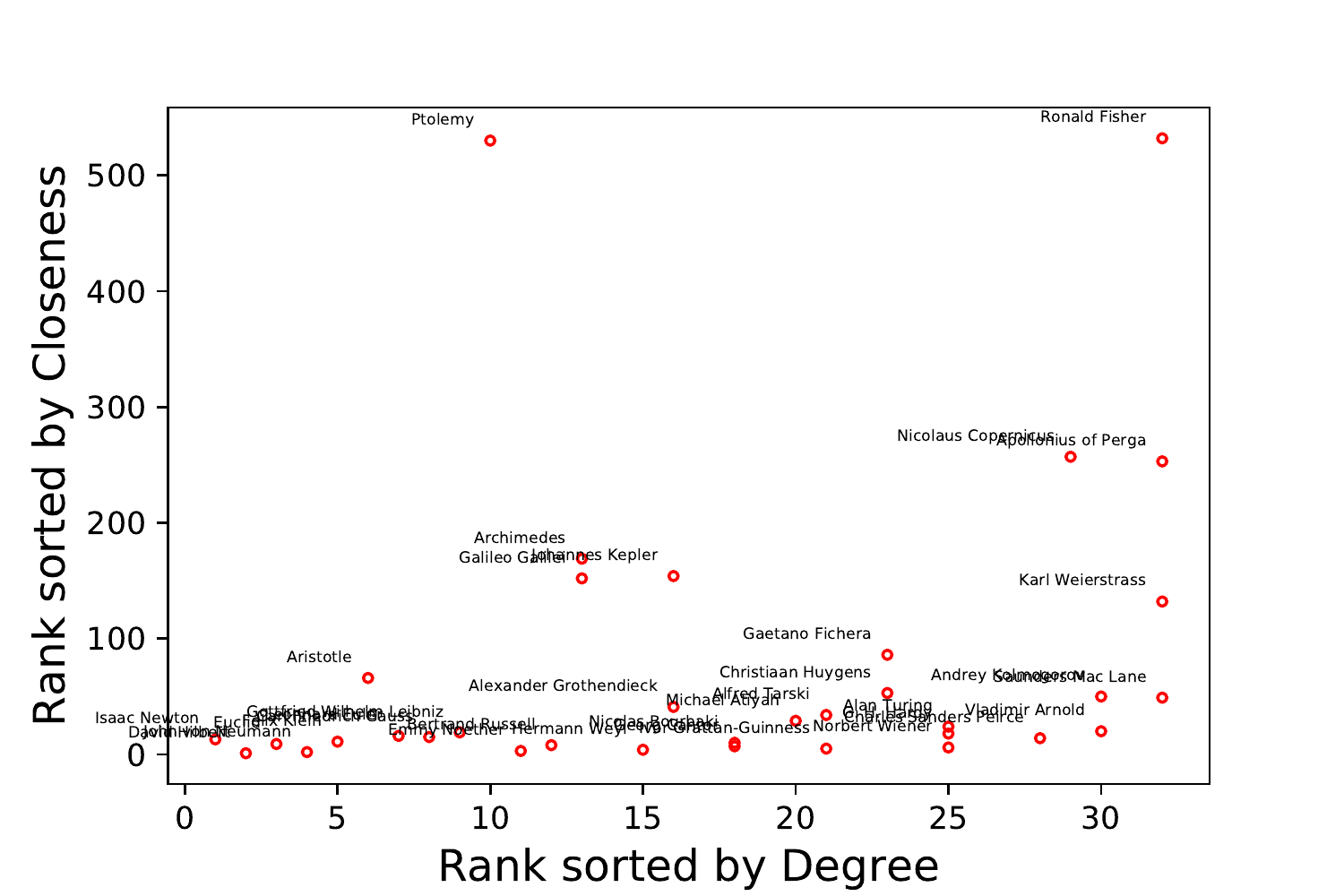}
\caption{A comparison of the rank of mathematicians by degree and by closeness.  the top 35 mathematicians by their average score in the 2017 Wikipedia data are shown under different centrality measures.}
\label{degclo35}
\end{figure}

\begin{figure}[htb!]
\centering
\includegraphics[width=0.75\textwidth]{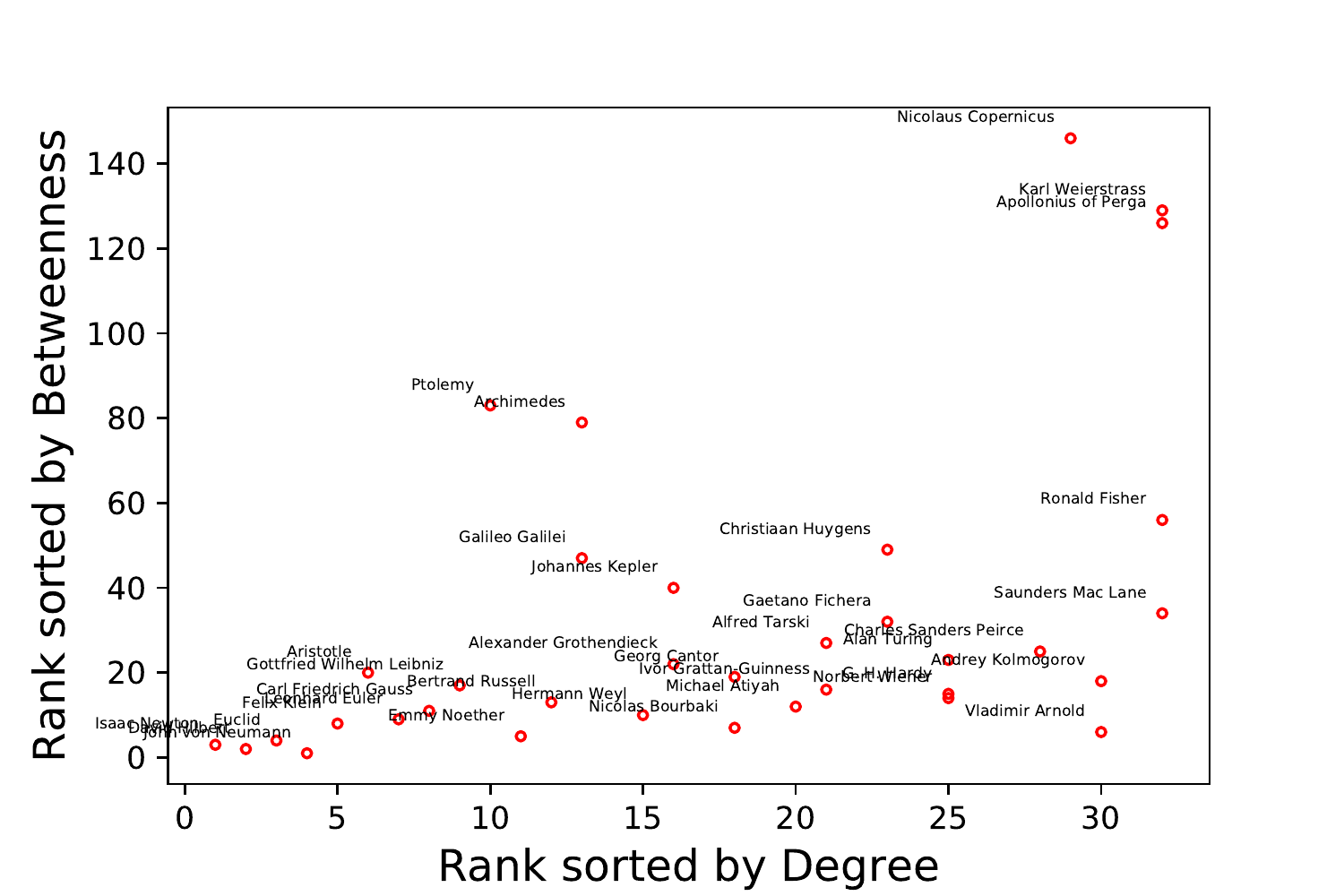}
\caption{A comparison of the rank of mathematicians by degree and by betweenness.  The top 35 mathematicians by their average score in the 2017 Wikipedia data are shown under different centrality measures.}
\label{degbet35}
\end{figure}

\clearpage
\subsubsection{Wikipedia 2018 Results}

\begin{table}[h]
\centering
\begin{tabular}{r||c|cc|c}
Quantity                & 2013    & 2018    & \%       & 2018                  \\
                        &         &         & Increase & After Rewiring        \\ \hline \hline
Mathematicians/Vertices & $6050$  & $8317$  & +37.4\%  &    $8317$             \\ \hline
Hyperlinks              & $15120$ & $22669$ & +49.9\%  &   $22669$             \\ \hline
Undirected Edges        & $9701$  & $14292$ & +47.3\%  &   $14291.42 \pm 0.8$ \\ \hline
Average Degree          & $3.21$  & $3.44$  &  +7.2\%  &    $3.44\pm 0.0001$              \\ \hline
Vertices in largest component         & $4096$  & $5829$  & +42.6\%  &  $5710.17 \pm 17.7$    \\ \hline
Edges in largest component            & $9573$  & $14115$ & +47.4\%  & $14152.41 \pm 10.0$     \\ \hline
Average Degree in largest component   & $4.71$  & $4.84$  &  +2.7\%  &    $4.96 \pm 0.01$    \\ \hline
Network Diameter        & $13$    & $15$    &  +15.3\%  &    $14.12 \pm 0.96$     \\ \hline
Average Path Length     &  $5.07$ & $5.14$  &  +1.4\%  &    $4.90 \pm 0.01$    \\ \hline
Clustering Coefficient  &  $0.13$ & $0.12$  &  -7.7\%  &    $0.09 \pm 0.002$   \\
\end{tabular}
\caption{Network parameters for the 2013 and 2018 dataset, the percentage change between 2013 and 2017 data, and the mean values found for an ensemble of 1000 rewired 2018 data sets (with one standard deviation uncertainty quoted) as defined by our noise model of \sref{snoise} with $p=0.1$. }
\label{para_18}
\end{table}

\begin{figure}[htb!]
\centering
\includegraphics[width=0.45\textwidth]{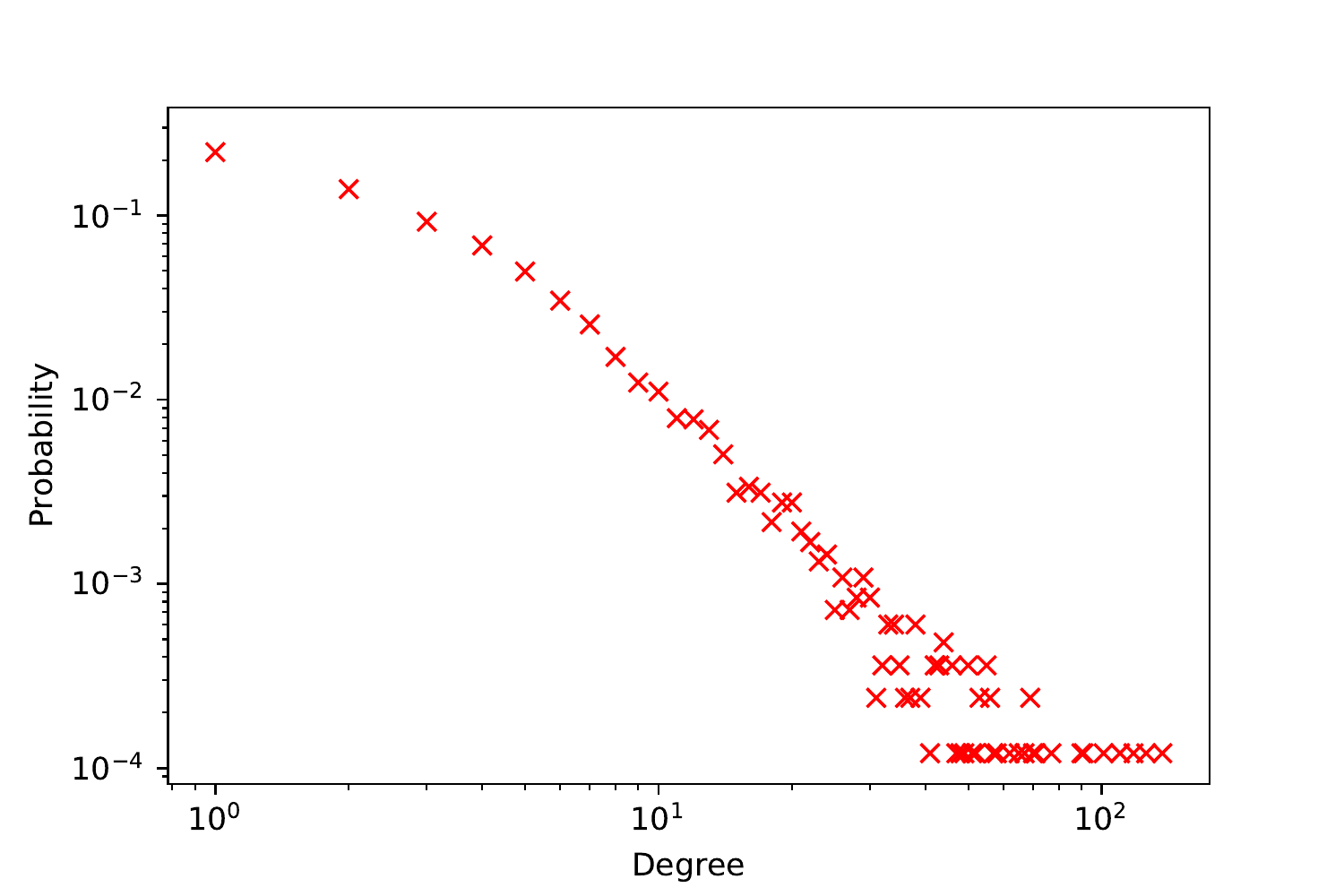}
\vspace*{0.05\textwidth}
\includegraphics[width=0.45\textwidth]{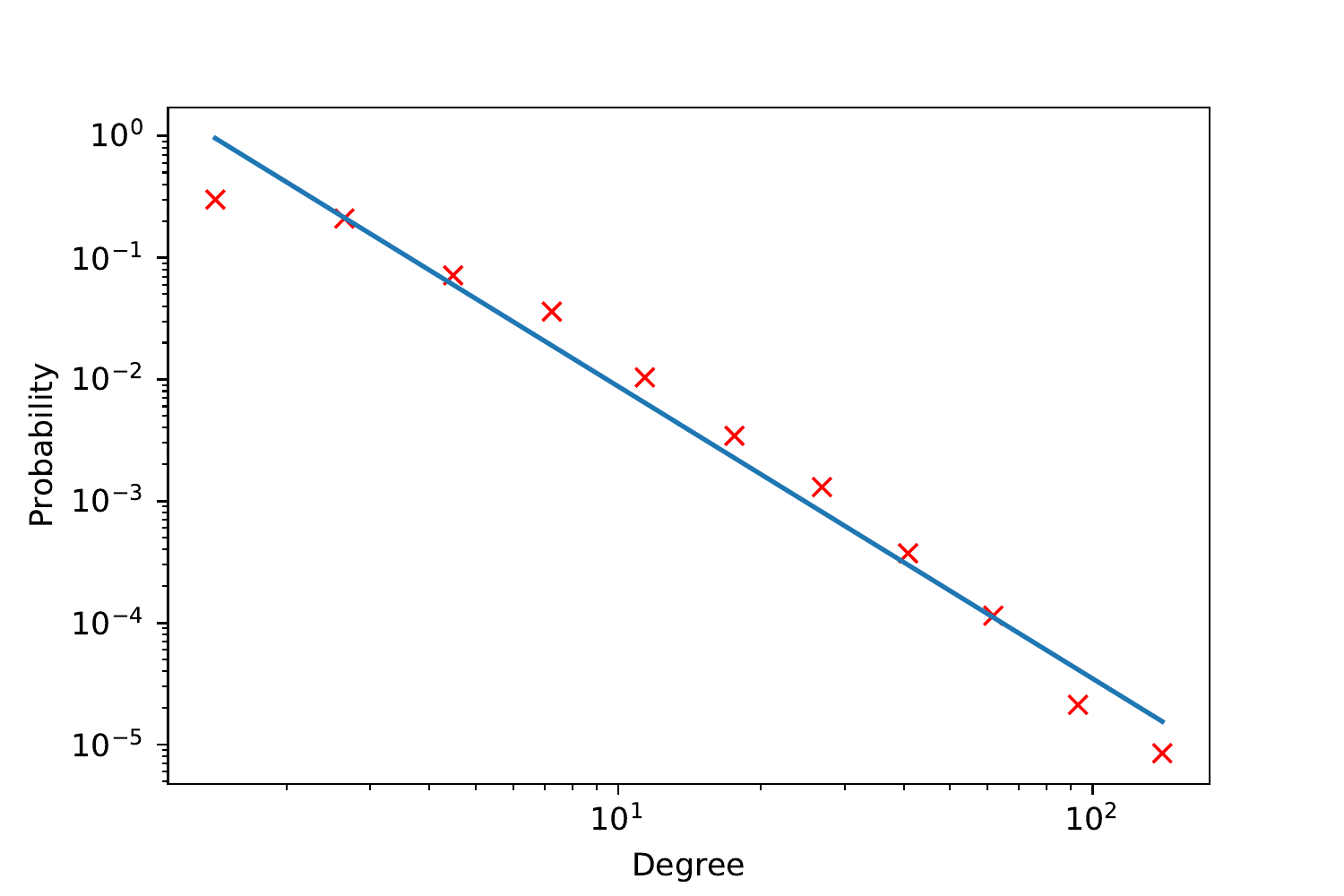}
\caption{On the left, the degree distribution for the 2018 network of mathematicians.  On the right the data is binned (using log binning with the ratio of consecutive bin edges set to be 1.5) and a best fit straight line to this data is shown added (slope is $-2.70 \pm 0.14$).}
\label{degreedisv2}
\end{figure}

\begin{figure}[htb!]
\centering
\includegraphics[width=0.75\textwidth]{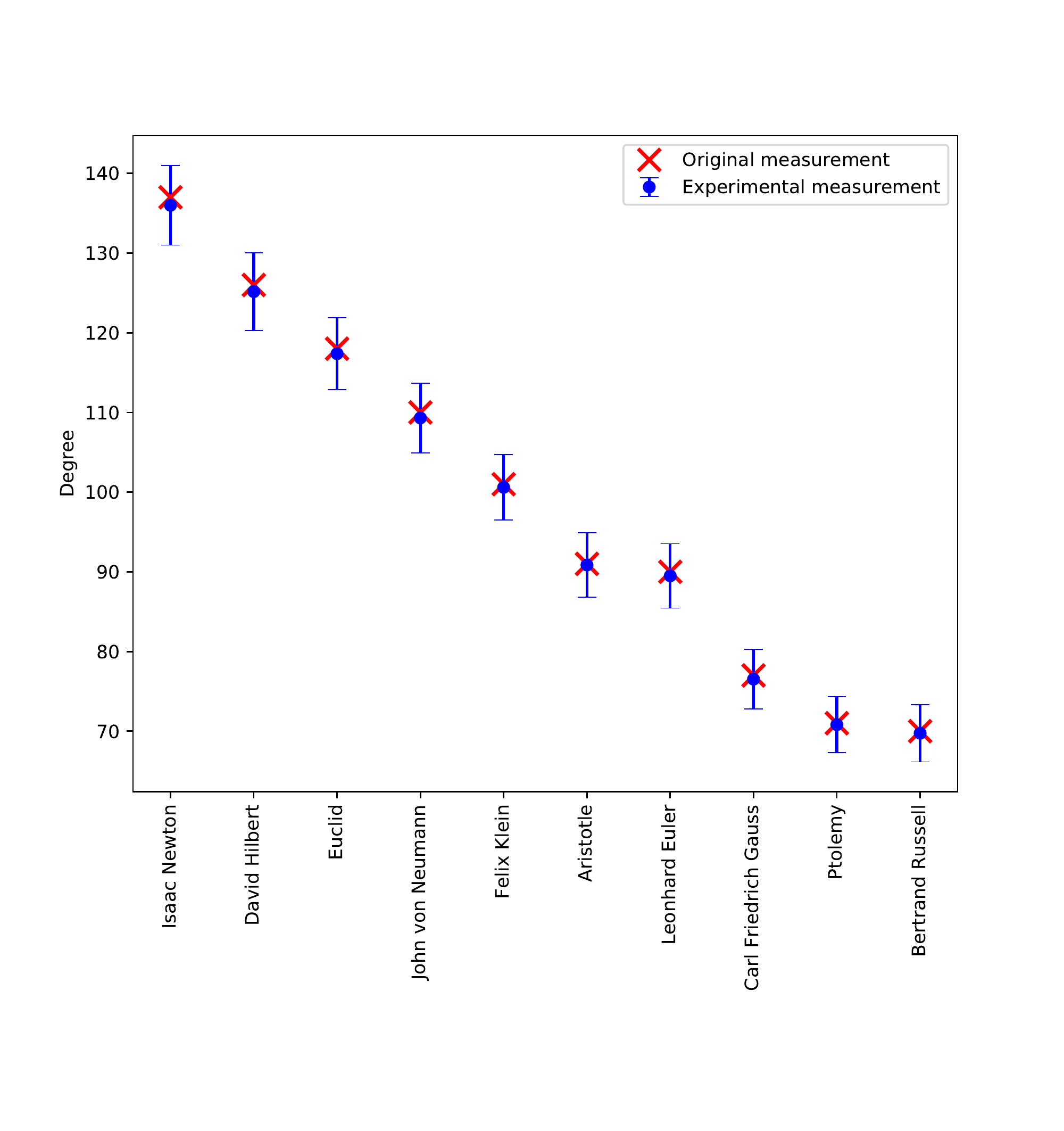}
\caption{Degree distribution for the ten mathematicians whose Wikipedia biographies have the largest degree in the 2018 data (crosses).  The circles give the mean degree for the same mathematicians as measured over 1000 simulations using our noise model of \sref{snoise} where the error bars are specified by one standard deviation.}
\label{zipfv2}
\end{figure}

\begin{figure}[htb!]
\centering
\includegraphics[width=0.6\textwidth]{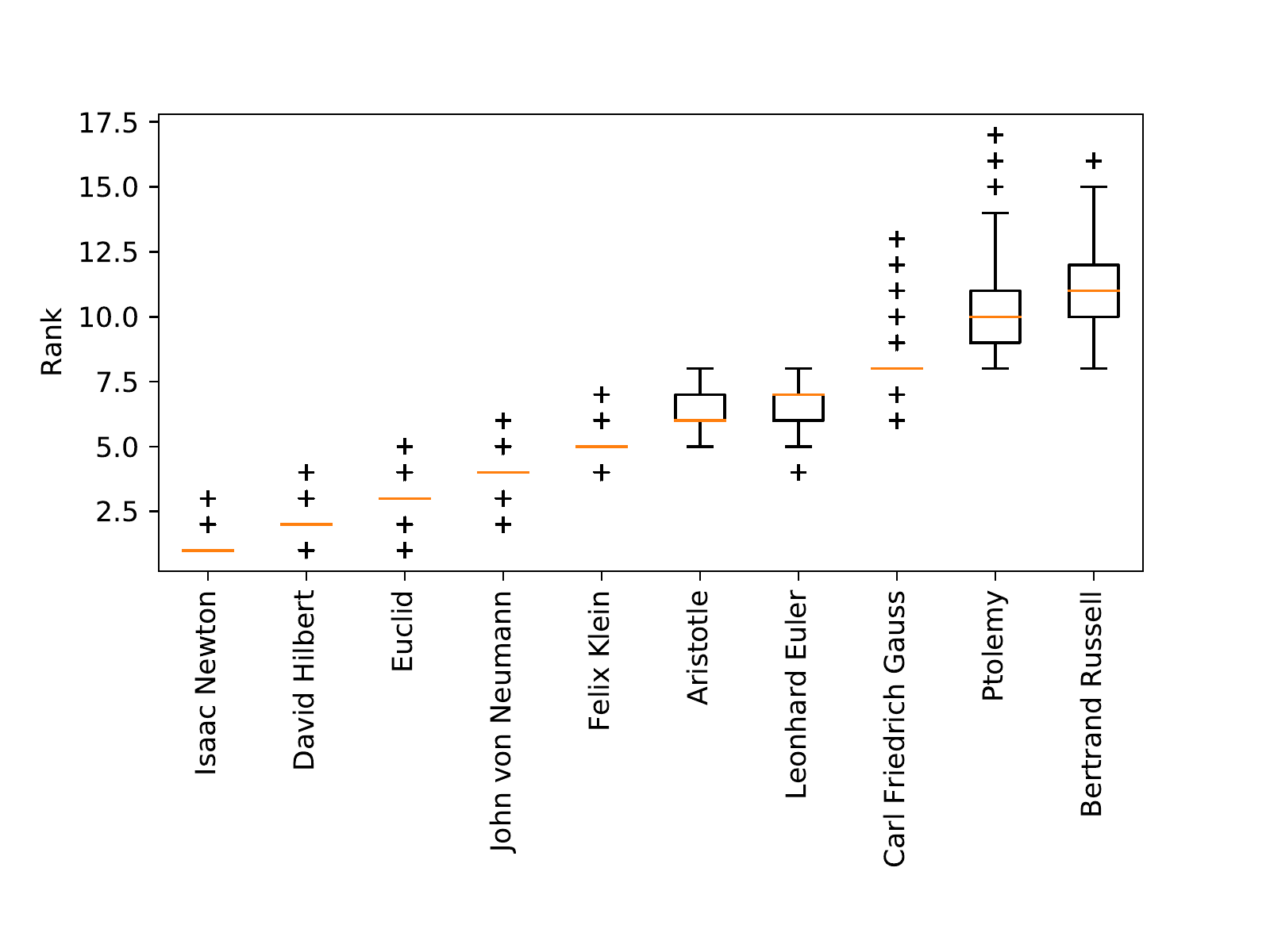}
\caption{Whisper box plot for degree rank of mathematicians from 1000 simulations of our noise model from \sref{snoise} applied to the 2018 data.
The lower and upper edges of blue box show the $25$ percentile ($Q_1$) and the $75$ percentile ($Q_3$) of the rank of each mathematician, the red line in the middle of the box is the median. Given the small variation here, these lines often coincide. The black lines, at the end of the whiskers connected to the box, are defined to be at $Q_1 - 1.5 (Q_3-Q_1)$ and  $Q_3 + 1.5 (Q_3-Q_1)$. The remaining black crosses beyond the whiskers indicate outliers beyond the whiskers.}
\label{degreeboxv2}
\end{figure}

\begin{figure}[htb!]
\centering
\includegraphics[width=0.6\textwidth]{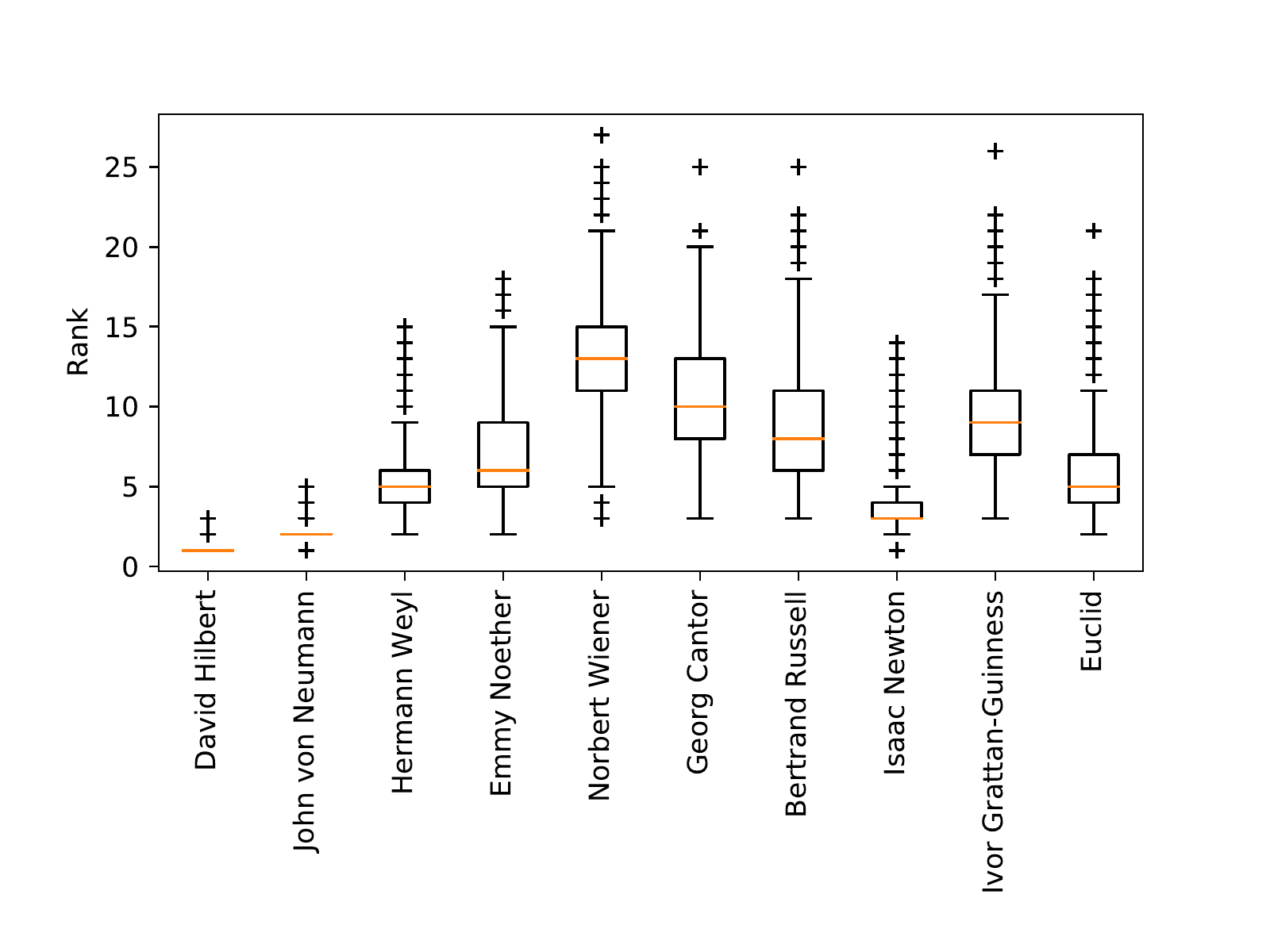}
\caption{Whisper box plot for the rank of mathematicians by closeness, for the ten mathematicians with largest closeness.  The closeness centrality is calculated for the largest component of the 2018 data and the uncertainties are estimated using 1000 simulations using the noise model of \sref{snoise} with $p=0.1$. The criteria used to place the boxes and other features of the plot are as in \fref{degreebox}.}
\label{cloboxv2}
\end{figure}

\begin{figure}[htb!]
\centering
\includegraphics[width=0.6\textwidth]{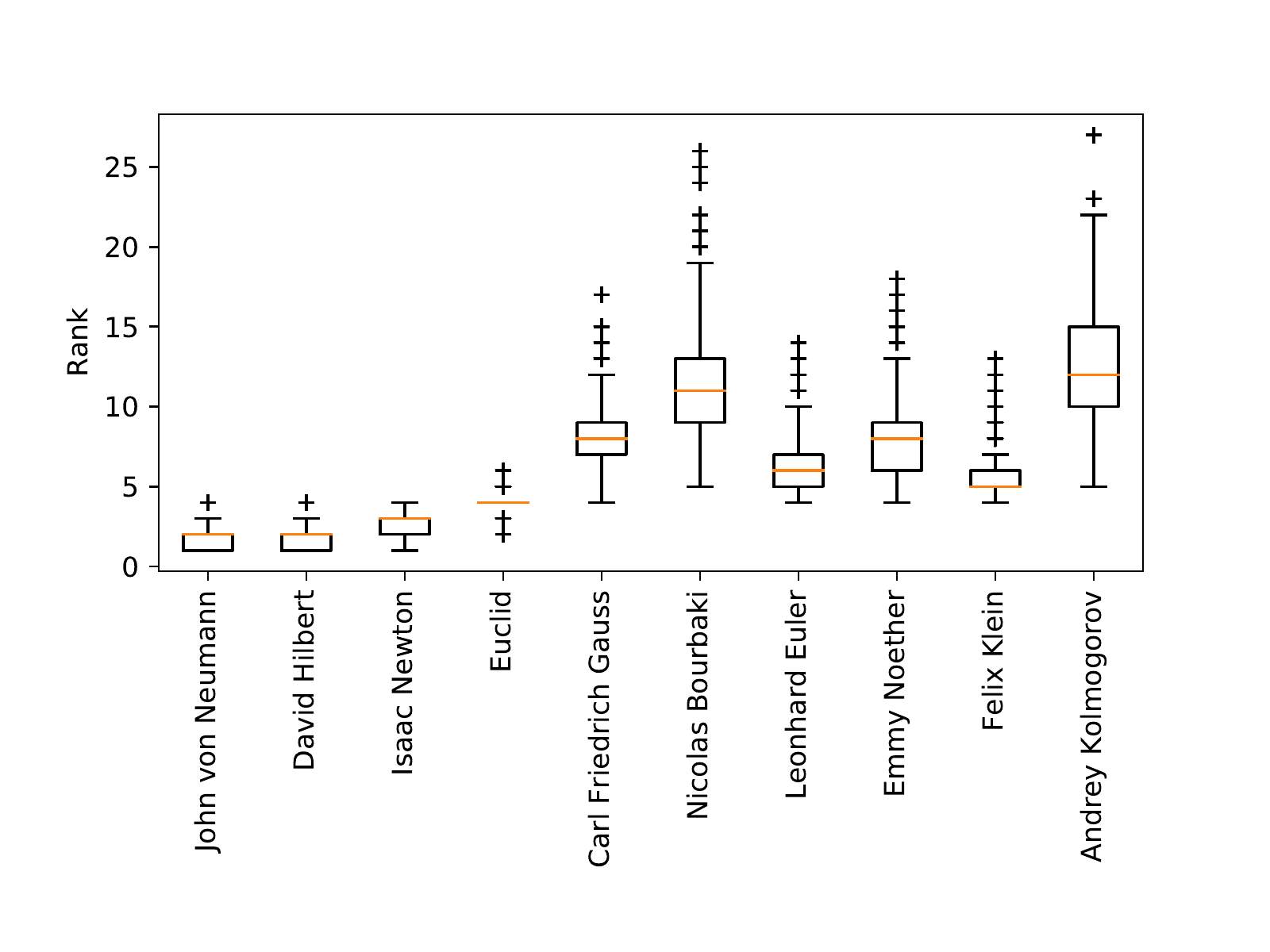}
\caption{Whisper box plot for rank by betweenness of the ten mathematicians with highest betweenness.  This is for the largest component of the 2018 data based on 1000 simulations using the noise model of \sref{snoise}.
The criteria used to place the boxes and other features of the plot are as in \fref{degreebox}.}
\label{betboxv2}
\end{figure}

\begin{figure}[htb!]
\centering
\includegraphics[width=0.6\textwidth]{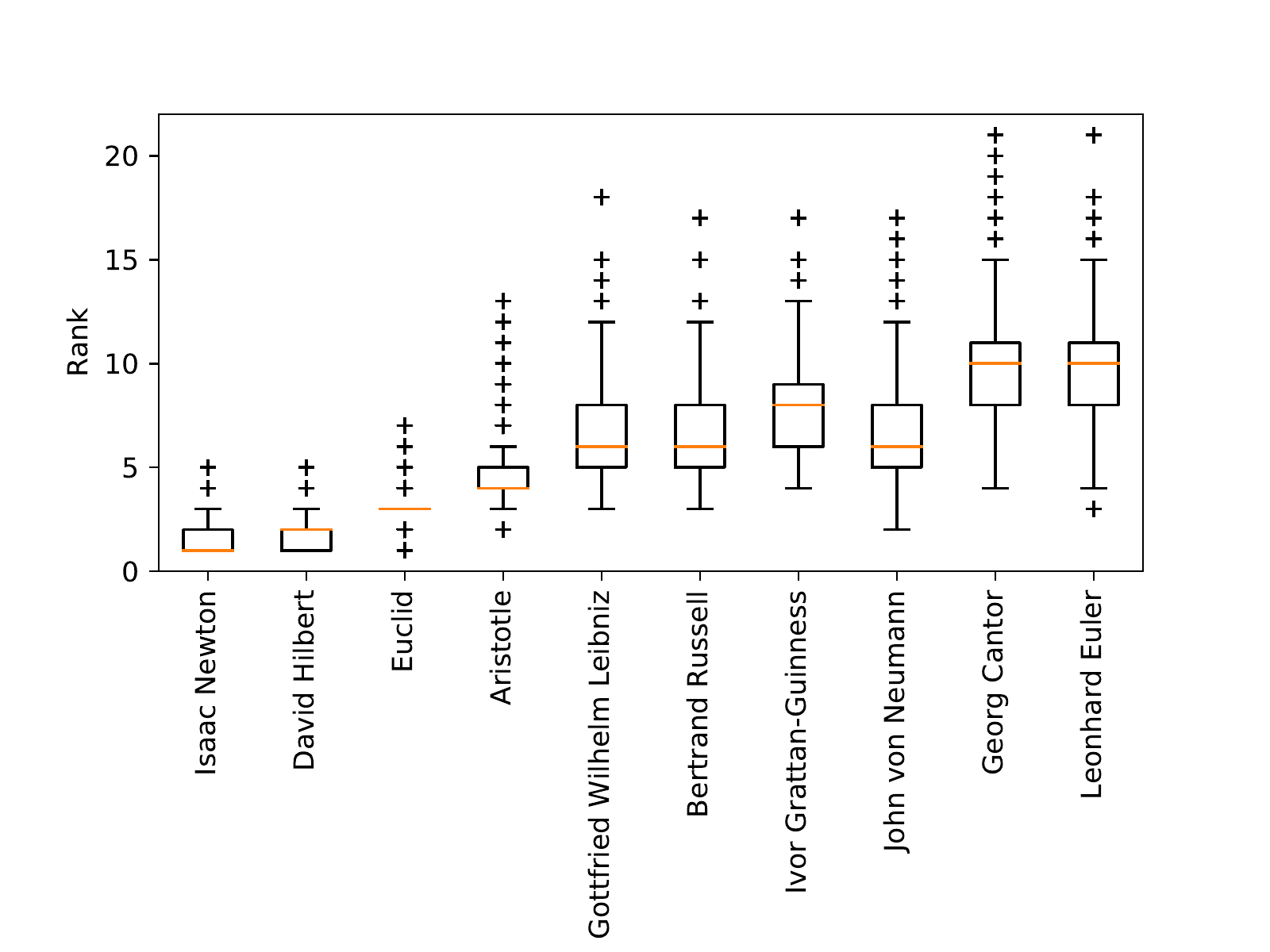}
\caption{Whisper box plot for rank of mathematicians derived from their Eigenvalue centrality.
This is for the largest component of the 2017 data based on 1000 simulations using the noise model of \sref{snoise}.
The criteria used to place the boxes and other features of the plot are as in \fref{degreebox}.}
\label{eigboxv2}
\end{figure}

\begin{figure}[htb!]
\centering
\includegraphics[width=0.6\textwidth]{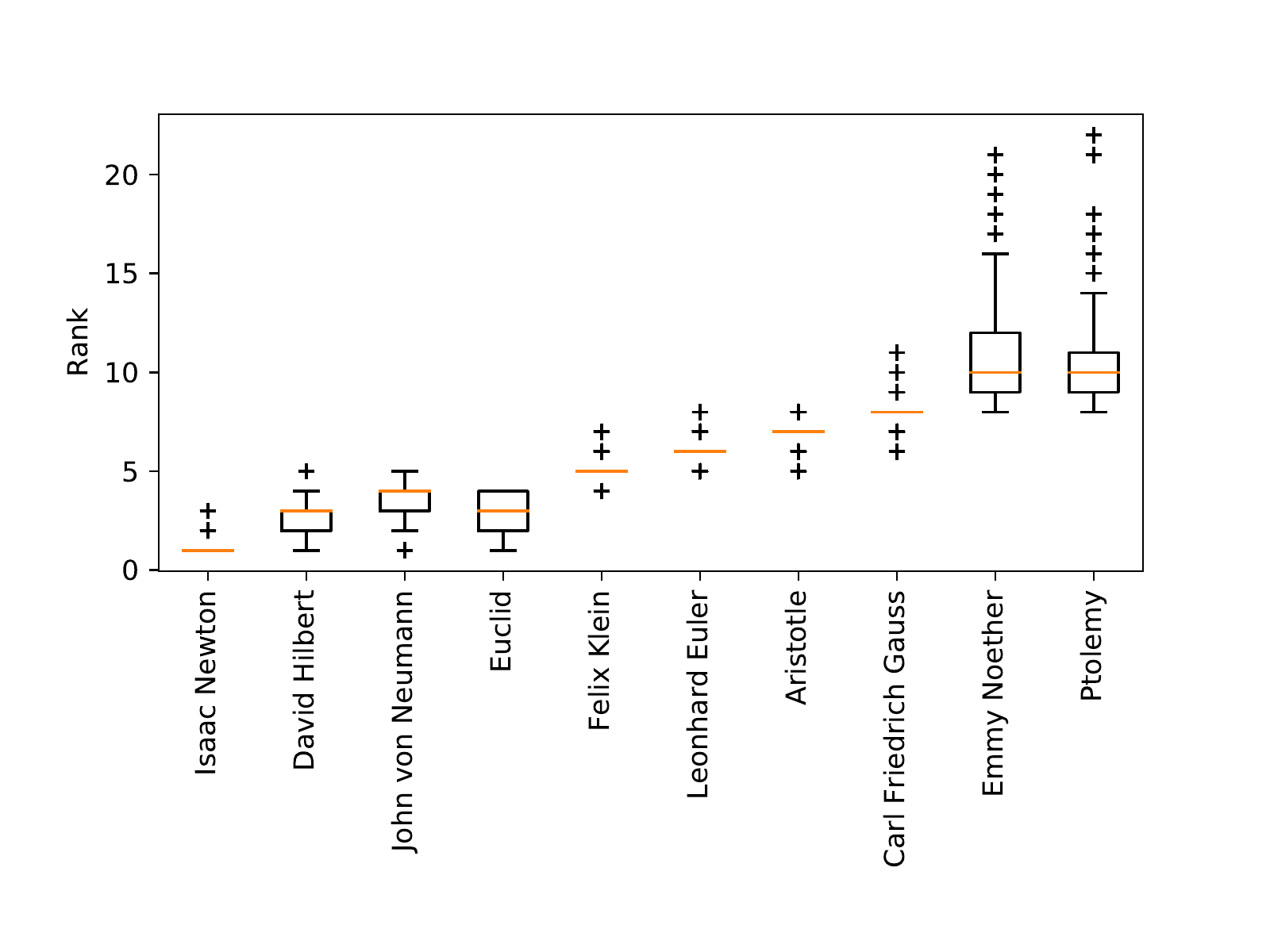}
\caption{Whisper box plot for rank of mathematicians derived from their PageRank ratings.
This is for the largest component of the 2018 data based on 1000 simulations using the noise model of \sref{snoise}.
The criteria used to place the boxes and other features of the plot are as in \fref{degreebox}.}
\label{pageboxv2}
\end{figure}

\begin{figure}[htb!]
\centering
\includegraphics[width=0.75\textwidth]{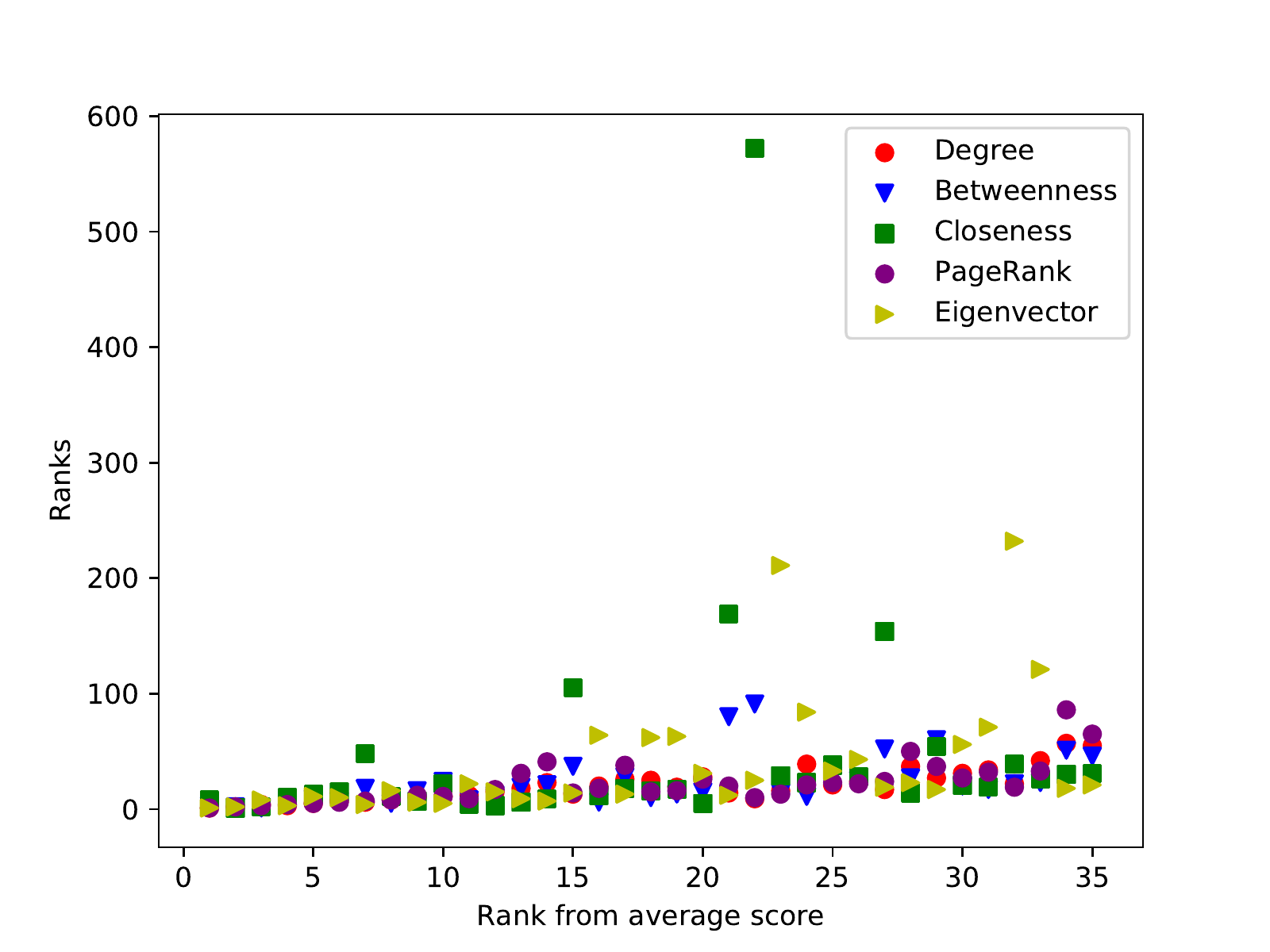}
\caption{A comparison of the rank of mathematicians under different centrality measures.  The horizontal axis is the rank of each mathematician by their average score; the top 35 are shown. Note that as the rank gets higher, there is a small but increasing variation in the ranks by different centrality measures for each mathematician.}
\label{overall35v2}
\end{figure}

\begin{figure}[htb!]
\centering
\includegraphics[width=0.75\textwidth]{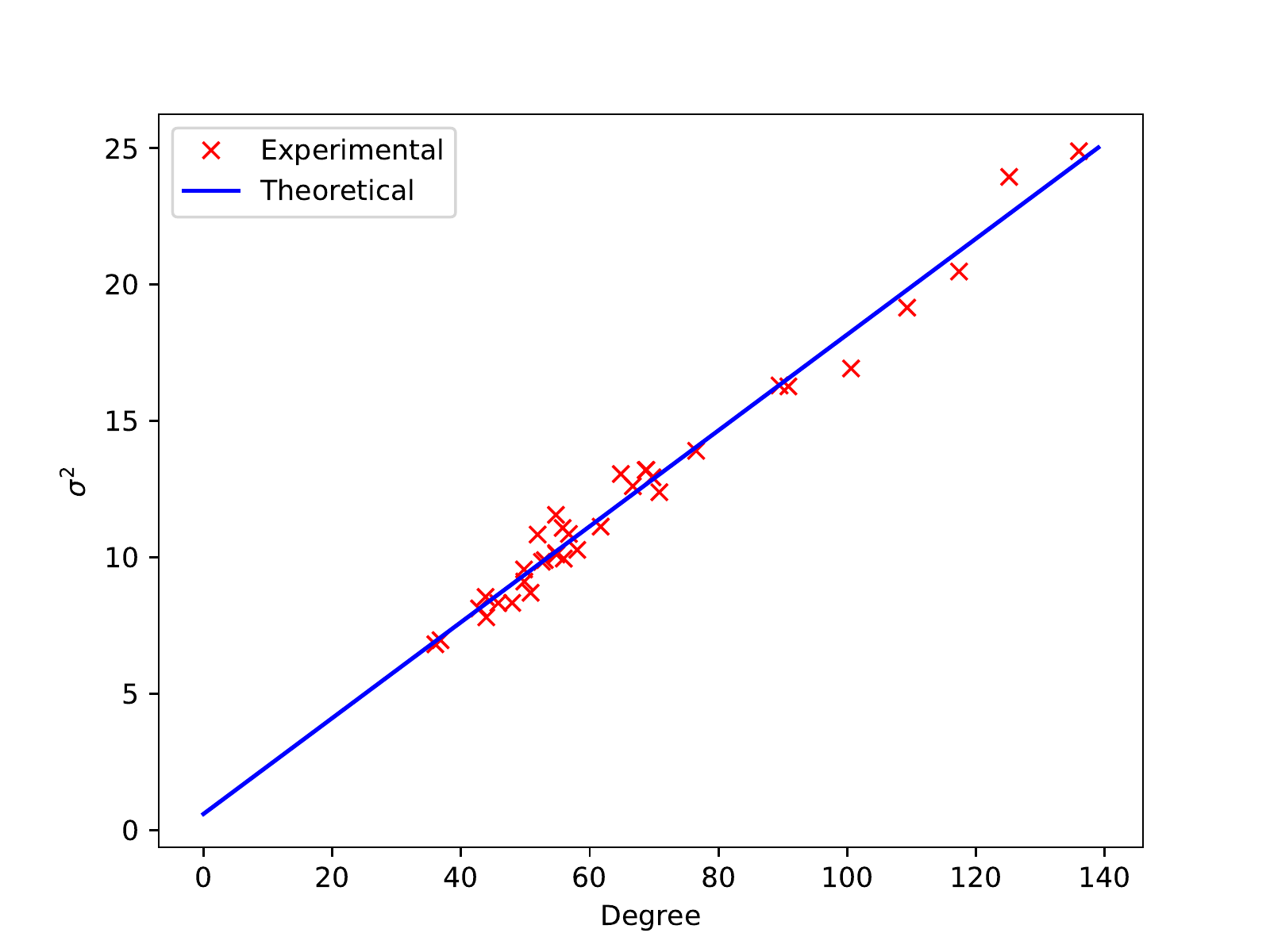}
\caption{Each cross indicates the standard deviation in degree of one node after $1000$ simulations for top 35 mathematician . The theoretical result that $\sigma \approx 0.42\sqrt{k_\mathrm{orig}}$ is compatible with this numerical result as the linear fit between variance and degree shows (an adjusted-r square value of $0.981$).}
\label{degstdv2}
\end{figure}

\begin{table}[htb!]
\centering
\resizebox{\textwidth}{!}{\begin{tabular}{cccccccc}
\hline \hline
Name                      & Degree & Betweenness & Closeness & Eigenvector & PageRank & Average mark & Rank \\ \hline
Isaac Newton              & $100$   & $77.81$     & $92.46$   & $100$       & $100$    & $99.84$      & $1$  \\
David Hilbert             & $92$    & $91.76$     & $100$     & $87.29$     & $94.15$  & $91.99$      & $2$  \\
Euclid                    & $86.1$  & $65.24$     & $92.32$   & $84.88$     & $83.27$  & $86.28$      & $3$  \\
John von Neumann          & $80.3$  & $100$       & $97.19$   & $87.28$     & $62.25$  & $80.35$      & $4$  \\
Felix Klein               & $73.7$  & $42$        & $91.65$   & $71.48$     & $55.62$  & $73.95$      & $5$  \\ \hline
Aristotle                 & $66.4$  & $26.93$     & $85.24$   & $62.14$     & $76.55$  & $66.78$      & $6$  \\
Leonhard Euler            & $65.7$  & $43.89$     & $91.35$   & $68.52$     & $60.06$  & $65.78$      & $7$  \\
Carl Friedrich Gauss      & $56.2$  & $44.41$     & $92.02$   & $58.59$     & $49.02$  & $56.26$      & $8$  \\
Ptolemy                   & $51.82$ & $8.82$      & $75.29$   & $48.67$     & $39.15$  & $52.05$      & $9$  \\
Bertrand Russell          & $51.09$ & $29.41$     & $92.47$   & $47.46$     & $69.87$  & $51.27$      & $10$ \\ \hline
Emmy Noether              & $50.36$ & $43.82$     & $93.62$   & $50.3$      & $42.41$  & $50.58$      & $11$ \\
Gottfried Wilhelm Leibniz & $50.36$ & $24.9$      & $89.43$   & $48.36$     & $70.02$  & $50.49$      & $12$ \\
Galileo Galilei           & $48.91$ & $16.45$     & $82.78$   & $46.02$     & $53.27$  & $49.02$      & $13$ \\
Archimedes                & $47.45$ & $9.45$      & $80.31$   & $41.72$     & $55.51$  & $47.66$      & $14$ \\
Hermann Weyl              & $45.26$ & $38.98$     & $94.93$   & $44.7$      & $49.74$  & $45.35$      & $15$ \\ \hline
Michael Atiyah            & $42.34$ & $33.76$     & $87.89$   & $46.76$     & $11.71$  & $42.68$      & $16$ \\
Johannes Kepler           & $41.61$ & $13.35$     & $80.79$   & $40.03$     & $43.29$  & $41.73$      & $17$ \\
G. H. Hardy               & $40.88$ & $35.26$     & $90.28$   & $45.61$     & $22.92$  & $41.14$      & $18$ \\
Georg Cantor              & $40.88$ & $26.24$     & $92.48$   & $37.35$     & $61.61$  & $41.01$      & $19$ \\
Alfred Tarski             & $40.15$ & $21.44$     & $86.52$   & $40.63$     & $32.19$  & $40.3$       & $20$ \\ \hline
Nicolas Bourbaki          & $40.15$ & $43.93$     & $91.82$   & $42.87$     & $22.58$  & $40.25$      & $21$ \\
Alexander Grothendieck    & $40.15$ & $25.73$     & $86.29$   & $42.31$     & $10.82$  & $40.25$      & $22$ \\
Alan Turing               & $38.69$ & $23.23$     & $87.94$   & $40.73$     & $29.29$  & $38.98$      & $23$ \\
Ivor Grattan-Guinness     & $38.69$ & $25.75$     & $92.44$   & $34.33$     & $66.9$   & $38.63$      & $24$ \\
Andrey Kolmogorov         & $37.96$ & $40.01$     & $90.5$    & $45.82$     & $23.5$   & $38.15$      & $25$ \\ \hline
Charles Sanders Peirce    & $37.23$ & $21.76$     & $89.98$   & $34.98$     & $54.02$  & $37.35$      & $26$ \\
Christiaan Huygens        & $36.5$  & $11.74$     & $85.01$   & $35.37$     & $47.6$   & $36.63$      & $27$ \\
Norbert Wiener            & $36.5$  & $32.58$     & $92.69$   & $39.1$      & $34.02$  & $36.6$       & $28$ \\
Richard Courant           & $35.04$ & $25.76$     & $89.47$   & $38.19$     & $23.77$  & $35.25$      & $29$ \\
Emil Artin                & $33.58$ & $28.58$     & $89.88$   & $37.22$     & $21.58$  & $33.67$      & $30$ \\ \hline
Vladimir Arnold           & $32.12$ & $39.89$     & $89.31$   & $41.16$     & $19.61$  & $32.29$      & $31$ \\
Bernhard Riemann          & $32.12$ & $21.79$     & $91.61$   & $29.56$     & $41.88$  & $32.21$      & $32$ \\
Srinivasa Ramanujan       & $31.39$ & $25.22$     & $88.38$   & $36.97$     & $16.17$  & $31.46$      & $33$ \\
Alfred North Whitehead    & $27.01$ & $14.23$     & $87.37$   & $25.73$     & $42.95$  & $27.07$      & $34$ \\
Pierre de Fermat          & $26.28$ & $13.53$     & $87.89$   & $23.31$     & $46.77$  & $26.47$      & $35$\\
\hline
\end{tabular}}
\caption{Centrality scores for top 35 mathematicians from 2018 data (without noise) given on a common scale with 100 for the largest value according to \tref{rescale}. Ordered in terms of their average score rating.}
\label{score2018}
\end{table}

\begin{table}[htb!]
\centering
\resizebox{\textwidth}{!}{\begin{tabular}{cccccccc}
\hline \hline
Name                      & Degree      & Betweenness & Closeness   & Eigenvector  & PageRank     & Average & Rank \\ \hline
Isaac Newton              & $99.84\pm0.87$ & $88.5\pm9.12$  & $95.95\pm1.37$ & $99.97\pm0.39$ & $95.99\pm6.39$ & $96.05\pm2.96$ & $1$  \\
David Hilbert             & $91.99\pm4.68$ & $93.96\pm7.34$ & $100\pm0.04$   & $87.38\pm4.95$ & $94.81\pm7.11$ & $93.63\pm3.93$ & $2$  \\
John von Neumann          & $80.35\pm4.33$ & $94.28\pm7.09$ & $97.37\pm1.12$ & $84.1\pm4.8$   & $64.15\pm6.57$ & $84.05\pm3.8$  & $3$  \\
Euclid                    & $86.28\pm4.51$ & $65.56\pm7.88$ & $94.41\pm1.23$ & $86.45\pm4.84$ & $79.48\pm7.73$ & $82.44\pm4.22$ & $4$  \\
Felix Klein               & $73.95\pm4.04$ & $51.32\pm7.14$ & $93.36\pm1.25$ & $72.17\pm4.18$ & $57.14\pm6.16$ & $69.59\pm3.7$  & $5$  \\ \hline
Leonhard Euler            & $65.78\pm3.77$ & $48\pm6.62$    & $92.91\pm1.33$ & $68.1\pm4.07$  & $57.94\pm5.81$ & $66.55\pm3.53$ & $6$  \\
Aristotle                 & $66.78\pm3.68$ & $37.76\pm5.86$ & $90.44\pm1.47$ & $64.2\pm3.86$  & $70.39\pm7.24$ & $65.91\pm3.53$ & $7$  \\
Carl Friedrich Gauss      & $56.26\pm3.44$ & $43.92\pm5.98$ & $92.76\pm1.27$ & $57.83\pm3.68$ & $47.83\pm5.11$ & $59.72\pm3.19$ & $8$  \\
Bertrand Russell          & $51.27\pm3.19$ & $33.47\pm5.39$ & $93.49\pm1.24$ & $48.17\pm3.17$ & $64.87\pm6.17$ & $58.25\pm3.11$ & $9$  \\
Gottfried Wilhelm Leibniz & $50.49\pm3.17$ & $30.17\pm4.76$ & $91.74\pm1.21$ & $48.91\pm3.27$ & $64.08\pm5.87$ & $57.08\pm2.98$ & $10$ \\ \hline
Emmy Noether              & $50.58\pm3.2$  & $44.23\pm6.51$ & $94.07\pm1.22$ & $49.8\pm3.25$  & $44.08\pm5.22$ & $56.55\pm3.26$ & $11$ \\
Hermann Weyl              & $45.35\pm2.88$ & $37.65\pm5.68$ & $94.65\pm1.13$ & $44.28\pm2.96$ & $49.35\pm5.2$  & $54.26\pm2.93$ & $12$ \\
Georg Cantor              & $41.01\pm2.85$ & $26.2\pm4.37$  & $92.98\pm1.13$ & $38.01\pm2.81$ & $57.65\pm5.61$ & $51.17\pm2.76$ & $13$ \\
Galileo Galilei           & $49.02\pm3.09$ & $23.16\pm4.21$ & $87.52\pm1.48$ & $47.44\pm3.19$ & $48.54\pm5.44$ & $51.14\pm2.78$ & $14$ \\
Ivor Grattan-Guinness     & $38.63\pm2.63$ & $25.96\pm4.61$ & $93.38\pm1.13$ & $34.85\pm2.52$ & $61.48\pm5.47$ & $50.86\pm2.71$ & $15$ \\ \hline
Archimedes                & $47.66\pm3.13$ & $17.55\pm3.89$ & $86.16\pm1.61$ & $43.65\pm3.06$ & $50.47\pm5.92$ & $49.1\pm2.87$  & $16$ \\
Ptolemy                   & $52.05\pm3.15$ & $18.21\pm3.93$ & $83.47\pm1.91$ & $50.37\pm3.22$ & $36.73\pm5.01$ & $48.17\pm2.77$ & $17$ \\
Charles Sanders Peirce    & $37.35\pm2.53$ & $23.5\pm4.12$  & $91.54\pm1.24$ & $35.5\pm2.52$  & $49.84\pm5.34$ & $47.55\pm2.57$ & $18$ \\
Nicolas Bourbaki          & $40.25\pm2.9$  & $38.06\pm5.35$ & $91.63\pm1.24$ & $41.52\pm3.05$ & $25.35\pm3.9$  & $47.36\pm2.74$ & $19$ \\
G. H. Hardy               & $41.14\pm2.71$ & $34.34\pm4.84$ & $90.34\pm1.35$ & $44.09\pm2.98$ & $24.5\pm3.64$  & $46.88\pm2.51$ & $20$ \\ \hline
Andrey Kolmogorov         & $38.15\pm2.75$ & $35.8\pm4.91$  & $90.4\pm1.34$  & $43.34\pm3.15$ & $24.71\pm3.44$ & $46.48\pm2.52$ & $21$ \\
Norbert Wiener            & $36.6\pm2.55$  & $30.45\pm4.5$  & $92.28\pm1.2$  & $38.06\pm2.72$ & $34.31\pm4.1$  & $46.34\pm2.49$ & $22$ \\
Johannes Kepler           & $41.73\pm2.84$ & $18.74\pm3.81$ & $86.14\pm1.66$ & $40.92\pm2.92$ & $39.88\pm4.88$ & $45.48\pm2.64$ & $23$ \\
Michael Atiyah            & $42.68\pm2.78$ & $32.69\pm4.96$ & $88.73\pm1.46$ & $45.29\pm3.01$ & $15.73\pm3.14$ & $45.02\pm2.53$ & $24$ \\
Alfred Tarski             & $40.3\pm2.69$  & $23.26\pm3.78$ & $87.75\pm1.35$ & $40.38\pm2.82$ & $31.56\pm4.28$ & $44.65\pm2.46$ & $25$ \\ \hline
Alan Turing               & $38.98\pm2.73$ & $23.44\pm4.09$ & $88.53\pm1.35$ & $40.01\pm2.88$ & $29.47\pm4$    & $44.09\pm2.44$ & $26$ \\
Christiaan Huygens        & $36.63\pm2.55$ & $15.06\pm3.18$ & $87.39\pm1.39$ & $35.95\pm2.59$ & $43.16\pm4.59$ & $43.64\pm2.29$ & $27$ \\
Bernhard Riemann          & $32.21\pm2.38$ & $20.69\pm3.64$ & $91.88\pm1.13$ & $29.94\pm2.32$ & $40.48\pm4.34$ & $43.04\pm2.27$ & $28$ \\
Alexander Grothendieck    & $40.25\pm2.78$ & $29.47\pm4.98$ & $88.17\pm1.57$ & $41.1\pm2.92$  & $15.16\pm3.07$ & $42.83\pm2.53$ & $29$ \\
Richard Courant           & $35.25\pm2.47$ & $25.87\pm4.06$ & $89.44\pm1.25$ & $37.06\pm2.67$ & $25.38\pm3.48$ & $42.6\pm2.27$  & $30$ \\ \hline
Vladimir Arnold           & $32.29\pm2.32$ & $31.19\pm4.48$ & $89.4\pm1.49$  & $38.7\pm2.77$  & $20.74\pm3.33$ & $42.46\pm2.31$ & $31$ \\
Emil Artin                & $33.67\pm2.47$ & $26.6\pm4.12$  & $89.74\pm1.31$ & $35.9\pm2.67$  & $23.28\pm3.55$ & $41.84\pm2.35$ & $32$ \\
Srinivasa Ramanujan       & $31.46\pm2.39$ & $23.45\pm3.43$ & $88.12\pm1.34$ & $35.41\pm2.69$ & $17.56\pm3.09$ & $39.2\pm2.14$  & $33$ \\
Pierre de Fermat          & $26.47\pm2.09$ & $13.18\pm2.89$ & $89.29\pm1.3$  & $24.16\pm2.02$ & $42.52\pm4.28$ & $39.12\pm2.07$ & $34$ \\
Alfred North Whitehead    & $27.07\pm2.18$ & $14.93\pm2.95$ & $88.61\pm1.22$ & $25.92\pm2.19$ & $38.91\pm4.29$ & $39.09\pm2.07$ & $35$\\
\hline
\end{tabular}}
\caption{Centrality scores for top 35 mathematicians derived from the the noise model described of \sref{snoise} applied to the 2018  data with $p=0.1$ for 1000 simulations. The mean value and one standard deviation is quoted for each centrality measure for each mathematician. As the scores for each run are always rescaled so that the largest value is 100, explaining why the value quoted for any one centrality measure is always less than 100. The column marked average gives the average over the five named centrality measures with associated standard deviation. Mathematicians are ordered in terms of this average and the ranks given  are in terms of this average over centrality values.}
\label{tcentralitynoise2018}
\end{table}

\begin{table}[htb!]
\centering
\begin{tabular}{l||c|c|c|c|c|c}
Top 35          & Degree          & PageRank          & Eigenvector          & Betweenness          & Closeness          & Average          \\ \hline \hline
Degree          & 1.00            & \pvalue{0.98}     & \pvalue{0.75}        & \pvalue{0.78}        & \pvalue{0.36}      & \pvalue{0.96}    \\ \hline
PageRank        & \svalue{0.92}   & 1.00              & \pvalue{0.64}        & \pvalue{0.67}        & \pvalue{0.32}      & \pvalue{0.95}    \\ \hline
Eigenvector     & \svalue{0.66}   & \svalue{0.44}     & 1.00                 & \pvalue{0.20}        & \pvalue{0.35}      & \pvalue{0.79}    \\ \hline
Betweenness     & \svalue{0.50}   & \svalue{0.85}     & \svalue{0.44}        & 1.00                 & \pvalue{0.74}      & \pvalue{0.87}    \\ \hline
Closeness       & \svalue{0.31}   & \svalue{0.41}     & \svalue{0.30}        & \svalue{0.78}        & 1.00               & \pvalue{0.56}    \\ \hline
Average         & \svalue{0.87}   & \svalue{0.81}     & \svalue{0.75}        & \svalue{0.69}        & \svalue{0.62}      & 1.00
\end{tabular}
\caption{The correlation values for the 35 top mathematicians as defined by the average of centrality scores in the 2018 data. The upper right triangle contains the Pearson correlation values (in blue) while the lower left triangle contains the Spearman correlation values (in red italics). Note that for both cases the degree and PageRank are particularly well correlated as are Betweenness and Closeness.}
\label{tPearsonSpearman35_2018}
\end{table}

\begin{table}[htb!]
\centering
\begin{tabular}{l||c|c|c|c|c|c}
largest component         & Degree          & PageRank          & Eigenvector          & Betweenness          & Closeness          & Average          \\ \hline \hline
Degree          & 1.00            & \pvalue{0.98}     & \pvalue{0.82}        & \pvalue{0.86}        & \pvalue{0.57}      & \pvalue{0.95}    \\ \hline
PageRank        & \svalue{0.95}   & 1.00              & \pvalue{0.74}        & \pvalue{0.92}        & \pvalue{0.51}      & \pvalue{0.91}    \\ \hline
Eigenvector     & \svalue{0.63}   & \svalue{0.42}     & 1.00                 & \pvalue{0.49}        & \pvalue{0.94}      & \pvalue{0.86}    \\ \hline
Betweenness     & \svalue{0.88}   & \svalue{0.87
}     & \svalue{0.70}        & 1.00                 & \pvalue{0.39}      & \pvalue{0.82}    \\ \hline
Closeness       & \svalue{0.70}   & \svalue{0.52}     & \svalue{0.56}        & \svalue{0.59}        & 1.00               & \pvalue{0.78}    \\ \hline
Average         & \svalue{0.85}   & \svalue{0.69}     & \svalue{0.90}        & \svalue{0.69}        & \svalue{0.96}      & 1.00
\end{tabular}
\caption{The correlation values for mathematicians in largest component as defined by the average of centrality scores in the 2018 data. The upper right triangle contains the Pearson correlation values (in blue) while the lower left triangle contains the Spearman correlation values (in red italics). Note that for both cases the degree and PageRank are particularly well correlated as are Betweenness and Closeness.}
\label{tPearsonSpearmanFull_2018}
\end{table}

\end{document}